# Exploiting Structure in the Boolean Weighted Constraint Satisfaction Problem: A Constraint Composite Graph-Based Approach

Hong Xu

Sven Koenig

T. K. Satish Kumar

**University of Southern California**

This is a copy of Hong Xu's PhD dissertation submitted to the University of Southern California (Department of Physics) in December 2018. It is presented here as an article coauthored by him and his PhD advisers Sven Koenig and T. K. Satish Kumar.

# EXPLOITING STRUCTURE IN THE BOOLEAN WEIGHTED CONSTRAINT SATISFACTION PROBLEM: A CONSTRAINT COMPOSITE GRAPH-BASED APPROACH

by

Hong Xu

---

A Dissertation Presented to the

FACULTY OF THE GRADUATE SCHOOL

UNIVERSITY OF SOUTHERN CALIFORNIA

In Partial Fulfillment of the

Requirements for the Degree

DOCTOR OF PHILOSOPHY

(PHYSICS)

May 2019

# Acknowledgments

I would like to thank my advisors Dr. T. K. Satish Kumar and Dr. Sven Koenig for their persistent guidance and support on research, academic principles, and especially their willingness to accept me as their student, who had approached them with a very different academic background. I also would like to thank them for helping me revise this document.

I also would like to thank my family members for their understanding, support, and encouragement. Specifically, I would like to thank my wife for her extraordinary patience.

I also would like to thank junior PhD, master's, and undergraduate students who have been advised by me, including Cheng Cheng, Dylan Johnke, Masaru Nakajima, Kexuan Sun, Zhi Wang, and Ka Wa Yip. Many ideas that excited me would not be published without their help.

I also would like to thank other group members and collaborators, including Liron Cohen, Ferdinando Fioretto, Itay Hen, Craig A. Knoblock, Anoop Kumar, Jiaoyang Li, Hang Ma, Craig Milo Rogers, Tansel Uras, and Tang Zheng, for their intriguing discussions from time to time.

I also would like to thank qualifying and defense committee members Gene Bickers, Stephan Haas, Itay Hen, and Aiichiro Nakano for their valuable comments and suggestions. In particular, I would like to thank Dr. Aiichiro Nakano for his extra patience and help with my qualifying exam and defense.

I would like to thank the National Science Foundation (NSF). The research at the University of Southern California was supported by NSF under grant numbers 1724392, 1409987, and 1319966. The views and conclusions contained in this document are those of the authors and should not be interpreted as representing the official policies, either expressed or implied, of the sponsoring organizations, agencies or the U.S. government.

*The order in which I have acknowledged everyone is no indication of any order in which I feel thankful to them.*

# Contents

# List of Figures

# List of Tables

# Abstract

What is “structure”? And how can we exploit it in combinatorial optimization? These are the fundamental questions addressed in this dissertation for many reasoning tasks on complex physical and non-physical systems.

Reasoning tasks involving system design, state estimation, and prediction can be cast as combinatorial optimization problems (COPs). Traditionally, different kinds of COPs have been solved using dedicated algorithms. While such algorithms are certainly valuable, they have some important drawbacks. First, the algorithms developed for very specific subclasses are not applicable to real-world instances if they don’t belong to these subclasses. Second, different research communities working on very specific COPs could be oblivious of each other’s works and end up developing different terminologies and techniques for solving the same problem. Therefore, a general mathematical framework that captures a wide variety of COPs facilitates informedness of different research communities, cross-fertilization of different perspectives, and a wider applicability to real-world domains.

The weighted constraint satisfaction problem (WCSP) is a general mathematical framework for COPs. It not only subsumes important COPs studied in many different research communities but also has a strong representational power useful for reasoning about complex physical and non-physical systems. Such systems include classical spin glass systems, percolation theory-characterized systems, and social networks, among many other examples.

Yes, the WCSP is representationally very powerful. But how can we design algorithms for solving it efficiently? Isn't the very generality of the WCSP a curse? Are we up against a very general and intractable problem? While these questions are certainly valid for the general WCSP, the proposal in this dissertation is to exploit "structure". Imagine two subclasses of COPs, Class A and Class B. If Class B is more general than Class A, it is likely that we can build specialized algorithms for solving instances from Class A more efficiently. But the hallmark of a good algorithm for solving instances from Class B is its ability to imitate the specialized algorithm if the input is in fact from Class A. Such an algorithm is said to exploit "structure." Although the formal definition of "structure" is elusive, the idea is to create a general-purpose algorithm for solving the WCSP that automatically simulates more specialized algorithms for subclasses of the WCSP.

We can talk about two types of structure in the WCSP. The macro structure or the graphical structure represents which variables interact with each other. The micro structure or the numerical structure represents how the variables interact with each other. Two completely different schools of thought have led to algorithmic techniques that exploit either the macro structure or the micro structure, but not both simultaneously. In 2008, the quest for a unifying mathematical framework that represents the macro structure as well as the micro structure of a WCSP was settled by the novel idea of the constraint composite graph (CCG). The CCG of a WCSP instance is an undirected graph that uses the same variables as the WCSP instance and an auxiliary set of variables to capture structure. Solving the WCSP is equivalent to solving the minimum weighted vertex cover (MWVC) problem on its associated CCG. Although the CCG can be constructed very efficiently, not much work was done until now in exploiting this transformation for theoretical or practical gains.

In this dissertation, we raise and answer three research questions: Are there any theoretical advantages of the CCG other than for identifying tractable classes of the WCSP? Is there any practical usefulness of the CCG? Is it promising to extend the CCG to the WCSP with non-Boolean variables? We answer all three questions affirmatively. We answer the first question by proving new theoretical properties of the CCG. We answer the second question by efficiently implementing the CCG construction procedure and conducting experiments. We answer the third question by proposing new encodings for non-Boolean variables and preliminarily demonstrate their promisingness.

On the one hand, the generality of the WCSP is intended to make it widely applicable and bring together researchers from different research communities. On the other hand, our theory of the CCG reduces it to a very specific COP, i.e., the MWVC problem. This transformation not only holds the remarkable promise of a general-purpose algorithm that can exploit structure in the WCSP but also emphasizes the importance of the MWVC problem as a substrate COP. Specifically, in this dissertation, we show how the CCG-based transformation can be used to: (a) kernelize a WCSP instance, i.e., fix the optimal values of a subset of its variables using a maxflow procedure even before search is initiated, (b) improve the efficiency of the min-sum message passing algorithm, (c) make use of integer linear programming (ILP) solvers, and (d) solve COPs on quantum annealers more effectively. In addition, because our algorithms solve general COPs in the WCSP framework more efficiently on classical computers, we provide better baselines for comparison against quantum computers, whose true efficiency over classical computers is still debated.

# Chapter 1

# Introduction

## 1.1 Motivation

*Combinatorial optimization problems* (COPs) use discrete variables and the interactions between them to characterize real-world and abstract systems. They usually cannot be solved scalably by simply applying exhaustive search, since the amount of time required by such search increases exponentially as the number of variables increases, and therefore intelligent algorithms are usually required for solving them. Fortunately, after decades of efforts by researchers, we know that there is a rich class of COPs in P, meaning that there are known algorithms to solve them in polynomial time. However, on the other hand, we also know that there is another rich class of COPs that are known to be NP-hard, meaning that no algorithm can solve them in polynomial time under the assumption of P≠NP. Yet, in practice, we can still often solve many of them quickly, thanks to intelligent algorithms that exploit their structure.

What is "structure"? And how can we exploit it in combinatorial optimization? These are the fundamental questions addressed in this dissertation for many reasoning tasks on complex physical and non-physical systems. In this section, we present the motivation to study these questions. We first present two reasons for studying COPs: (a) Many problems in complex physical and non-physical systems have a deep nexus to COPs, and (b) improving algorithms for solving COPs helps advance the state of the debates on whether the quantum annealer has a true advantage over classical computers. After that, we discuss the specificity of COPs, i.e., how COPs have been categorized and studied individually. Then we discuss the generality of the *weighted constraint satisfaction problem* (WCSP) and why we should study it. We later discuss the ways to solve the WCSP and two types of structure—macro and micro structure—in the WCSP. Finally, we discuss the *constraint composite graph* (CCG) for the WCSP and how it simultaneously captures the two types of aforementioned structure.

### 1.1.1 Why Studying COPs is Important

Many classical complex physical and non-physical systems, such as classical spin glass systems, percolation theory-characterized systems, and social networks, have a deep nexus to COPs due to their discrete nature. The task of computing ground energy states of a spin glass system can be modeled as a COP in which each discrete variable represents a spin and the optimization goal characterizes interactions between them. In fact, many discrete physics models, such as the *random Ising model*, have been frequently used as test beds for techniques that solve COPs (De Simone et al. 1995). COPs such as the *minimum spanning tree problem* are common tools to study percolation theory (Alexander 1995; Bezuidenhout, Grimmett, and Löffler 1998). In social networks and citation networks, two categories of complex systems that are commonly studied in physics (Golosovsky 2017; Wu et al. 2018), and many problems such as the maximum influence problem (Kempe, Kleinberg, and Tardos 2003) and community detection (Kanawati 2014) can be modeled as COPs. Therefore, improving COP solving in general may help us solve and understand these systems.

On the other hand, *quantum annealers*, the only commercially available physical realization of quantum computers nowadays, by their nature solve *quadratic* *unconstrained binary optimization* (QUBO) problems, a subset of COPs. A common way to use quantum annealers for solving COPs in general is via *hybrid quantum-classical algorithms* (HQCAs), a class of algorithms that interleave the use of quantum annealers and classical computers. However, despite their quantum nature, their true efficiency is controversial: Shor's algorithm (Shor 1994), the only known polynomial-time algorithm on quantum computers for solving NP-complete problems until today, cannot be implemented on quantum annealers. It is doubtful that there may exist an HQCA that has a super-polynomial speedup compared to existing algorithms on classical computers. Indeed, none of such algorithms have been confirmed to exist. For this reason, to disprove its efficiency, a lot of research has been put into solving COPs using algorithms on classical computers more efficiently.

### 1.1.2 Specificity of COPs

Traditionally, different COPs have been solved independently using dedicated algorithms. For example, the (weighted) Max-Cut problem, which is equivalent to the problem of finding a minimum energy state of an Ising system, has its dedicated algorithms such as (Gao, Zeng, and Dong 2008; Kochenberger et al. 2013; Krishnan and Mitchell 2006; Rendl, Rinaldi, and Wiegele 2008); the Max-SAT problem, which can be used to model problems in logic with uncertainty, has its dedicated algorithms such as EvaSolver (Narodytska and Bacchus 2014), OpenWBO (Martins, Manquinho, and Lynce 2014), and LMHS (Saikko, Berg, and Järvisalo 2016); and the maximum (weighted) clique problem, which has been used to help solve many important problems, such as graph coloring, has its dedicated algorithms such as Cliquer (Niskanen and Östergård 2003), FastWClq (Cai and Lin 2016), MWCLQ (Fang, Li, and K. Xu 2016), and OTClique (Shimizu et al. 2017).

While such algorithms are valuable, they have important drawbacks. First, these algorithms are designed to solve a specific problem—they are not applicable to a different problem, often even if it is only slightly different. Secondly, different research communities working on different specific COPs could be oblivious of each other's works and often end up developing different terminologies and techniques for solving the same or similar problems. Therefore, a mathematical framework that captures a wide variety of COPs not only leads to wider applicability, but also facilitates informedness of different research communities and cross-fertilization of different perspectives.

### 1.1.3 Generality of the Weighted Constraint Satisfaction Problem

The *weighted constraint satisfaction problem* (WCSP) is a general mathematical framework of COPs. It subsumes many important COPs such as the aforementioned (weighted) Max-SAT, (weighted) Max-Cut, and maximum (weighted) clique problems. The *constraint satisfaction problem* (CSP) is a classic combinatorial problem. It consists of a set of discrete variables of finite domains and a set of constraints, each of which allows and forbids certain assignments of values to a subset of variables. The WCSP can be viewed as an optimization variant of the CSP where constraints are no longer "hard," but associated with non-negative costs (weights). The goal of the WCSP is to find an assignment of values to the variables that minimizes the sum of the costs (Bistarelli et al. 1999).

Studying the WCSP brings together different research communities. While the terminology was proposed in the *constraint programming* (CP) community, the problem itself is also known in different communities by different names and has been solved and understood using different algorithms. In CP, *branch-and-bound* (BnB) search has been traditionally used to solve the WCSP (Hurley et al. 2016; Marinescu and Dechter 2006; Marinescu and Dechter 2007). In physics, the WCSP can be seen as a general framework that characterizes classical systems with $n$-body interactions. Simulated annealing and Monte Carlo algorithms, due to their embodiment of physical principles, are commonly used to understand physical systems (e.g., (Ferrenberg, J. Xu, and Landau 2018; Heim et al. 2015; Wauters et al. 2017)). In probabilistic reasoning, the WCSP is known to be equivalent to the *maximum-a-posteriori* (MAP) problem on a Markov random field. Message passing algorithms are commonly used to solve this problem (Koller and Friedman 2009). In multi-agent systems, the WCSP is known as the *distributed constraint optimization problem* (DCOP). Here, distributed versions of BnB (e.g., (Yeoh, Felner, and Koenig 2010)) and message passing algorithms (Cohen and Zivan 2018; Farinelli et al. 2008) have been used to solve this problem. Studies on the WCSP, therefore, are of interest for many different fields and facilitate the bond between physics and computer science.

The WCSP has a strong representational power—it is a general powerful tool that has been used to model many important COPs in many complex physical and non-physical systems. For example, in condensed matter physics, the WCSP can be used to find the ground state in Potts model and its generalizations such as the multi-body $p$-spin model (Mézard and Montanari 2009, p. 155) (as illustrated in Figure 1.1); in biophysics, it can be used to locate motifs in RNA sequences (Zytnicki, Gaspin, and Schiex 2008) (as illustrated in Figure 1.2); in information theory, it can be used to reconstruct a message sent through a noisy channel using error correcting codes (Yedidia, Freeman, and Weiss 2003); in social science, it can be used to solve Schelling's model of segregation (Easley and Kleinberg 2010); and in computer vision applications, it can be used to solve energy minimization problems towards tasks such as image restoration, total variation minimization, and panoramic image stitching (Kolmogorov 2005).

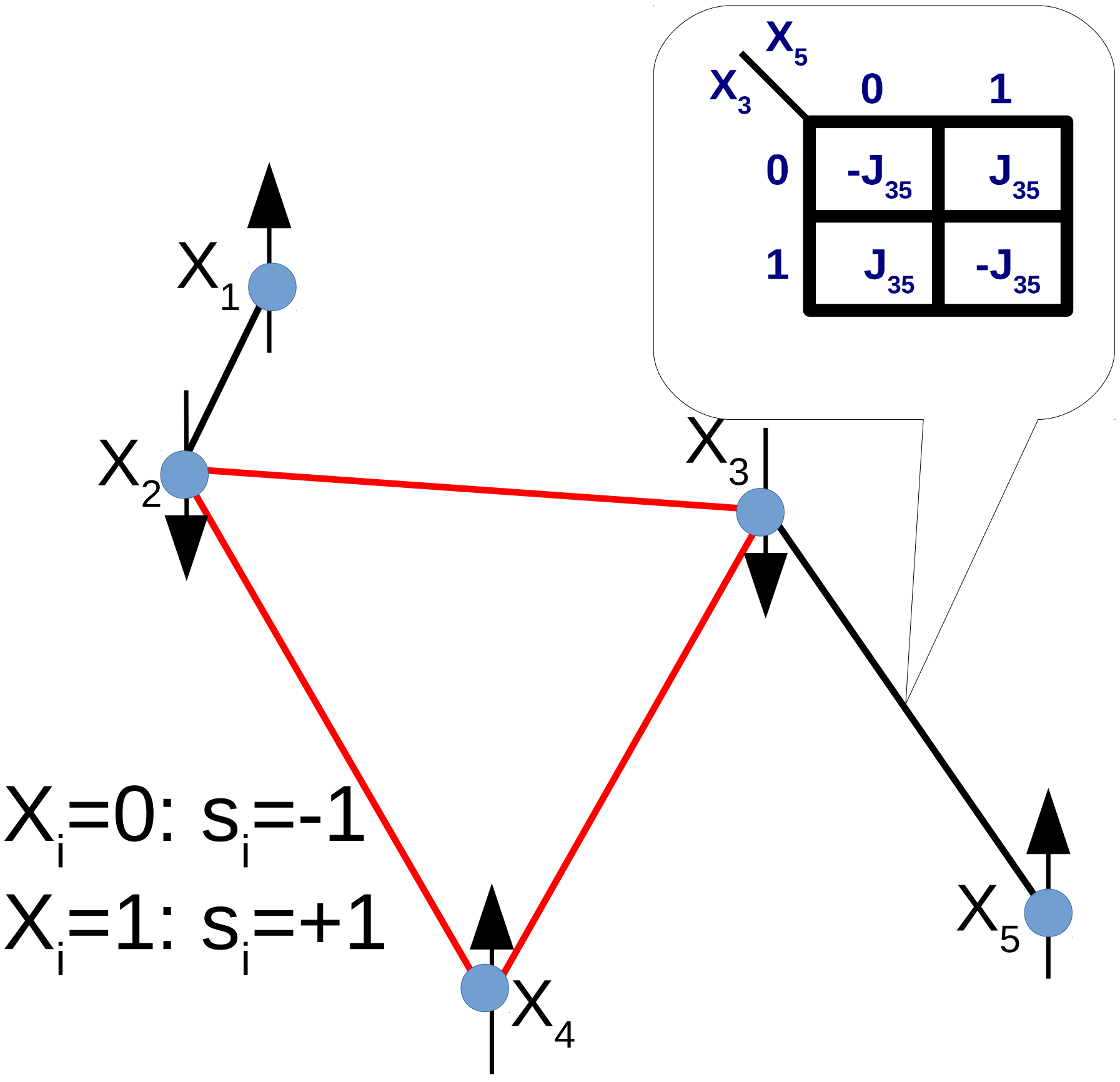

Figure 1.1: Illustrates the problem of finding the ground state energy of a random Potts model seen as a WCSP instance. Each solid circle represents a spin. Each black edge represents a two-body interaction. The three red edges represent a three-body interaction.

### 1.1.4 Solving the WCSP

How can we design algorithms for solving the WCSP efficiently? Isn't the very generality of the WCSP a curse? Are we up against a hard intractable problem?

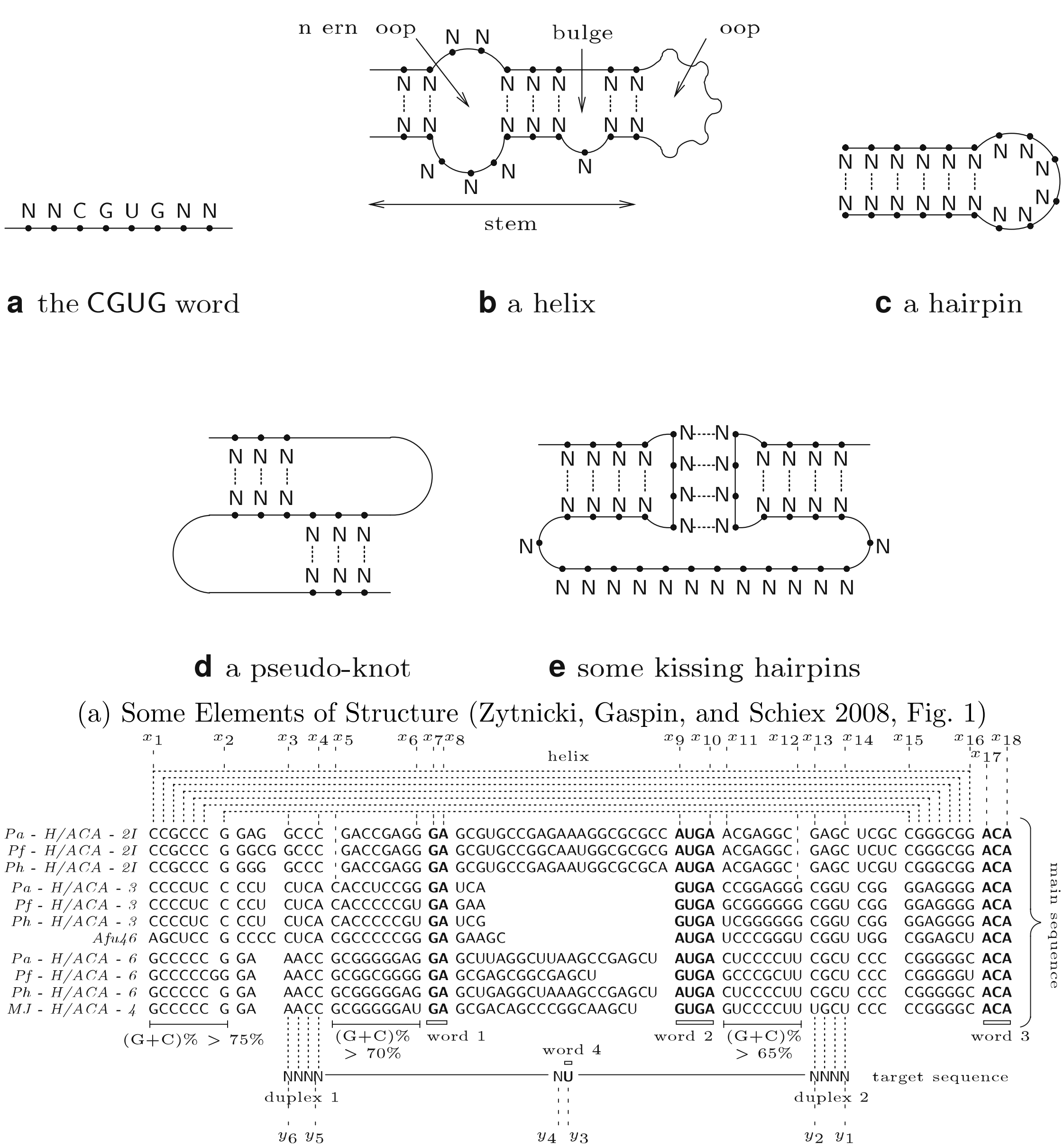

(b) RNA Motif Localization as the WCSP (Zytnicki, Gaspin, and Schiex 2008, Fig. 2a)

Figure 1.2: Illustrates locating motifs in RNA sequences. Figure (a)a shows an example word in an RNA sequence. Figures (a)b-e show some elements of structure that have different stabilities. (b) illustrates the formulation of the RNA motif localization problem as the WCSP. In the WCSP formulation, each variable is the location of a key position in an element of structure in an RNA sequence. Costs in constraints characterize the stability associated with different elements of structure.

While these questions are certainly valid, in this dissertation, we exploit "structure." Imagine two subclasses of COPs, Class A and Class B. If Class B is more general than Class A, it is likely that we can build specialized algorithms for solving instances from Class A more efficiently. But the hallmark of a good algorithm for solving instances from Class B is its ability to imitate the specialized algorithm if the input is in fact from Class A. Such an algorithm is said to exploit "structure." Although the formal definition of "structure" is elusive, the idea is to create a general-purpose algorithm for solving the WCSP that automatically simulates more specialized algorithms for subclasses of the WCSP.

By its nature, the WCSP has two types of structure: the macro structure, or the graphical structure, and the micro structure, or the numerical structure. The graphical structure characterizes which variables interact locally and the numerical structure characterizes the details of each local interaction. For example, if the WCSP is used to characterize a spin glass system, then the graphical structure is a fully connected graph under a mean-field assumption and is sparse under a nearest-neighbor assumption; the numerical structure for spin interactions is symmetric for an Ising system and is asymmetric for a system with more than one type of spin.

Unfortunately, traditional algorithms do not exploit both types of structure simultaneously. Rather, they either focus on one of them or exploit them individually. For example, one traditional way in which this has been done is by studying the underlying *variable-interaction graphs* (Dechter 1992). The variable-interaction graph incorporates basic information about which variables are constrained with which other variables in the problem instance, and this "locality" information can be exploited in solution procedures that employ dynamic programming. Despite its apparent usefulness, the variable-interaction graph does not represent/capture information about the costs in the weighted constraints, and therefore cannot be used to characterize/exploit any important combinatorial structure that might be present in them. In fact, there are many fundamental combinatorial problems—like the hypergraph min-*st*-cut problem—that can be formulated as the Boolean WCSP, and that are tractable not by virtue of the graphical structure in their associated variable-interaction graphs, but by virtue of the numerical structure in their associated weighted constraints.

### 1.1.5 The Constraint Composite Graph

In 2008, the quest for a unifying mathematical framework that represents the macro structure as well as the micro structure of a WCSP was settled by the novel idea of the *constraint composite graph* (CCG). The CCG of a WCSP instance is an undirected graph that uses the same variables as the WCSP instance and an auxiliary set of variables to capture structure. For a WCSP instance, it is equivalent to solve the *minimum weighted vertex cover* (MWVC) problem on its CCG. It has many interesting properties: It can be constructed in polynomial time; it is always tripartite; and its construction can be done on individual constraints and then be merged (meaning that its construction can be easily made parallel and incremental). It has also been used to discover some tractable subclasses of the WCSP. However, since then, there has not been much work done until today in exploiting this transformation for theoretical or practical gains. Due to these reasons, the development and discovery of the usefulness of the CCG for the WCSP have become important and interesting.

## 1.2 Hypothesis and Research Questions

In this dissertation, we hypothesize that the CCG can help algorithms discover structure in the Boolean WCSP and therefore help solve it faster and with better theoretical guarantees. Under this hypothesis, we propose three research questions in this section.

While the CCG has been used to identify some tractable subclasses of the Boolean WCSP, just like other algorithmic techniques, its theoretical usefulness can be far beyond that. Hence, the first research question is:

**Q1**: Is there any theoretical usefulness of the CCG other than for identifying tractable subclasses of the Boolean WCSP?

Although the CCG has demonstrated some theoretical usefulness for the Boolean WCSP, there is no known implementation of it and its practical usefulness remains unknown and unexplored. Hence, the second research question is:

**Q2**: Is there any practical usefulness of the CCG for the Boolean WCSP?

While the CCG is promising, its applicability has been limited to the Boolean WCSP. However, many real-world problems can be more easily modeled as the WCSP with non-Boolean variables. The extension of the CCG to the WCSP with non-Boolean variables is understudied and also inefficient (Kumar 2008b). Hence, the third research question is:

**Q3**: Can the CCG be efficiently extended to the WCSP with non-Boolean variables?

## 1.3 Overview and Contributions of This Dissertation

In this dissertation, we attempt to answer the three aforementioned research questions:

1. We answer **Q1** by demonstrating the theoretical benefits that the CCG brings. In particular, we show that it enables the *Nemhauser-Trotter reduction* (NT reduction) and prove that it improves the ILP encoding of the Boolean WCSP.

2. We answer **Q2** by implementing the CCG construction algorithm as described in (Kumar 2008a)—along with various improvements—and experimentally evaluate various algorithms for solving the (Boolean) WCSP with the help of the CCG. In particular, we experimentally demonstrate that several CCG-based algorithms are more advantageous than their counterparts that work directly on the (Boolean) WCSP, including the NT reduction, the *min-sum message passing* (MSMP) algorithm, and the HQCA.

3. We answer **Q3** by proposing three new non-Boolean variable encodings, namely the *binary number-based encoding*, the *direct symmetric encoding*, and the *clique-based encoding*, for the WCSP with non-Boolean variables. We also compare them using theoretical arguments and preliminary experimental results. We pose it as a promising future work of this dissertation.

This dissertation is organized as follows. In Chapter 2, we introduce background material, including that on the CCG. In Chapter 3, we demonstrate that the CCG enables the use of the NT reduction, a polynomial-time procedure that reduces problem sizes for the MWVC problem, on the Boolean WCSP. In Chapter 4, we experimentally demonstrate that the MSMP algorithm is more efficient when applied to the CCG of a Boolean WCSP instance than on the Boolean WCSP instance itself. In Chapter 5, we demonstrate the theoretical advantage of the CCG-based ILP encoding of a Boolean WCSP instance over other ILP encodings, and experimentally compare three different ILP encodings. In Chapter 6, we show that the CCG-based HQCA for solving the Boolean WCSP is more advantageous than a few other baseline HQCAs. In Chapter 7, we point out that extending the CCG for the WCSP with non-Boolean variables can be promising. We do this by proposing three non-Boolean variable encodings and demonstrating the usefulness of the CCG on the WCSP with non-Boolean variables in preliminary experiments. Finally, in Chapter 8, we draw our conclusions and discuss other potential future research directions. In each of Chapters 3 to 7, we also point out the potential impact of improving the specific algorithm presented in that chapter.

In summary, in this dissertation, we address **Q1** and **Q2** in Chapters 3 to 6, and **Q3** in Chapter 7.

# Chapter 2

# Background

## 2.1 Basics in Graph Theory

We denote an undirected graph using $G = \langle V, E \rangle$, where $V$ is a set of vertices and $E$ is a set of edges. A vertex-weighted undirected graph is an undirected graph with a non-negative weight (integer or real number) associated with each vertex. We denote a vertex-weighted undirected graph using $G = \langle V, E, w \rangle$, where $V$ and $E$ have the same meaning as before and $w$ is a function that maps a vertex to a non-negative integer or real number. (For notational simplicity, we also write $w_i$ short for $w(v_i)$, where $v_i$ is a vertex in $V$.) The weight of a subset of vertices $S \subseteq V$ is the sum of all weights of vertices in $S$.

A set of vertices $S \subseteq V$ is an *independent set* (IS) of an undirected graph $G = \langle V, E \rangle$ if and only if no two vertices in $S$ are connected by an edge, i.e., $\forall u, v \in S : (u, v) \notin E$. A set of vertices $S \subseteq V$ is a *vertex cover* (VC) of a graph $G = \langle V, E \rangle$ if and only if every edge has at least one endpoint vertex in $V$, i.e., $\forall (u, v) \in E : u \in S \vee v \in S$. A VC $S$ of an undirected graph $G$ is a *minimum VC* (MVC) if and only if $|S|$ is no greater than the cardinality of any other VC of $G$. A VC $S$ of a vertex-weighted undirected graph $G$ is a *minimum weighted VC* (MWVC) if and only if the weight of $S$ is no greater than the weight of any other VC of $G$. Figure 2.1 illustrates the concept of MWVCs.

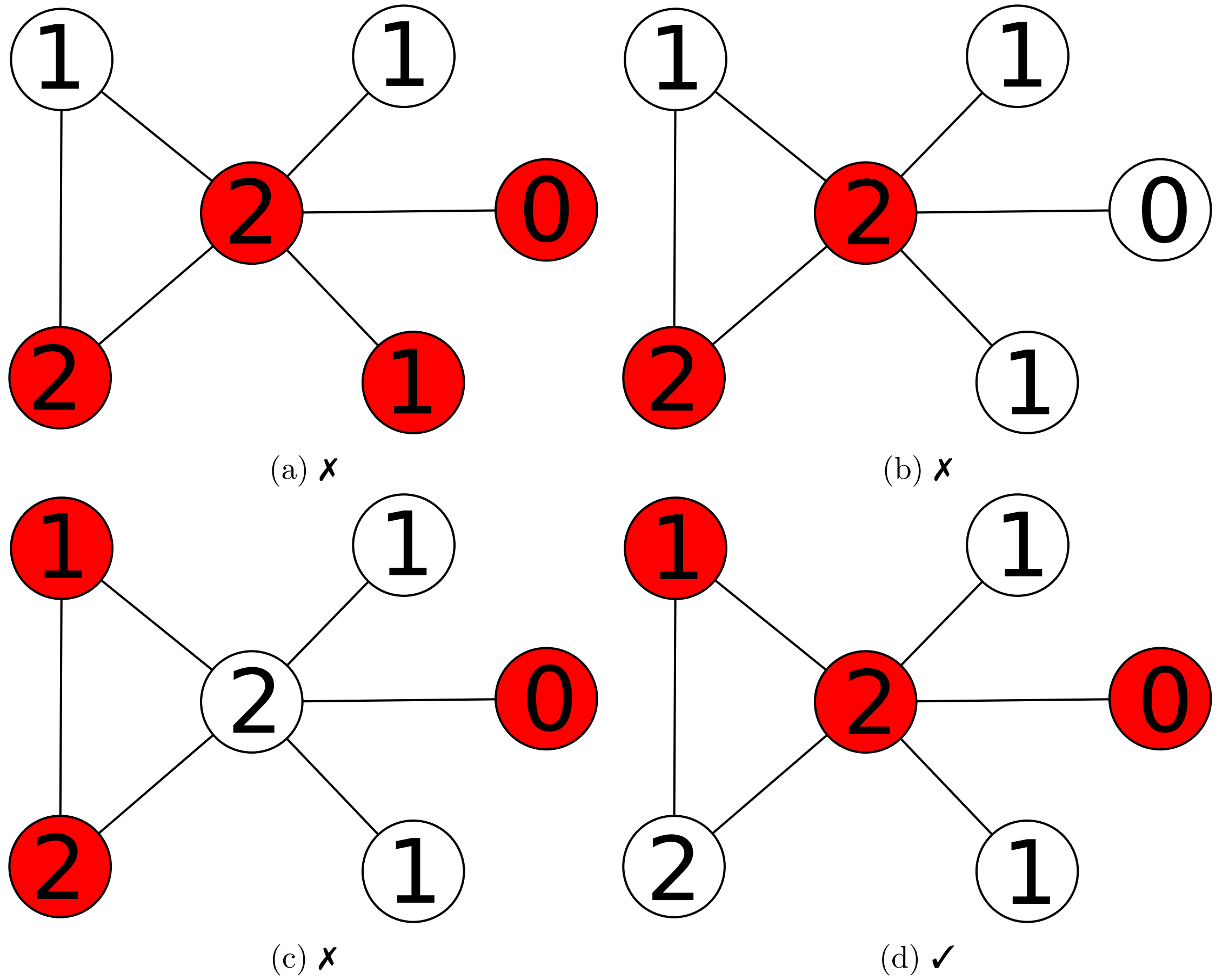

Figure 2.1: Illustrates MWVCs. Each circle represents a vertex. The number in each circle represents the weight of the corresponding vertex. The red circles represent vertices in $S$. ✓ and ✗ mean that $S$ in the corresponding figures are and are not MWVCs, respectively. $S$ in (a) and (b) are not MWVCs because their weights are not minimized. $S$ in (c) is not an MWVC because it is not a VC. $S$ in (d) is an MWVC, although it is not an MVC.

## 2.2 The Weighted Constraint Satisfaction Problem

Formally, the *weighted constraint satisfaction problem* (WCSP) is a triplet $\langle X, D, C\rangle$, where $X = \{X_1, \cdots, X_N\}$ is a set of variables, $D = \{D_1, \cdots, D_N\}$ is the set of discrete-valued domains that specify the set of values that each variable can take,

**Algorithm 2.1:** Solve the WCSP using branch-and-bound search.

1 **Function** `SolveWCSP`($P$)
  - **Input:** $P$: A WCSP instance.
  - **Output:** The optimal solution of $P$ and its total weight.

2   **return** `BranchAndBound`($P$, $\emptyset$, 0, $\emptyset$, $+\infty$);

3 **Function** `BranchAndBound`($P = \langle \mathcal{X}, \mathcal{D}, \mathcal{C} \rangle$, $a$, $w_a$, $a^\dagger$, $w^\dagger$)
  - **Input:** $P = \langle \mathcal{X}, \mathcal{D}, \mathcal{C} \rangle$: A WCSP instance.
  - **Input:** $a$: A partial or complete assignment of values to variables.
  - **Input:** $w_a$: The total weight associated with $a$.
  - **Input:** $a^\dagger$: The current best solution.
  - **Input:** $w^\dagger$: The weight of the current best solution.
  - **Output:** Updated current best solution and its total weight.

4   **if** $\mathcal{X} = \emptyset$ **then**

5     **if** $w_a < w^\dagger$ **then**

6       **return** $a, w_a$;

7   **else**

8     $(P' = \langle \mathcal{X}', \mathcal{D}', \mathcal{C}' \rangle), global_consist :=$ `EnforceLocalConsistency`($P$, $w^\dagger - w_a$);

9     **if** $\neg global_consist$ **then**

10       **return** $a^\dagger, w^\dagger$;

11     $X :=$ `ChooseVariable`($\mathcal{X}'$);

12     $D :=$ `OrderDomain`($\mathcal{D}'(X)$);

13     **foreach** $x \in D$ **do**

14       $a' := a \cup \{X = x\}$;

15       $w_{a'} := w_a + E_{C'_X}(\{X = x\})$;

16       $P'' :=$ `ConstructWCSPSubInstance`($X$, $x$, $P'$);

17       $a^\dagger, w^\dagger :=$ `BranchAndBound`($P''$, $a'$, $w_{a'}$, $a^\dagger$, $w^\dagger$);

18   **return** $a^\dagger, w^\dagger$;

and $C = \{C_1, \cdots, C_N\}$ is a set of constraints. Each constraint $C_i$ is defined on a subset of variables $S_i \in X$ and specifies a non-negative cost for each possible assignment of values to the varibles in $S_i$. An optimal solution is an assignment of values to all variables such that the sum of the costs is minimized. If $\forall D_i \in D : |D_i| = 2$, then the WCSP is called a Boolean WCSP (Kumar 2008a).

| $X_1$ \ $X_2$ | 0 | 1 |
|---|---|---|
| 0 | 0.5 | 0.6 |
| 1 | 0.7 | 0.3 |

Figure 2.2: Shows an example binary constraint.

The most mainstream class of algorithms for solving the WCSP is based on branch-and-bound search, which explores a search tree with each node representing an assignment of values to a subset of variables (Larrosa and Schiex 2004). In a search tree, internal nodes represent partial assignments, whereas leaf nodes represent complete assignments. During search, a currently known best solution $a^\dagger$, which we refer to as the *current best solution*, is maintained along with its total weight $w^\dagger$. At each node, the search algorithm computes the total weight $w_a$ corresponding to the assignment of that node. If $w_a > w^\dagger$, the subtree below this node is pruned. The details of this algorithm are depicted in Algorithm 2.1. While this approach works well in practice, it does not (intend to) explicitly discover structure in WCSP instances.

A representative state-of-the-art solver that falls into this class of algorithms is `toulbar2` (Hurley et al. 2016). It is centralized, single-threaded, and CPU-based. It is known to solve all 715 benchmark instances on CVPR/Scene Decomposition, with a maximum number of variables being 208 and a maximum domain size being 8, within 0.07 seconds (Hurley et al. 2016).

## 2.3 The Constraint Composite Graph

The *constraint composite graph* (CCG) for a Boolean WCSP instance $\langle \mathcal{X}, \mathcal{D}, \mathcal{C} \rangle$ is defined using a construction procedure (Kumar 2008a). It proceeds in 3 stages:

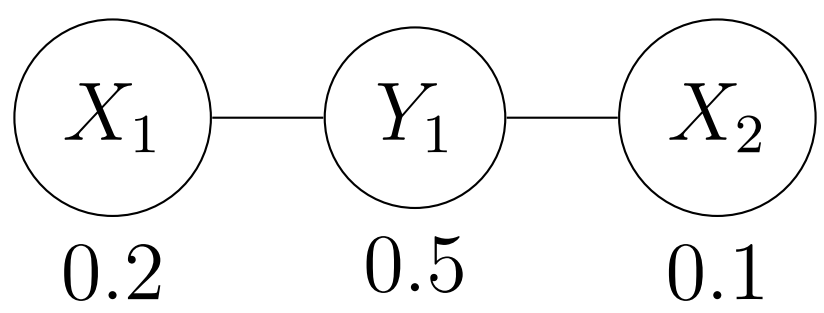

Figure 2.3: Shows that the projection of MWVCs on the IS $\{X_1, X_2\}$ of this vertex-weighted undirected graph leads to Figure 2.2. The weights on $x_1$, $x_2$, and $y_1$ are 0.2, 0.1, and 0.5, respectively. The entry 0.6 in cell $(X_1 = 0, X_2 = 1)$ in Figure 2.2, for example, indicates that, when $X_1$ is necessarily excluded from the MWVC but $X_2$ is necessarily included in it, then the weight of the MWVC—$\{X_2, Y_1\}$—is 0.6.

**1. Expressing Constraints as Polynomials** In this stage, each constraint $C \in \mathcal{C}$ is converted into a polynomial $p_C$ using standard Gaussian elimination. Consider the example constraint in Figure 2.2, which involves the variables $X_1$ and $X_2$. It can be written as a polynomial $p_C(X_1 = x_1, X_2 = x_2)$ in $x_1$ and $x_2$ of degree 1 each:

$$p_C(X_1 = x_1, X_2 = x_2) = c_{00} + c_{01}x_1 + c_{10}x_2 + c_{11}x_1x_2. \tag{2.1}$$

The coefficients $c_{00}, c_{01}, c_{10}$, and $c_{11}$ of the polynomial can be computed by solving a system of linear equations, where each equation corresponds to an entry in the constraint table, using standard Gaussian elimination. In our example, we have

$$\begin{array}{rrrr} p_C(0,0) = 0.5 & p_C(1,0) = 0.6 & p_C(0,1) = 0.7 & p_C(1,1) = 0.3 \\ c_{00} = 0.5 & c_{01} = 0.1 & c_{10} = 0.2 & c_{11} = -0.5. \end{array}$$

**2. Decomposing the Terms of the Polynomials** In this stage, for the polynomial constructed from each constraint, we construct a CCG gadget, a subgraph of the CCG. Before describing this procedure, we describe the *projection of MWVCs on an IS*, a cornerstone concept for the notion of the CCG.

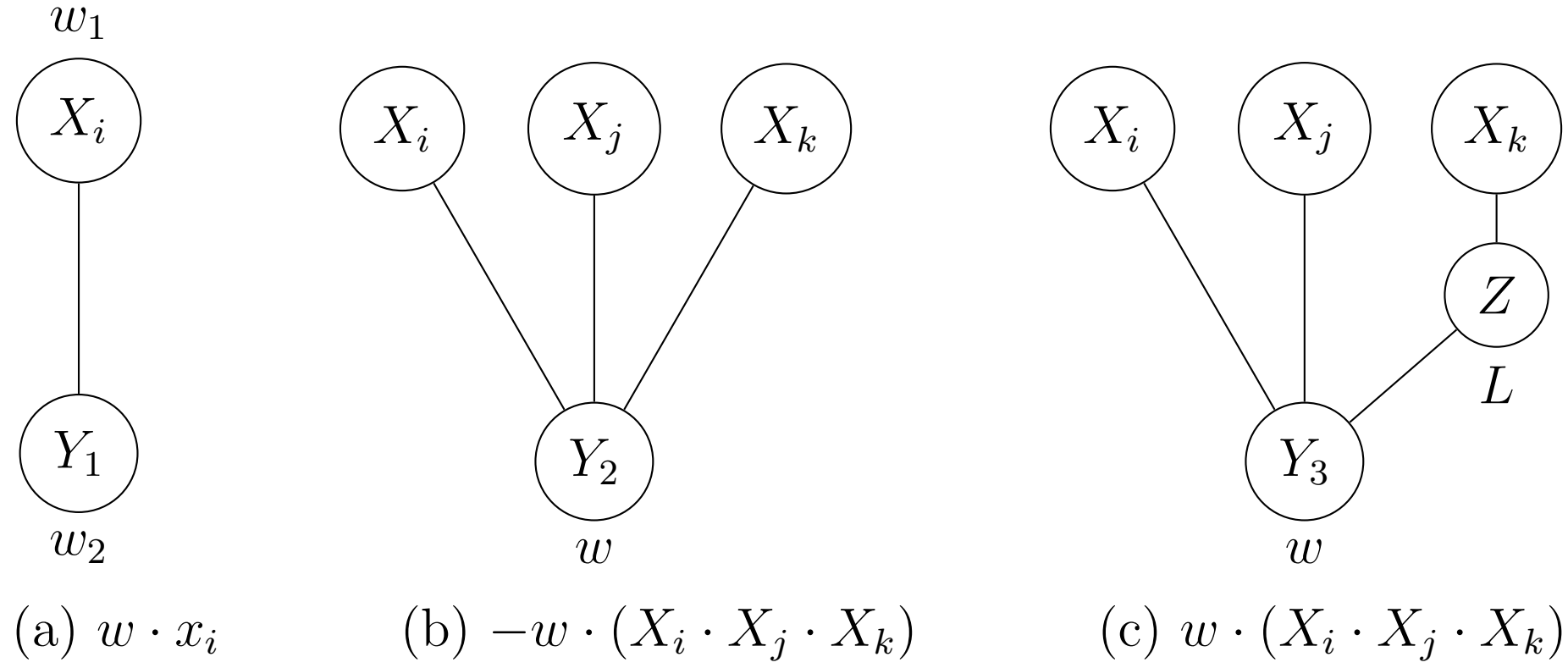

Figure 2.4: Shows the lifted graphical representations of (a) linear, (b) negative nonlinear, and (c) positive nonlinear terms in a polynomial. We assume that $w > 0$ in (b) and (c) (but not necessarily in (a)). A vertex has a zero weight if no weight is shown. In (a), $w_1$ and $w_2$ satisfy $w_1 - w_2 = w$.

For a given graph $G$, one can project MWVCs on a given IS $U \subseteq V$. The input to such a projection is the graph $G$ as well as an IS $U = \{u_1, u_2, \dots, u_k\}$ on $G$. The output is a table of $2^k$ numbers. Each entry in this table corresponds to a $k$-bit vector. We say that a $k$-bit vector $t$ imposes the following restrictions: **(a)** If the $i^{\text{th}}$ bit $t_i$ is 0, then vertex $u_i$ has to be excluded from the MWVC; and **(b)** if the $i^{th}$ bit $t_i$ is 1, then the vertex $u_i$ has to be included in the MWVC. The projection of an MWVC on the IS $U$ is then defined to be a table with entries corresponding to each of the $2^k$ possible $k$-bit vectors $t^{(1)}, t^{(2)}, \dots, t^{(2^k)}$. The value of the entry that corresponds to $t^{(j)}$ is the weight of the MWVC conditioned on the restrictions imposed by $t^{(j)}$. Figure 2.3 illustrates this projection for the subgraph of our example constraint in Figure 2.2.

The table produced by projecting an MWVC on the IS $U$ can be viewed as a constraint over $|U|$ Boolean variables. Conversely, given a constraint (consisting of Boolean variables), we design a lifted representation for it so as to be able to view it as the projection of an MWVC on an IS for some intelligently constructed vertex-weighted undirected graph (Kumar 2008a; Kumar 2008b). The lifted graphical representation of a constraint depends on the nature of the terms in the polynomial that describes the constraint. We distinguish three classes of terms: *linear terms*, *negative nonlinear terms*, and *positive nonlinear terms*. We can construct a lifted graphical representation, i.e., a *CCG gadget*, for each term in the polynomial of each constraint as follows.

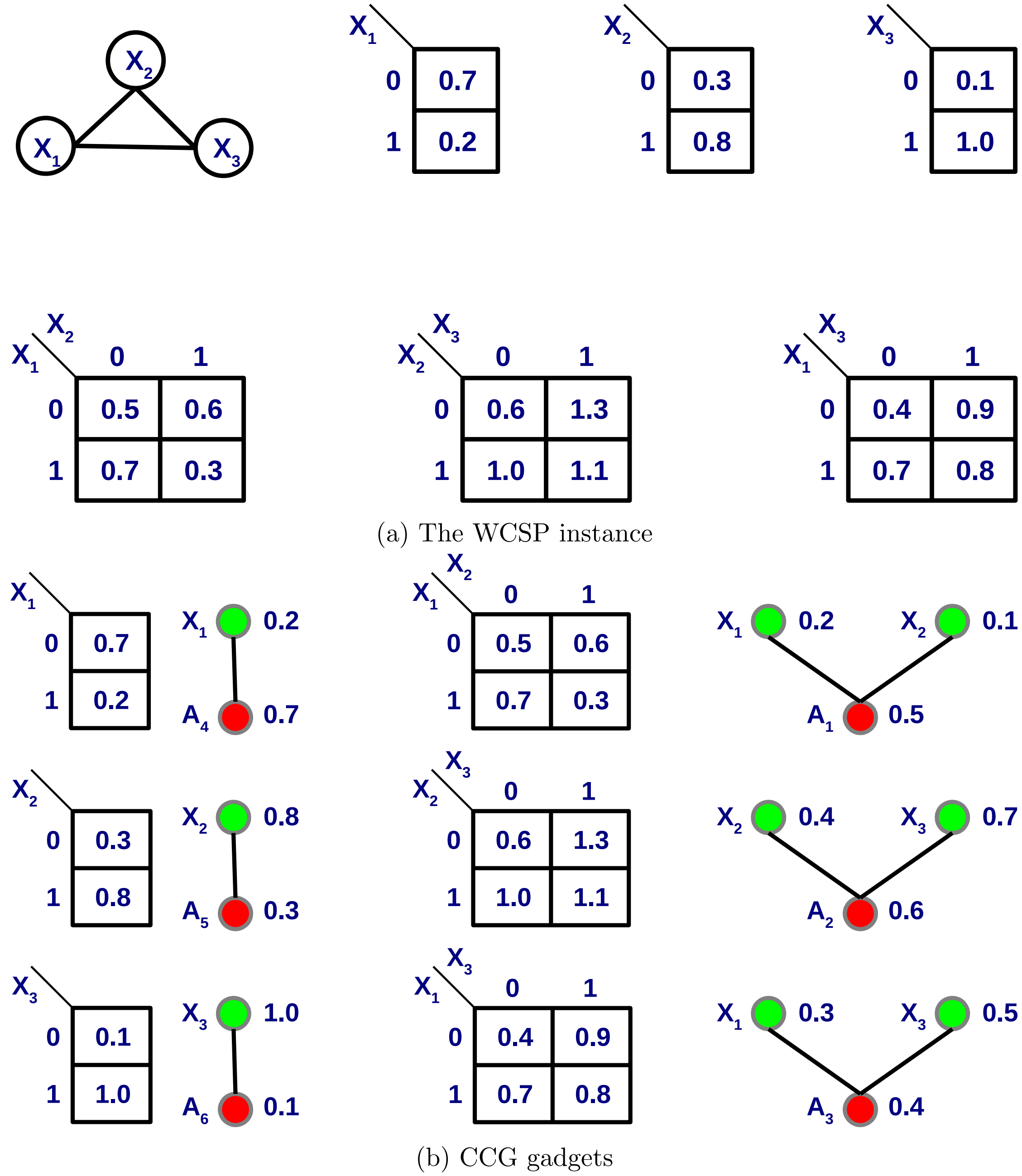

(a) The WCSP instance

(b) CCG gadgets

Figure 2.5: Illustrates the construction of the CCG for a given WCSP instance. Green vertices represent variable vertices and red vertices represent auxiliary vertices. The green tick marks represent the vertices in an MWVC. In this example, the constructed CCG is a bipartite graph, meaning that the original WCSP instance falls in a tractable subclass and can be efficiently solved. (to be continued)

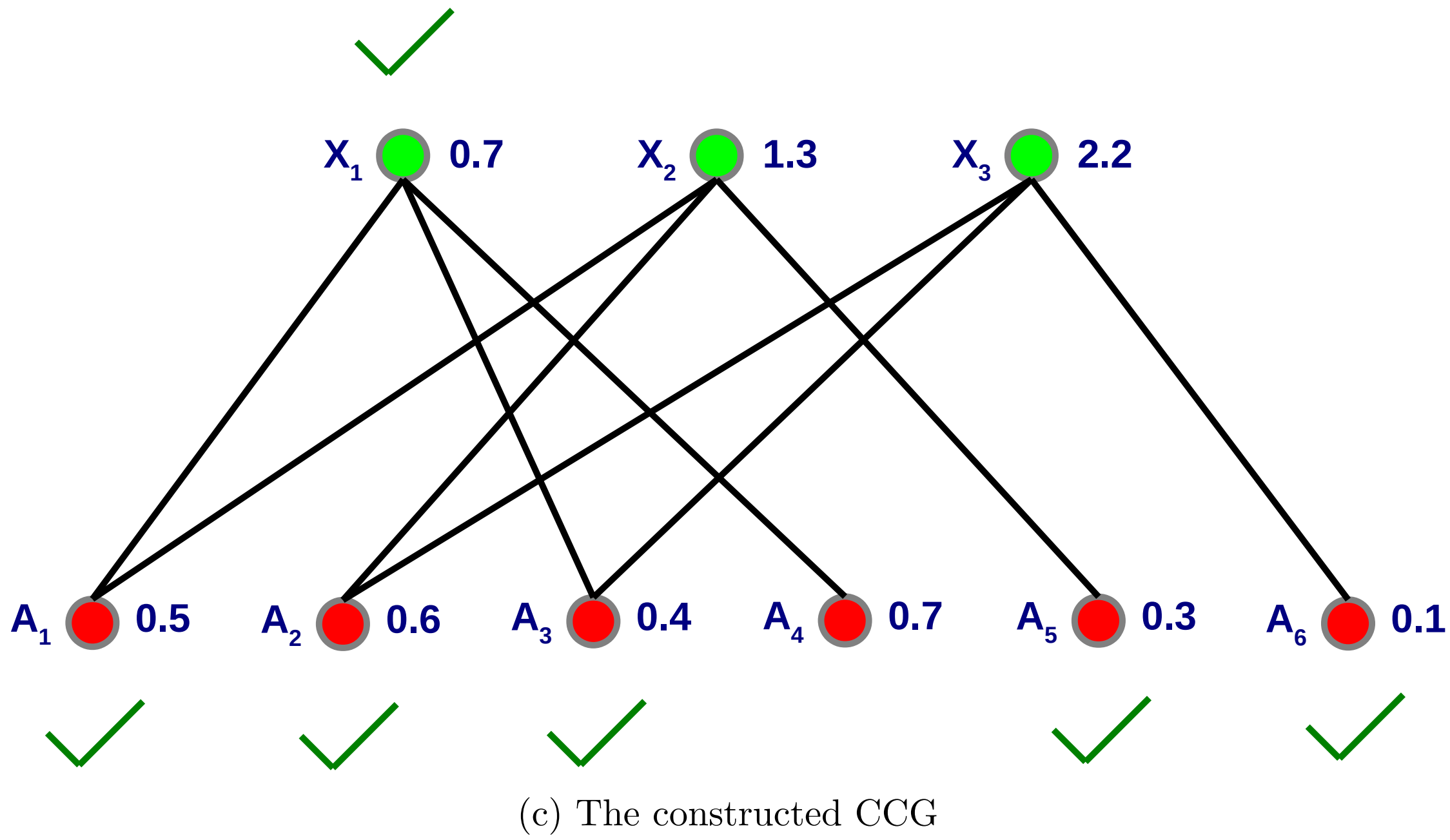

(c) The constructed CCG

Figure 2.5: Continued.

- A linear term is represented with the two-vertex graph shown in Figure 2.4(a) by connecting the variable vertex with an auxiliary vertex.

- A negative nonlinear term is represented with the "flower" structure as shown in Figure 2.4(b). Consider the term $-w\cdot(X_i\cdot X_j\cdot X_k)$ where $w>0$. Projecting an MWVC on the "flower" structure on the variable vertices represents $w - w\cdot(X_i\cdot X_j\cdot X_k)$. The constant term $w$ does not affect the optimality of the solution.

- **A positive nonlinear term** is represented using the "flower+thorn" structure as shown in Figure 2.4(c). Consider the term $w \cdot (X_i \cdot X_j \cdot X_k)$ where $w > 0$. The projection of an MWVC on the "flower+thorn" structure on the variable vertices represents $L \cdot (1 - X_k) + w - w \cdot (X_i \cdot X_j \cdot (1 - X_k))$, where $L > w + 1$ is a large real number. By constructing CCG gadgets that cancel out the lower order terms as shown before, we arrive at a lifted graphical representation of the positive nonlinear term.

**3. Merging CCG Gadgets into a CCG** Finally, we construct the CCG by merging their CCG gadgets: We merge vertices representing the same variables by adding their weights and keep all edges connecting them to all other vertices. Computing the MWVC for the CCG yields a solution for the Boolean WCSP: If variable $X \in \mathcal{X}$ is in the MWVC, then it is assigned the value 1 in the Boolean WCSP, otherwise it is assigned the value 0.

This construction procedure is also illustrated in Figure 2.5.

### 2.3.1 Theoretical Properties of the CCG

The CCG has the following known theoretical properties:

- It can always be constructed in polynomial time. This is due to the fact that each CCG gadget can be constructed in polynomial time and that the total number of CCG gadgets is polynomial with respect to the problem size of the WCSP instance. Efficient CCG construction makes the applicability of the CCG more practical.

- It is always tripartite and can be bipartite for a subclass of the Boolean WCSP. Since the MWVC problem on a bipartite graph is tractable, a Boolean WCSP instance can be solved efficiently if its CCG is bipartite. This can be used to discover tractable subclasses of the Boolean WCSP.

- It is decomposable. The construction procedure (a) constructs a CCG gadget for each individual constraint and then (b) assembles them together to form the CCG. This construction procedure is decomposable: If we add or remove a constraint from the original Boolean WCSP instance, we can obtain the CCG of the new Boolean WCSP instance by modifying the CCG of the original Boolean WCSP instance using the CCG gadget of the added or removed constraint.

# Chapter 3

# The Nemhauser-Trotter Reduction on the CCG

## 3.1 Introduction

Many interesting combinatorial problems are NP-hard. Despite many sophisticated search algorithms dedicated to solving them, the search spaces still remain intractable for large instances. Therefore, a polynomial-time procedure that reduces the sizes of problem instances and identifies a combinatorial core can be beneficial as a preprocessing step. Such a procedure is called a *kernelization algorithm*, and the combinatorial core is called a kernel (illustrated in Figure 3.1).

The *Nemhauser-Trotter Reduction* (NT reduction) is one such algorithm for the MWVC problem (Nemhauser and Trotter 1975). Hence, it can be applied on the MWVC problem on the CCG as well. It is based on the observation that the MWVC problem is a half-integral problem. This means that its *Integer Linear Programming* (ILP) formulation exhibits the following property. We consider a vertex-weighted graph $G = \langle V, E, w \rangle$. In the ILP formulation of the MWVC problem instance on $G$, a Boolean decision variable $Z_i$ is first associated with the presence of vertex $v_i$ in the MWVC. Then, the ILP formulation is

$$\begin{aligned} \text{minimize} \quad & \sum_{i=1}^{|V|} w_i Z_i, \\ \forall\ v_i \in V: \quad & Z_i \in \{0,1\}, \\ \forall\ (v_i, v_j) \in E: \quad & Z_i + Z_j \geq 1. \end{aligned} \tag{3.1}$$

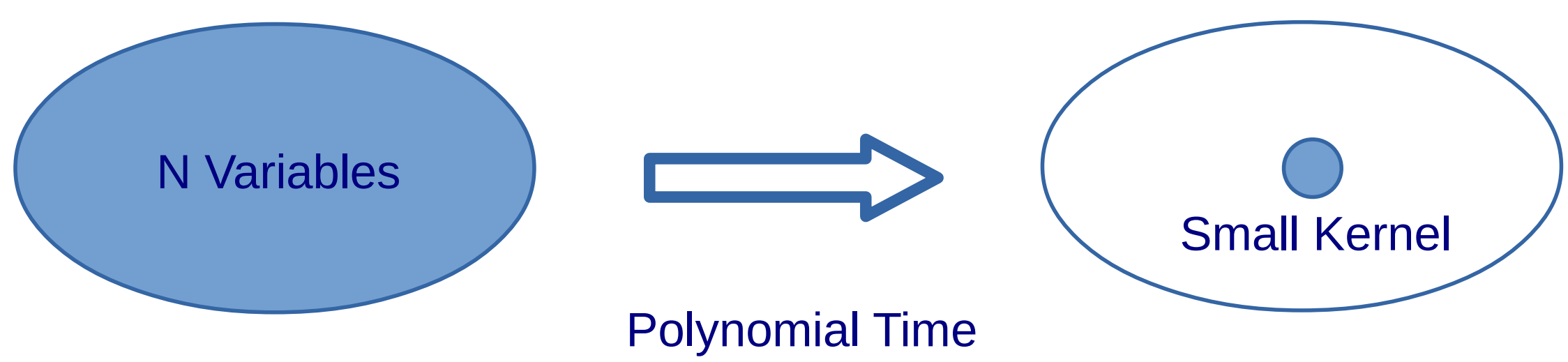

Figure 3.1: Illustrates kernelization algorithms. After a polynomial-time procedure, the problem instance of $N$ variables is reduced to a small combinatorial kernel.

If we relax the integrality constraints $Z_i \in \{0,1\}$ for all $i \in \{1,2,\ldots,|V|\}$ and solve the relaxed LP, the optimal solution of the LP is guaranteed to be half-integral—i.e., $\forall i \in \{1,2,\ldots,|V|\} : Z_i \in \{0, \frac{1}{2}, 1\}$. There then exists an MWVC on $G$ that includes $v_i$ if $Z_i = 1$ and excludes $v_i$ if $Z_i = 0$. Therefore, one can kernelize the MWVC problem instance on $G$ to an MWVC problem instance on a subgraph of $G$ by retaining only those vertices whose Boolean variables in an optimal solution of the LP are $\frac{1}{2}$.

The half-integrality property can be further exploited to solve the LP relaxation of the MWVC problem with a maxflow algorithm instead of a general LP solver (Kumar 2003). We first transform $G$ to a vertex-weighted undirected bipartite graph $G_b = \langle V^L_{G_b}, V^R_{G_b}, E_{G_b}, w\rangle$ as follows. For each vertex $v_i \in V$, we create two vertices $v^L_i \in V^L_{G_b}$ and $v^R_i \in V^R_{G_b}$, both with weight $w_i$. For each edge $(v_i, v_j) \in E$, we create two edges $(v_i^L, v_j^R) \in E_{G_b}$ and $(v_j^L, v_i^R) \in E_{G_b}$. The MWVC problem can be solved in polynomial time on the bipartite graph $G_b$ using a maxflow algorithm (Kumar 2003); and the half-integral solution of the above LP relaxation can be retrieved as follows. If both $v_i^L$ and $v_i^R$ are in the MWVC of $G_b$, then $Z_i = 1$ and $v_i$ can be safely included in the MWVC of $G$; if neither $v_i^L$ nor $v_i^R$ is in the MWVC of $G_b$, then $Z_i = 0$ and $v_i$ can be safely excluded from the MWVC of $G$; if exactly one of $v_i^L$ or $v_i^R$ is in the MWVC of $G_b$, then $Z_i = \frac{1}{2}$ and $v_i$ is retained in the kernel of the MWVC problem instance posed on $G$. Figure 3.2 illustrates this procedure.

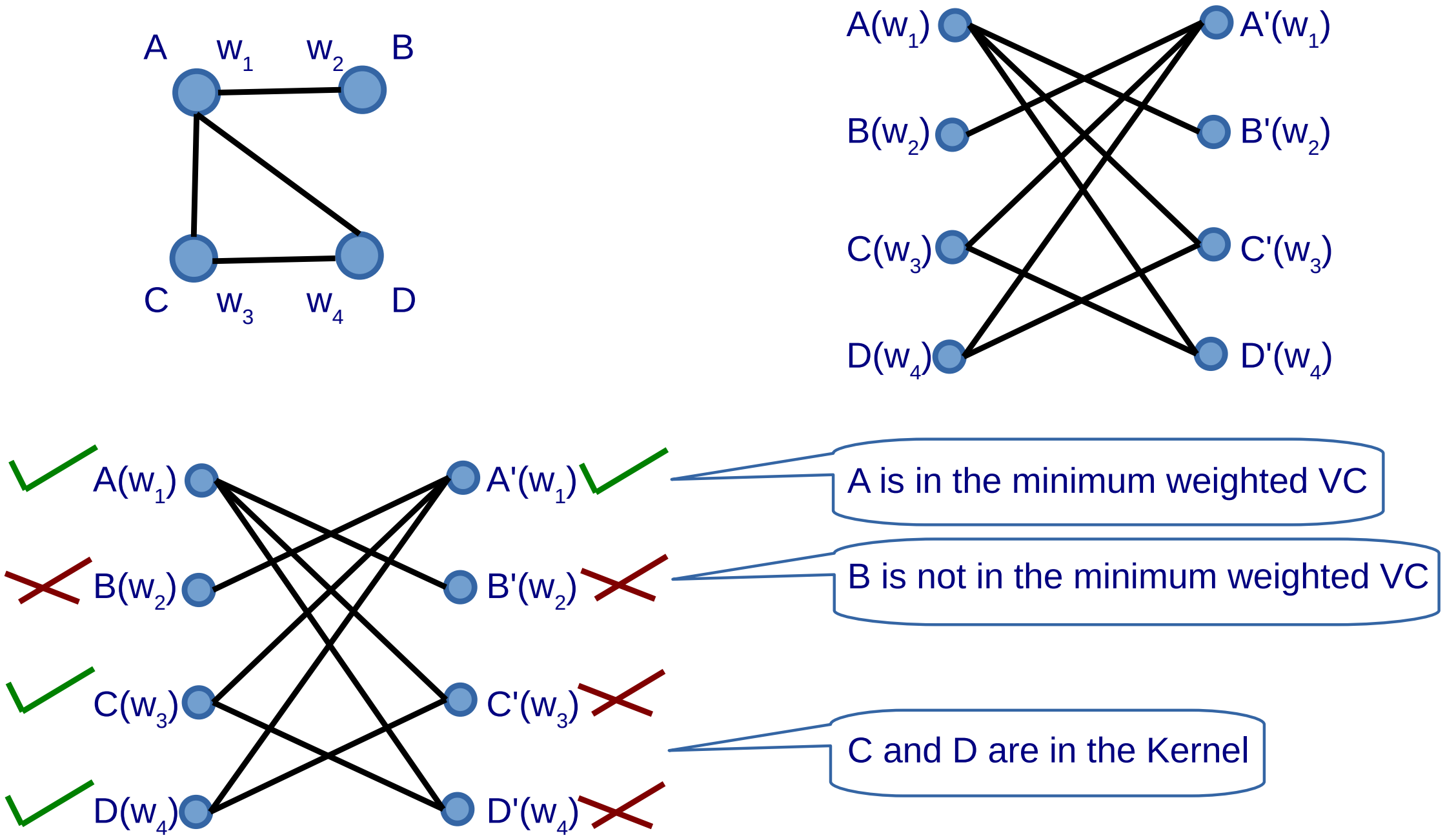

Figure 3.2: Illustrates the NT reduction. The left-upper panel shows the graph $G$. The right-upper panel shows its corresponding vertex-weighted bipartite graph $G_b$ where each vertex in $G$ has a corresponding vertex in each partition of $G_b$. The NT reduction then computes an MWVC on $G_b$. The lower-left panel illustrates one possible computed MWVC of $G_b$. The call-outs in the lower-right panel show the result of the NT reduction by inspecting the presences of vertices in the computed MWVC in $G_b$.

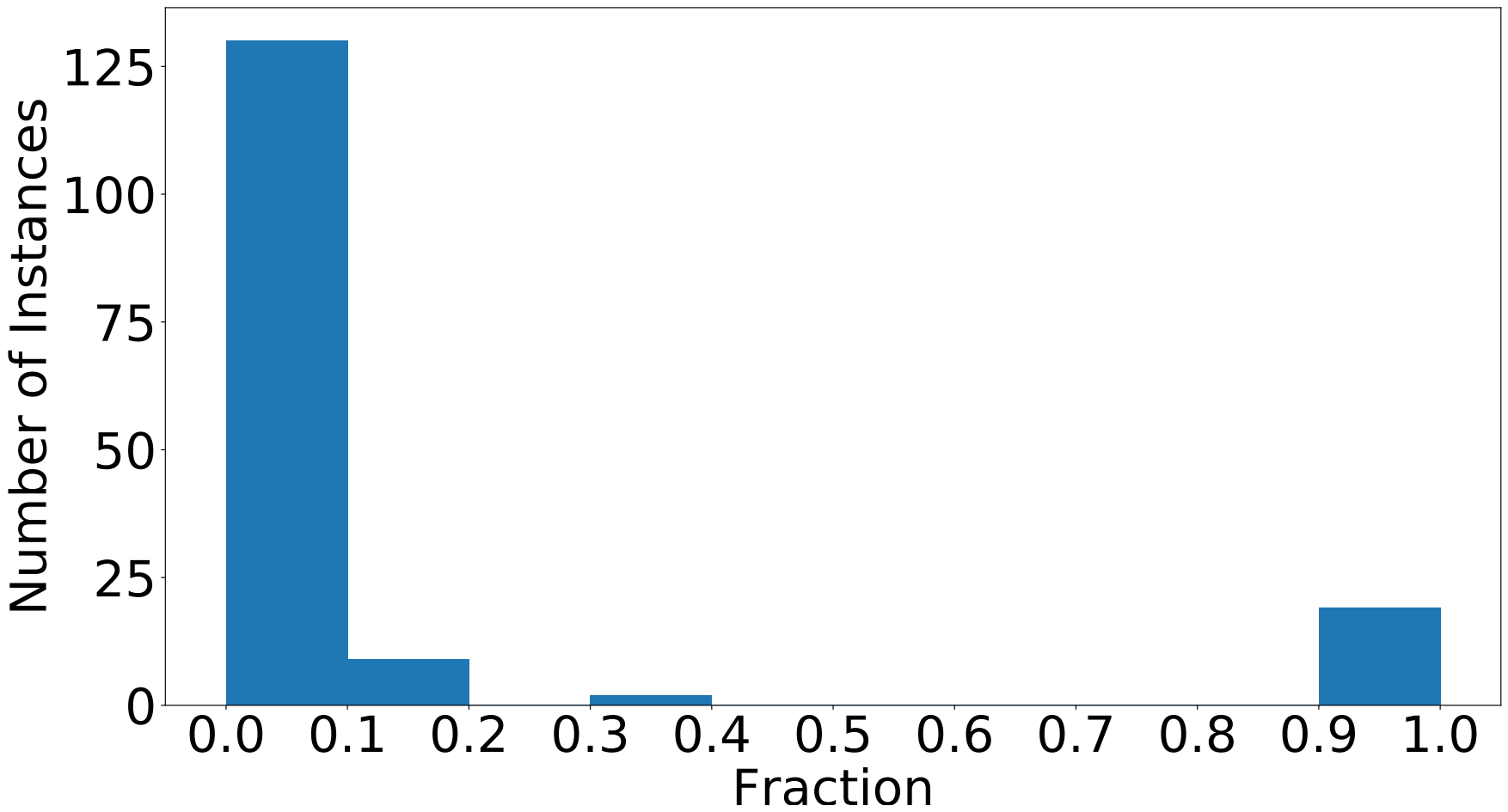

(a) Benchmark instances from the UAI 2014 Inference Competition: 19 out of 160 benchmark instances solved by the NT reduction

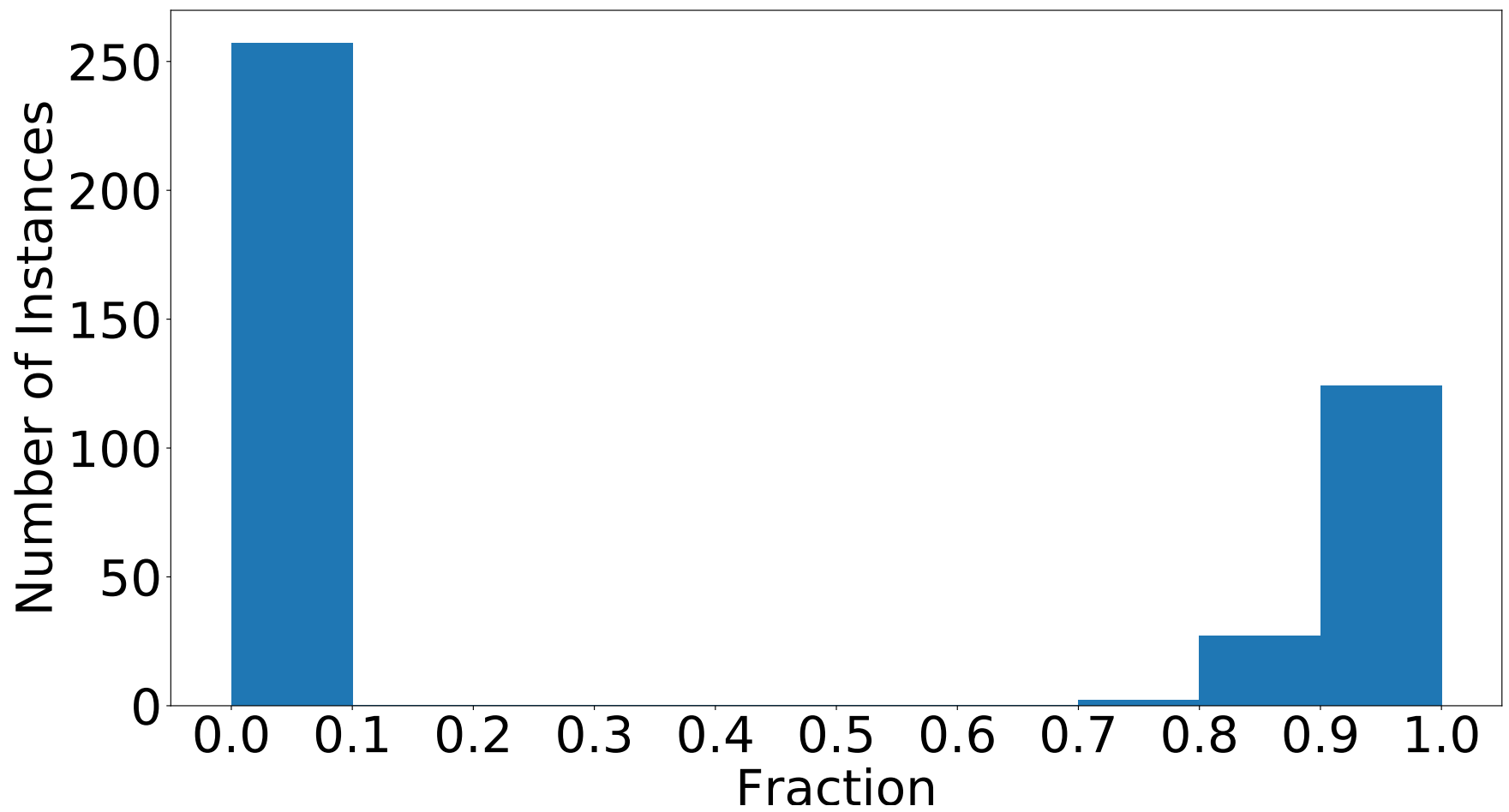

(b) Benchmark instances from (Hurley et al. 2016): 53 out of 410 benchmark instances solved by the NT reduction

Figure 3.3: Shows the effectiveness of the NT reduction. The x-axes show the fraction of variables that are eliminated by the NT reduction. The y-axes show the number of benchmark instances on which this happens for a fraction range.

## 3.2 Experimental Evaluation

In this section, we present experimental evaluations of the NT reduction. We used two sets of Boolean WCSP benchmark instances for our experiments. The first set of benchmark instances is from the UAI 2014 Inference Competition[1]. Here, *maximum-a-posteriori* (MAP) inference queries with no evidence on the PR and MMAP benchmark instances can be reformulated as Boolean WCSP instances by first taking the negative logarithms of the probabilities in each factor and then normalizing them. The second set of benchmark instances is from (Hurley et al. 2016)[2]. This set includes the Probabilistic Inference Challenge 2011, the Computer Vision and Pattern Recognition OpenGM2 benchmark, the Weighted Partial MaxSAT Evaluation 2013, the MaxCSP 2008 Competition, the MiniZinc Challenge 2012 & 2013, and the CFLib (a library of cost function networks). The experiments were performed on those benchmark instances that have only Boolean variables.

The optimal solutions of the benchmark instances in (Hurley et al. 2016) were computed using toulbar2 (Hurley et al. 2016). Since toulbar2 cannot solve WCSP instances with non-integral weights, the optimal solutions of the benchmark instances from the UAI 2014 Inference Competition were computed by finding MWVCs on their CCGs. For each benchmark instance, the MWVC problem was solved by first kernelizing it using the NT reduction, then reformulating it as an ILP (H. Xu, Kumar, and Koenig 2016) and finally solving the ILP using the Gurobi optimizer (Gurobi Optimization, Inc. 2018) with a running time limit of 5 minutes.

[1] `http://www.hlt.utdallas.edu/~vgogate/uai14-competition/index.html`

[2] `http://genoweb.toulouse.inra.fr/~degivry/evalgm/`

For our experiments, we implemented the NT reduction using the Gurobi optimizer (Gurobi Optimization, Inc. 2018) as the LP solver. The CCG construction algorithm was implemented in C++ using the Boost graph library (Siek, Lee, and Lumsdain 2002) and was compiled by gcc 4.9.2 with the "-O3" option. Our experiments were performed on a GNU/Linux workstation with an Intel Xeon processor E3-1240 v3 (8MB Cache, 3.4GHz) and 16GB RAM.

Figure 3.3 shows the effectiveness of the NT reduction on the benchmark instances. The polynomial-time NT reduction solved about $1/8^{\text{th}}$ of these benchmark instances yielding empty kernels. Being able to solve this many benchmark instances without search is indicative of the potential usefulness of the NT reduction for solving structured real-world problems.

We also observed that the NT reduction had very little or no effects for the majority of the rest of the benchmark instances. Here, we discuss an intuitive understanding of why the NT reduction is ineffective on many benchmark instances. $G_b = \langle V^L_{G_b}, V^R_{G_b}, E_{G_b}, w\rangle$ cannot have an MWVC with weight larger than $W = \sum_{v_i \in V^L_{G_b}} w_i$, since the VC consisting of all vertices in either partition of $G_b$ has weight $W$.

- If the NT reduction has no effect, then the weight of an MWVC of $G_b$ must equal $W$. This is because, for each $v_i \in G$, exactly one of $v^L_i$ and $v^R_i$ is in the computed MWVC of $G_b$.

- If the weight of an MWVC of $G_b$ is $W$, then the NT reduction has the option of simply choosing all vertices in one partition as the computed MWVC. The probability of the existence of an alternative MWVC that consist of vertices in both partitions is low if the weights and the additions thereof are mostly unique. In fact, this is true for many real-world applications and, in practice, we can achieve this by adding a random small number to each weight.

Combining the two points above, while it is not true in theory, practically, we can understand that the ineffectiveness of the NT reduction is mostly caused by the fact that the weight of an MWVC of $G_b$ is $W$. Therefore, if $G_b$ of a benchmark instance has an MWVC of weight $W$, the NT reduction is likely to be ineffective on it.

## 3.3 Conclusion

We showed that the NT reduction popularly used for kernelization of the MWVC problem can also be applied to the CCG of the WCSP. This leads to a polynomial-time preprocessing algorithm that fixes the optimal values of a subset of variables in a WCSP instance. This subset is often the set of all variables: We observed that the NT reduction could determine the optimal values of all variables for about $1/8^{\text{th}}$ of the benchmark instances without search.

Enabling the NT reduction for the WCSP can be potentially useful for improving branch-and-bound search for the WCSP: In principle, we can potentially replace `EnforceLocalConsistency` in Algorithm 2.1 with running the NT reduction and develop a new variant of branch-and-bound search algorithm for the WCSP. In other words, thanks to the enabling of the NT reduction for the WCSP, we can now conceptually view the NT reducibility of the WCSP as a new implicit form of local consistency. It is well known that local consistency is an important segue towards understanding the WCSP and our work in this chapter therefore can potentially help researchers deepen understanding of the WCSP.

# Chapter 4

# The Min-Sum Message Passing Algorithm on the CCG

## 4.1 Introduction

*Belief propagation* (BP) is a well-known technique for solving many combinatorial problems across a wide range of fields such as probabilistic reasoning, artificial intelligence, and information theory. It can be used to solve hard inference problems that arise in statistical physics, computer vision, error-correcting coding theory, or, more generally, on graphical models such as Bayesian Networks and Markov random fields (Yedidia, Freeman, and Weiss 2003). BP is an efficient algorithm that is based on local message passing. Although a complete theoretical analysis of its convergence and correctness is elusive, it works well in practice on many important combinatorial problems.

While BP performs message passing for the objective of marginalization over probabilities, the *min-sum message passing* (MSMP) algorithm is a variant of BP that is used to find an assignment of values to all variables in $\vec{X}$ that minimizes functions of the form

$$E(\vec{X}) = \sum_i E_i(\vec{X}_i), \qquad (4.1)$$

where $\vec{X}$ is the set of all variables in the global function $E$; $E_i$ is a local function constituting the $i^{\text{th}}$ term of $E$; and $\vec{X}_i$ is a subset of $\vec{X}$ containing all variables that participate in $E_i$.

To minimize the function $E(\vec{X})$, the MSMP algorithm first builds a factor graph, i.e., an undirected bipartite graph with one partition containing vertices that represent the variables in $\vec{X}$ and the other partition containing vertices that represent the local functions $E_i$ for all $i$. An edge represents the participation of a variable in a local function. Furthermore, a message is a vector associated with each direction of each edge. Intuitively, messages represent interactions between individual variables and local functions. The value of $E_i$ is the potential of its corresponding vertex because it is indicative of its "potential" to affect other vertices. Messages are updated iteratively until convergence. In each iteration, the message from vertex $u$ to vertex $v$ is influenced by incoming messages to $u$ as well as $u$'s potential if it represents a local function. Upon convergence, a solution can be extracted from the messages.

The MSMP algorithm converges and produces an optimal solution if the factor graph is a tree (Mézard and Montanari 2009). This is, however, not necessarily the case if the factor graph is loopy (Mézard and Montanari 2009). Although the clique tree algorithm alleviates this problem to a certain extent by first converting loopy graphs to trees (Koller and Friedman 2009), the technique only scales to graphs with low treewidths. If the MSMP algorithm operates directly on loopy graphs, the theoretical underpinnings of its convergence and optimality properties still remain poorly understood. Nonetheless, it works well in practice on a number of important combinatorial problems in artificial intelligence, statistical physics, and signal processing (Mézard and Montanari 2009; Moallemi and Roy 2010).

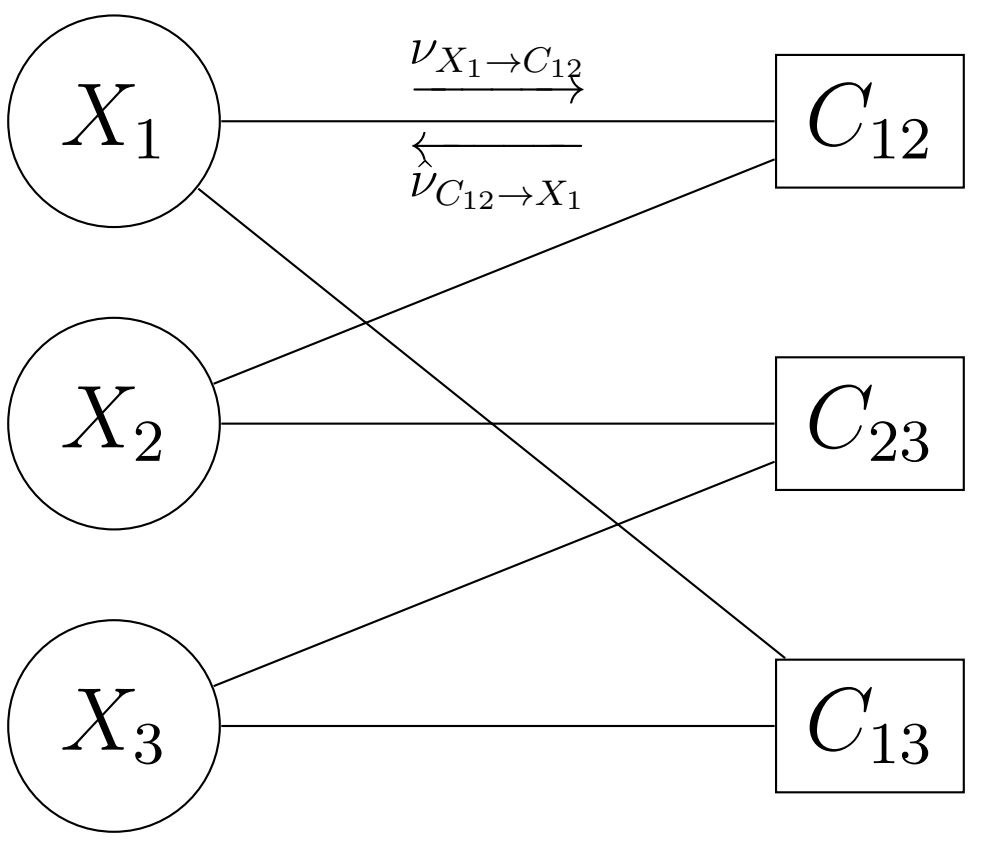

Figure 4.1: Illustrates the factor graph of a Boolean WCSP instance with 3 variables $\{X_1, X_2, X_3\}$ and 3 constraints $\{C_{12}, C_{13}, C_{23}\}$. Here, $X_1, X_2 \in C_{12}$, $X_1, X_3 \in C_{13}$, and $X_2, X_3 \in C_{23}$. The circles are variable vertices, and the squares are constraint vertices. $\nu_{X_1 \to C_{12}}$ and $\hat{\nu}_{C_{12} \to X_1}$ are the messages from $X_1$ to $C_{12}$ and from $C_{12}$ to $X_1$, respectively. Such a pair of messages annotates each edge (not all are explicitly shown).

Examples include the CSP (Montanari, Ricci-Tersenghi, and Semerjian 2007), $K$-satisfiability (Mézard and Zecchina 2002), and the MVC problem (Weigt and Zhou 2006). In this chapter, we show how we improve the MSMP algorithm for the Boolean WCSP by using the CCG.

## 4.2 The Min-Sum Message Passing Algorithm Applied Directly on the Boolean WCSP

We now describe how the MSMP algorithm can be applied directly to solving the Boolean WCSP defined by $\langle \mathcal{X}, \mathcal{D}, \mathcal{C} \rangle$. We refer to this as the *original MSMP algorithm*. As explained before, we first construct its factor graph. We create a vertex for each variable in $\mathcal{X}$ (variable vertex) and for each constraint in $\mathcal{C}$ (constraint vertex). A variable vertex $X_i$ and a constraint vertex $C_j$ are connected by an edge if and only if $C_j$ contains $X_i$. Figure 4.1 shows an example.

After the factor graph is constructed, a message (two real numbers) for each of the two directions along each edge is initialized, for instance, to zeros. A pair of messages $\nu_{X_1 \to C_{12}}$ and $\hat{\nu}_{C_{12} \to X_1}$ is illustrated in Figure 4.1. The messages are then updated iteratively by using the min-sum update rules given by

$$\nu^{(t)}_{X_i \to C_j}(X_i = x_i) = \sum_{C_k \in \partial X_i \setminus \{C_j\}} \left[ \hat{\nu}^{(t-1)}_{C_k \to X_i}(X_i = x_i) \right] + c^{(t)}_{X_i \to C_j} \tag{4.2}$$

$$\begin{aligned} \hat{\nu}^{(t)}_{C_j \to X_i}(X_i = x_i) = \min_{a \in \mathcal{A}(\partial C_j \setminus \{X_i\})} & \Bigg[ E_{C_j}(a \cup \{X_i = x_i\}) \\ & + \sum_{X_k \in \partial C_j \setminus \{X_i\}} \nu^{(t)}_{X_k \to C_j}(a|\{X_k\}) \Bigg] + \hat{c}^{(t)}_{C_j \to X_i} \end{aligned} \tag{4.3}$$

for all $X_i \in \mathcal{X}, C_j \in \mathcal{C}$ and $x_i \in \{0, 1\}$ until convergence (Mézard and Montanari 2009), where

- $\hat{\nu}^{(t)}_{C_j \to X_i}(X_i = x_i)$ for both $x_i \in \{0, 1\}$ are the two real numbers of the message that is passed from the constraint vertex $C_j$ to the variable vertex $X_i$ in the $t^{\text{th}}$ iteration,

- $\nu^{(t)}_{X_i \to C_j}(X_i = x_i)$ for both $x_i \in \{0, 1\}$ are the two real numbers of the message that is passed from the variable vertex $X_i$ to the constraint vertex $C_j$ in the $t^{\text{th}}$ iteration,

- $\partial X_i$ and $\partial C_j$ are the sets of neighboring vertices of $X_i$ and $C_j$, respectively, and

- $c^{(t)}_{X_i \to C_j}$ and $\hat{c}^{(t)}_{C_j \to X_i}$ are normalization constants such that

$$\min \left[ \nu^{(t)}_{X_i \to C_j}(X_i = 0), \nu^{(t)}_{X_i \to C_j}(X_i = 1) \right] = 0 \tag{4.4}$$

$$\min \left[ \hat{\nu}^{(t)}_{C_j \to X_i}(X_i = 0), \hat{\nu}^{(t)}_{C_j \to X_i}(X_i = 1) \right] = 0. \tag{4.5}$$

The message update rules can be understood as follows. Each message from a variable vertex $X_i$ to a constraint vertex $C_j$ is updated by summing up all $X_i$'s incoming messages from its other neighboring vertices. Each message from a constraint vertex $C_j$ to a variable vertex $X_i$ is updated by finding the minimum of the constraint function $E_{C_j}$ plus the sum of all $C_j$'s incoming messages from its other neighboring vertices. The messages can be updated in various orders.

We use the superscript $\infty$ to indicate the values of messages upon convergence. The final assignment of values to variables in $\mathcal{X} = \{X_1, X_2, \ldots, X_N\}$ can then be found by computing

$$E_{X_i}(X_i = x_i) \equiv \sum_{C_k \in \partial X_i} \hat{\nu}^{(\infty)}_{C_k \to X_i}(X_i = x_i) \tag{4.6}$$

for all $X_i \in \mathcal{X}$ and $x_i \in \{0, 1\}$. Here, $E_{X_i}(X_i = 0)$ and $E_{X_i}(X_i = 1)$ can be proven to be equal to the minimum values of the total weights conditioned on $X_i = 0$ and $X_i = 1$, respectively. By selecting the value of $x_i$ that leads to a smaller value of $E_{X_i}(X_i = x_i)$, we obtain the final assignment of values to all variables in $\mathcal{X}$.

## 4.3 The Min-Sum Message Passing Algorithm Applied on the CCG

To solve a given WCSP instance, we can first transform it to an MWVC problem instance on its CCG. We can then apply the MSMP algorithm on the CCG. We refer to this procedure as the *lifted MSMP algorithm*.

The MWVC problem on the vertex-weighted graph $\langle V, E, w \rangle$ is a subclass of the Boolean WCSP. Throughout this section, we use the variable $X_i$ to represent the $i^{\text{th}}$ vertex in $V$: $X_i = 1$ means the $i^{\text{th}}$ vertex is selected in the MWVC, and $X_i = 0$ means the $i^{\text{th}}$ vertex is not selected in the MWVC. The MWVC problem can therefore be rewritten as a subclass of the Boolean WCSP with only the following two types of constraints:

- Unary weighted constraints: Each of these weighted constraints corresponds to a vertex in the MWVC problem. We use $C_i^V$ to denote the weighted constraint that corresponds to the $i^{\text{th}}$ vertex. $C_i^V$ therefore only has one variable $X_i$. In the weighted constraint $C_i^V$, the tuple in which $X_i = 1$ has weight $w_i \geq 0$ and the other tuple has weight zero. This type of weighted constraints represents the minimization objective of the MWVC problem.

- Binary weighted constraints: Each of these weighted constraints corresponds to an edge in the MWVC problem. We use $C_j^E$ to denote the weighted constraint that corresponds to the $j^{\text{th}}$ edge. The indices of the endpoint vertices of this edge are denoted by $j(+1)$ and $j(-1)$. $C_j^E$ therefore has two variables $X_{j(+1)}$ and $X_{j(-1)}$. In the weighted constraint $C_j^E$, the tuple in which $X_{j(+1)} = X_{j(-1)} = 0$ has weight infinity and the other tuples have weight zero. This type of weighted constraints represents the requirement that at least one endpoint vertex must be selected for each edge.

Given that the MWVC problem is a subclass of the Boolean WCSP, Equations (4.2), (4.3) and (4.6) can be reused for the MSMP algorithm on it. For the MWVC problem, these equations can be further simplified. For notational convenience, we omit normalization constants in the following derivation.

For each of the unary weighted constraints $C_i^V$, we have

- the added weight for selecting a vertex:

$$E_{C_i^V}(X_i = x_i) = \begin{cases} w_i & x_i = 1 \\ 0 & x_i = 0 \end{cases}, \tag{4.7}$$

- and exactly one variable in $C_i^V$:

$$\partial C_i^V \setminus \{X_i\} = \emptyset. \tag{4.8}$$

By plugging Equations (4.7) and (4.8) into Equation (4.3) for $t = \infty$, we have

$$\hat{\nu}^{(\infty)}_{C_i^V \to X_i}(X_i = x_i) = \begin{cases} w_i & x_i = 1 \\ 0 & x_i = 0 \end{cases} \tag{4.9}$$

for all $C_i^V$. Note that here we do not need Equation (4.2) for $C_i^V$ since it has only one variable and thus the message passed to it does not affect the final solution.

For each of the binary weighted constraints $C_j^E$, we have

- the requirement that at least one endpoint vertex must be selected for each edge:

$$E_{C_j^E}(X_{j(+1)} = x_{j(+1)}, X_{j(-1)} = x_{j(-1)}) = \begin{cases} +\infty & x_{j(+1)} = x_{j(-1)} = 0 \\ 0 & \text{otherwise} \end{cases}, \tag{4.10}$$

- and exactly two variables in $C_j^E$:

$$\partial C_j^E \setminus \{X_{j(\ell)}\} = \{X_{j(-\ell)}\} \quad \forall \ell \in \{+1, -1\}. \tag{4.11}$$

By plugging Equations (4.9) to (4.11) into Equations (4.2) and (4.3) along with the fact that there exist only unary and binary weighted constraints, we have

$$\nu^{(t)}_{X_{j(\ell)}\to C^E_j}(X_{j(\ell)}=1) = \sum_{C_k\in\partial X_{j(\ell)}\setminus\{C^V_{j(\ell)},C^E_j\}} \left[\hat{\nu}^{(t-1)}_{C_k\to X_{j(\ell)}}(X_{j(\ell)}=1)\right] + w_{j(\ell)} \tag{4.12}$$

$$\nu^{(t)}_{X_{j(\ell)}\to C^E_j}(X_{j(\ell)}=0) = \sum_{C_k\in\partial X_{j(\ell)}\setminus\{C^V_{j(\ell)},C^E_j\}} \hat{\nu}^{(t-1)}_{C_k\to X_{j(\ell)}}(X_{j(\ell)}=0) \tag{4.13}$$

$$\hat{\nu}^{(t)}_{C^E_j\to X_{j(\ell)}}(X_{j(\ell)}=1) = \min_{a\in\{0,1\}} \nu^{(t)}_{X_{j(-\ell)}\to C^E_j}(X_{j(-\ell)}=a) \tag{4.14}$$

$$\hat{\nu}^{(t)}_{C^E_j\to X_{j(\ell)}}(X_{j(\ell)}=0) = \nu^{(t)}_{X_{j(-\ell)}\to C^E_j}(X_{j(-\ell)}=1) \tag{4.15}$$

for all $C^E_j$ and both $\ell\in\{+1,-1\}$. By plugging Equations (4.12) and (4.13) into Equations (4.14) and (4.15), we have

$$\begin{aligned}&\hat{\nu}^{(t)}_{C^E_j\to X_{j(\ell)}}(X_{j(\ell)}=1) = \\ &\min_{a\in\{0,1\}}\left[\sum_{C_k\in\partial X_{j(-\ell)}\setminus\{C^V_{j(-\ell)},C^E_j\}} \left[\hat{\nu}^{(t-1)}_{C_k\to X_{j(-\ell)}}(X_{j(-\ell)}=a)\right] + w_{j(-\ell)}\cdot a\right]\end{aligned} \tag{4.16}$$

$$\begin{aligned}&\hat{\nu}^{(t)}_{C^E_j\to X_{j(\ell)}}(X_{j(\ell)}=0) = \\ &\sum_{C_k\in\partial X_{j(-\ell)}\setminus\{C^V_{j(-\ell)},C^E_j\}} \left[\hat{\nu}^{(t-1)}_{C_k\to X_{j(-\ell)}}(X_{j(-\ell)}=1)\right] + w_{j(-\ell)}\end{aligned} \tag{4.17}$$

for all $C^E_j$ and both $\ell\in\{+1,-1\}$, where $\hat{\nu}^{(t)}_{C^E_j\to X_{j(\ell)}}(X_{j(\ell)}=b)$ for both $b\in\{0,1\}$ are the two real numbers of the message that is passed from the $j^{\text{th}}$ edge to the $j(\ell)^{\text{th}}$ vertex. Since each edge has exactly two endpoint vertices, the message from an edge to one of its endpoint vertices can be viewed as a message from the other endpoint vertex to it. Formally, for the $j^{\text{th}}$ edge, we define the message from the $j(+1)^{\text{th}}$ vertex to the $j(-1)^{\text{th}}$ vertex in the $t^{\text{th}}$ iteration as

$$\mu_{j(+1)\to j(-1)}^{(t)} \equiv \hat{\nu}_{C_j^E \to X_{j(-1)}}^{(t)}. \tag{4.18}$$

By plugging in Equation (4.18) and substituting $j(\ell)$ with $i$ and $j(-\ell)$ with $j$, Equations (4.16) and (4.17) can be rewritten (with normalization constants) in the form of messages between vertices as

$$\mu_{j\to i}^{(t)}(X_i = 1) = \min_{a\in\{0,1\}} \left[ \sum_{k\in N(j)\setminus\{i\}} \mu_{k\to j}^{(t-1)}(X_j = a) + w_j \cdot a \right] + c_{j\to i}^{(t)} \tag{4.19}$$

$$\mu_{j\to i}^{(t)}(X_i = 0) = \sum_{k\in N(j)\setminus\{i\}} \mu_{k\to j}^{(t-1)}(X_j = 1) + w_j + c_{j\to i}^{(t)} \tag{4.20}$$

for all $i$ and $j$ such that the $i^{\text{th}}$ and $j^{\text{th}}$ vertices are connected by an edge in $E$. Here, $N(j)$ is the set of neighboring vertices of the $j^{\text{th}}$ vertex in $V$ and $c_{j\to i}^{(t)}$ represents the normalization constant such that $\min\left[\mu_{j\to i}^{(t)}(X_i = 1), \mu_{j\to i}^{(t)}(X_i = 0)\right] = 0$. Equations (4.19) and (4.20) are the message update rules of the MSMP algorithm adapted to the MWVC problem.

If the messages converge, by plugging Equations (4.9) and (4.18) into Equation (4.6), the final assignment of values to variables can be found by computing

$$E_{X_i}(X_i = x_i) = \sum_{j\in N(i)} \left[\mu_{j\to i}^{(\infty)}(X_i = x_i)\right] + w_i x_i, \tag{4.21}$$

where the meaning of $E_{X_i}(X_i = x_i)$ is similar to that in Equation (4.6).

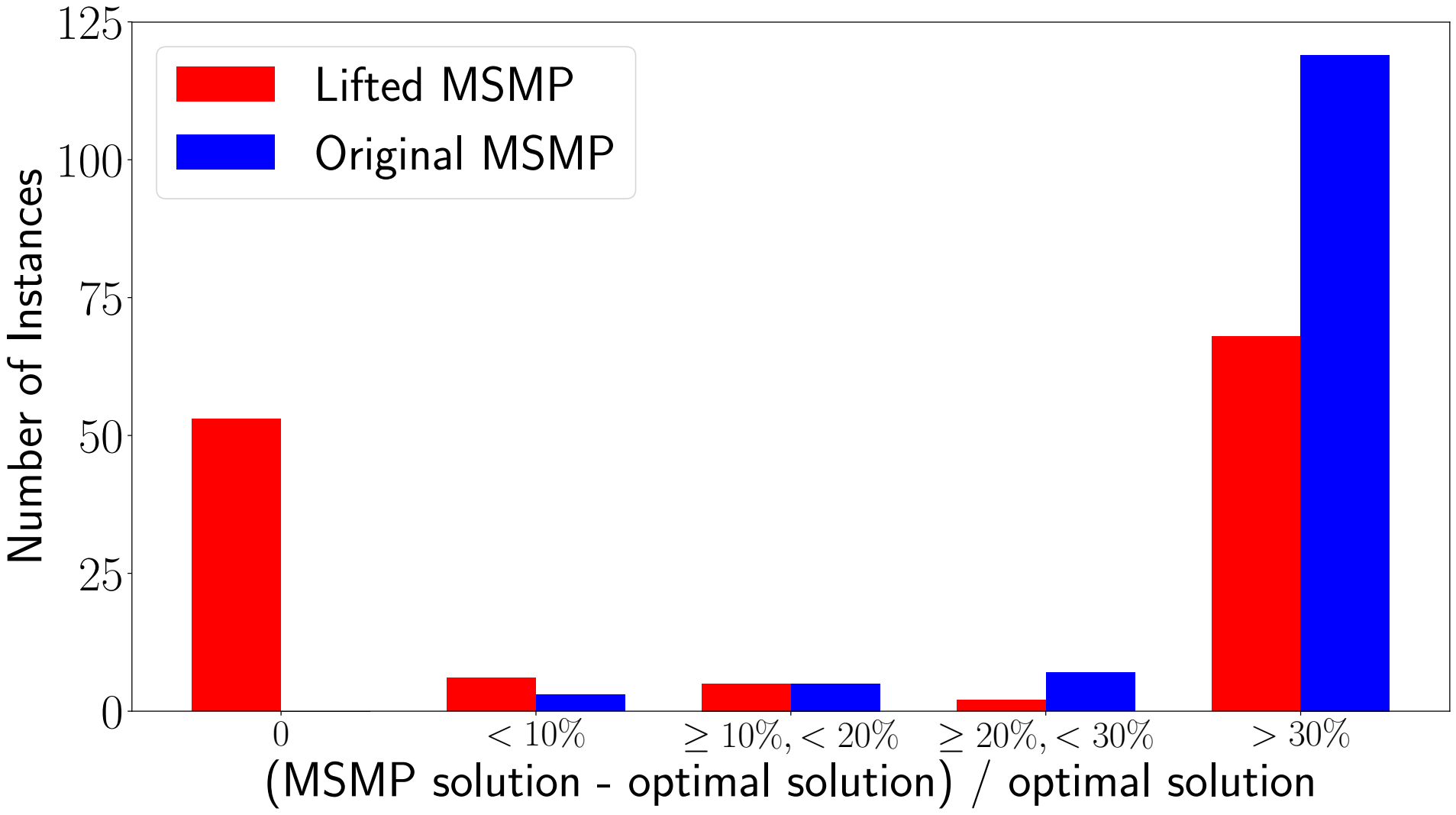

(a) Benchmark instances from the UAI 2014 Inference Competition

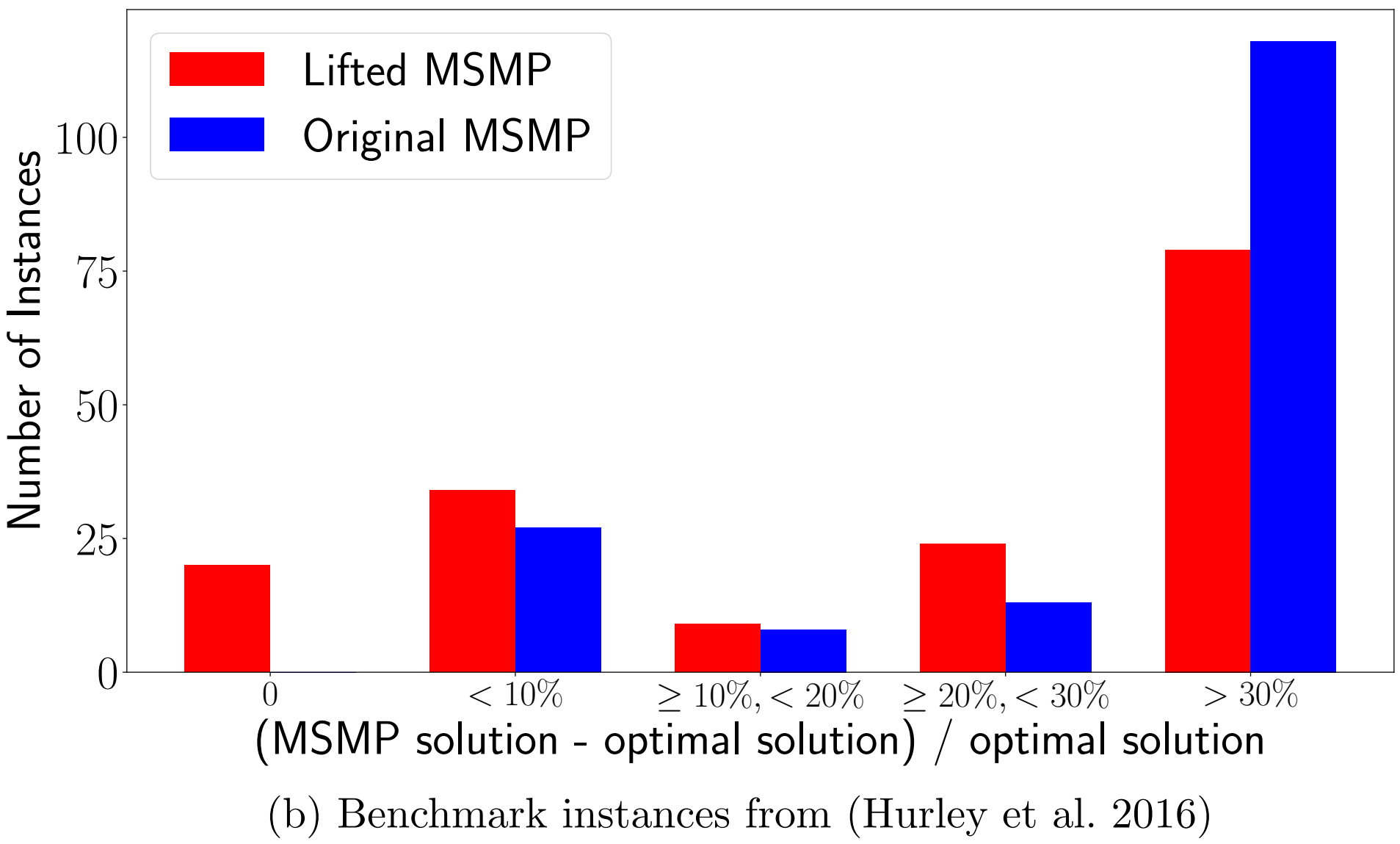

(b) Benchmark instances from (Hurley et al. 2016)

Figure 4.2: Shows the qualities of the solutions (total weights) produced by the original and the lifted MSMP algorithms in comparison to the optimal solutions (for benchmark instances with known optimal solutions). The x-axes show the suboptimality of the MSMP solutions. The y-axes show the number of benchmark instances for a range of suboptimality. Higher bars on the left are indicative of better solutions.

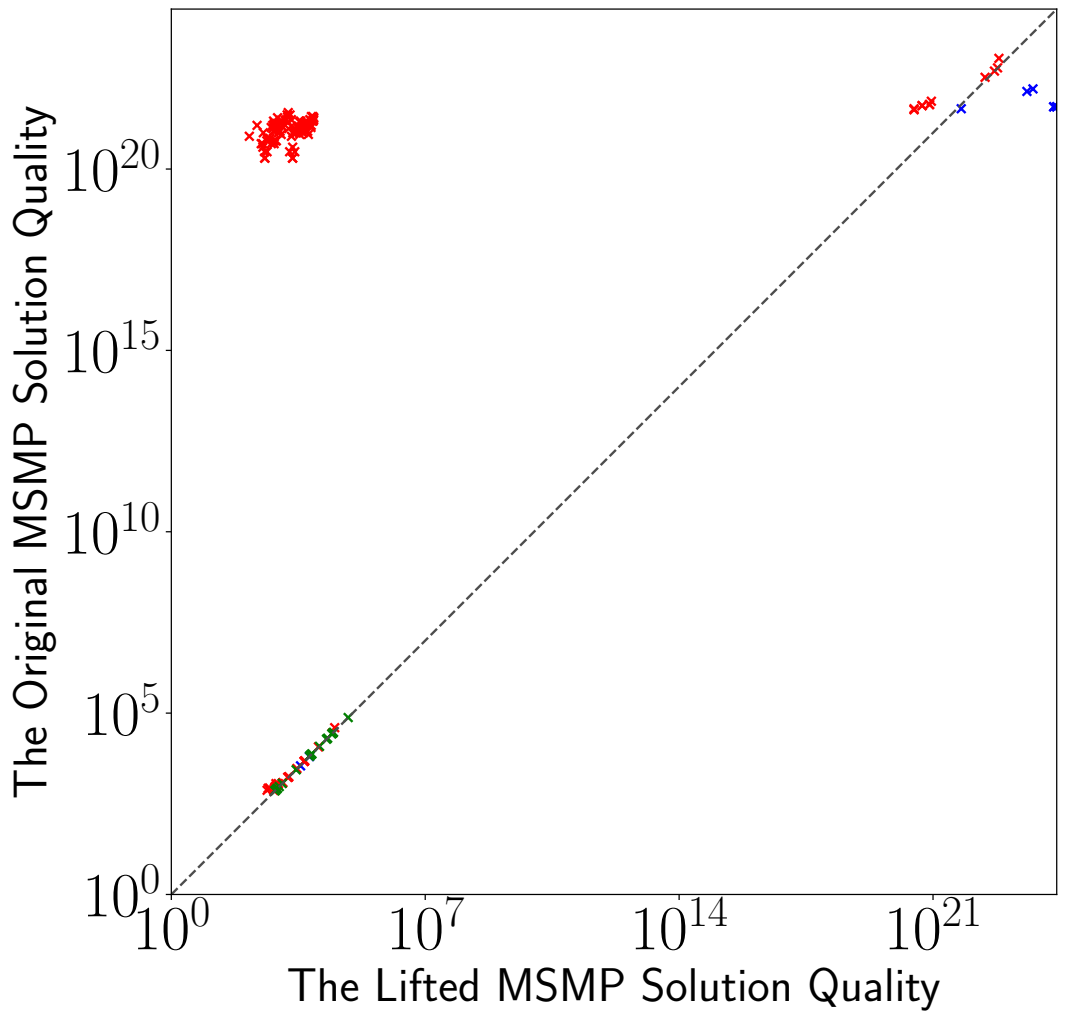

(a) Benchmark instances from the UAI 2014 Inference Competition: 126/9/18 above/below/close to the diagonal dashed line

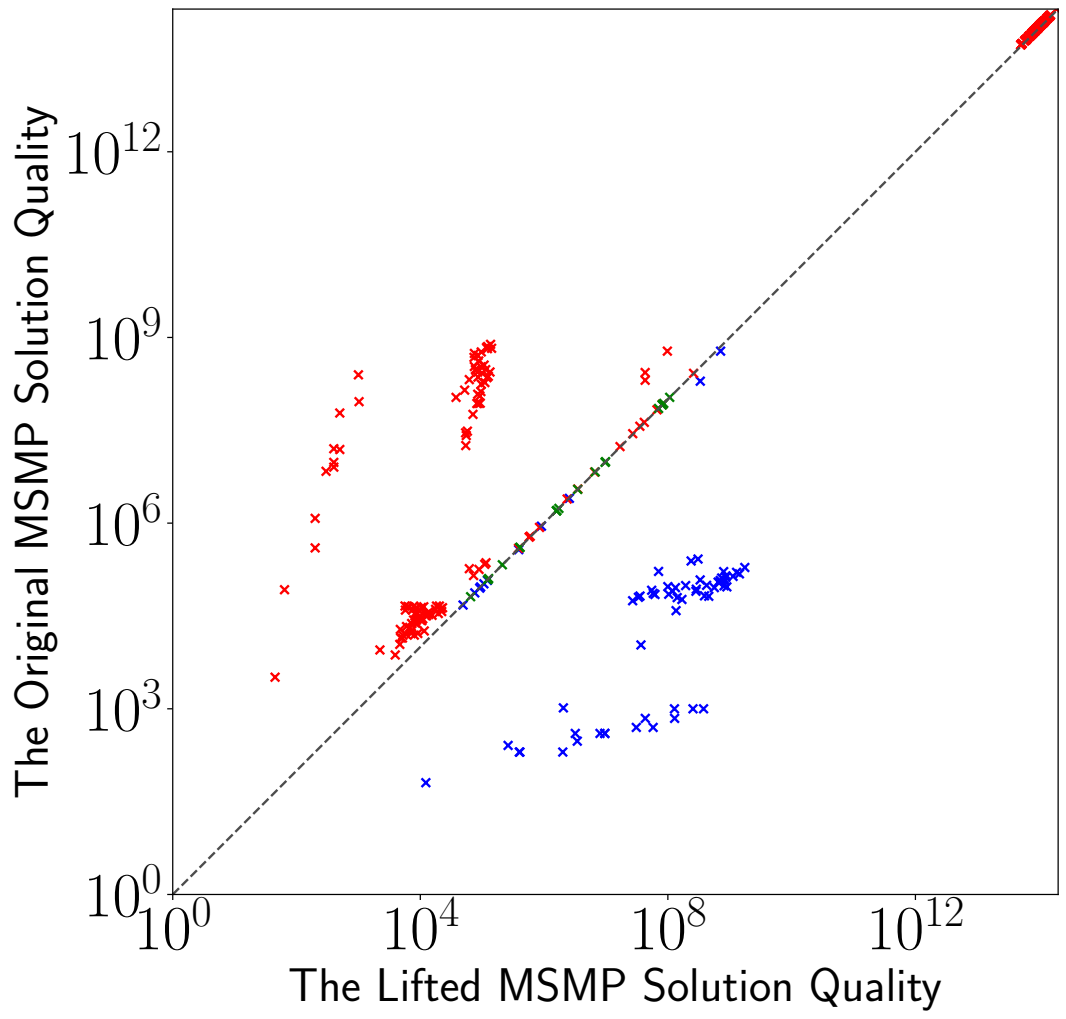

(b) Benchmark instances from (Hurley et al. 2016): 222/68/19 above/below/close to the diagonal dashed line

Figure 4.3: Shows the qualities of the solutions produced by the original MSMP algorithm in direct comparison to those produced by the lifted MSMP algorithm for both sets of benchmark instances. Each point in these plots represents a benchmark instance. The x and y coordinates of a benchmark instance represent the solution qualities produced by the lifted MSMP algorithm and the original MSMP algorithm, respectively. Benchmark instances above (red)/below (blue) the diagonal dashed line have better/worse solution qualities when using the lifted MSMP algorithm instead of the original MSMP algorithm. Benchmark instances whose MSMP solution qualities differ by only 1% are considered close (green) to the diagonal dashed line.

## 4.4 Experimental Evaluation

In this section, we present experimental evaluations of the lifted MSMP algorithm. We used two sets of Boolean WCSP benchmark instances for our experiments. The first set of benchmark instances is from the UAI 2014 Inference Competition[1]. Here, maximum a posteriori (MAP) inference queries with no evidence on the PR and MMAP benchmark instances can be reformulated as Boolean WCSP instances by first taking the negative logarithms of the probabilities in each factor and then normalizing them. The second set of benchmark instances is from (Hurley et al. 2016)[2]. This set includes the Probabilistic Inference Challenge 2011, the Computer Vision and Pattern Recognition OpenGM2 benchmark, the Weighted Partial MaxSAT Evaluation 2013, the MaxCSP 2008 Competition, the MiniZinc Challenge 2012 & 2013 and the CFLib (a library of cost function networks). The experiments were performed on those benchmark instances that have only Boolean variables.

[1] `http://www.hlt.utdallas.edu/~vgogate/uai14-competition/index.html`

The optimal solutions of the benchmark instances in (Hurley et al. 2016) were computed using toulbar2 (Hurley et al. 2016). Since toulbar2 cannot solve WCSP instances with non-integral weights, the optimal solutions of the benchmark instances from the UAI 2014 Inference Competition were computed by finding MWVCs on their CCGs. For each benchmark instance, the MWVC problem was solved by first kernelizing it using the NT reduction, then reformulating it as an ILP (H. Xu, Kumar, and Koenig 2016) and finally solving the ILP using the Gurobi optimizer (Gurobi Optimization, Inc. 2018) with a running time limit of 5 minutes.

For the MSMP algorithms, we set the initial values of all messages to zeros. If no message changed by an amount more than $10^{-6}$ in any iteration, we declared convergence. We used the synchronous message updating order, i.e., messages were updated in parallel in each iteration. This standardized the comparison between the two MSMP algorithms, factoring out the effects of different message updating orders within each iteration. In case of failure to converge within the time limit (5 minutes) for any benchmark instance, we reported the solution produced by the MSMP algorithm on that benchmark instance at the end of that time limit.

[2] `http://genoweb.toulouse.inra.fr/~degivry/evalgm/`

Table 4.1: Shows the number of benchmark instances on which each MSMP algorithm converged. The column "Neither"/"Both" indicates the number of benchmark instances on which neither/both of the MSMP algorithms converged within the time limit of 5 minutes. The column "Original"/"Lifted" indicates the number of benchmark instances on which only the original/lifted MSMP algorithm converged.

| Benchmark Instance Set | Neither | Both | Original | Lifted |
|---|---|---|---|---|
| UAI 2014 Inference Competition | 25 | 4 | 124 | 0 |
| (Hurley et al. 2016) | 258 | 7 | 44 | 0 |

The CCG construction algorithm and the MSMP algorithms were implemented in C++ using the Boost graph library (Siek, Lee, and Lumsdain 2002) and were compiled by gcc 4.9.2 with the "-O3" option. Our experiments were performed on a GNU/Linux workstation with an Intel Xeon processor E3-1240 v3 (8MB Cache, 3.4GHz) and 16GB RAM.

Figure 4.2 shows the qualities of the solutions (total weights) produced by the original MSMP algorithm versus the lifted MSMP algorithm in comparison to the optimal solutions. A significant fraction of the solutions produced by the lifted MSMP algorithm are very close to the optimal solutions. However, both MSMP algorithms produced solutions that are highly suboptimal in the $> 30\%$ suboptimality range. Therefore, Figure 4.3 presents a direct comparison of the qualities of the solutions produced by the two MSMP algorithms. From this figure, it is evident that solution qualities of the lifted MSMP algorithm are significantly better than those of the original MSMP algorithm.

Table 4.1 shows the number of benchmark instances on which each MSMP algorithm converged within the time limit. Table 4.2 shows the convergence time and number of iterations for those benchmark instances on which both algorithms converged. Although the original MSMP algorithm converged more frequently and faster, the lifted MSMP algorithm produced better solutions in general. In addition, both MSMP algorithms are anytime and can be easily implemented in distributed settings. Therefore, the comparison of the qualities of the solutions produced is more important than that of the frequency and speed of convergence.

Table 4.2: Shows the number of iterations and running time for each of the benchmark instances on which both MSMP algorithms converged within the time limit of 5 minutes. The column "Benchmark Instance" indicates the name of each benchmark instance. The "U:" and "T:" at the beginning of the names indicate that they are from the UAI 2014 Inference Competition and (Hurley et al. 2016), respectively. The columns "Iterations" and "Running Time" under "The Original MSMP" and "The Lifted MSMP" indicate the number of iterations and running time (in seconds) after which the original MSMP algorithm and the lifted MSMP algorithm converged, respectively. With a few exceptions, the number of iterations and running time for the original MSMP algorithm are in general smaller than those of the lifted MSMP algorithm.

| Benchmark Instance | The Original MSMP | | The Lifted MSMP | |
|---|---|---|---|---|
| | Iterations | Running Time | Iterations | Running Time |
| U:PR/relational_2 | 5 | 0.84 | 9 | 4.00 |
| U:PR/ra.cnf | 1 | 0.35 | 6 | 0.34 |
| U:PR/relational_5 | 5 | 1.18 | 3 | 0.76 |
| U:PR/Segmentation_12 | 9 | 0.04 | 44 | 0.14 |
| T:MRF/Segmentation/4_30_s.binary | 31 | 0.10 | 60 | 0.13 |
| T:MRF/Segmentation/2_28_s.binary | 9 | 0.05 | 44 | 0.11 |
| T:MRF/Segmentation/18_10_s.binary | 15 | 0.07 | 102 | 0.18 |
| T:MRF/Segmentation/12_20_s.binary | 31 | 0.13 | 50 | 0.14 |
| T:MRF/Segmentation/11_3_s.binary | 47 | 0.15 | 176 | 0.24 |
| T:MRF/Segmentation/1_28_s.binary | 35 | 0.11 | 60 | 0.14 |
| T:MRF/Segmentation/3_20_s.binary | 31 | 0.12 | 54 | 0.14 |

To further understand the lifted MSMP algorithm, we also experimented on small random benchmark instances. We generated 9 groups of benchmark instances. In each group, we generated 100 Boolean WCSP instances with 50 variables. For every pair of variables, we add a constraint between them with probability $p$ (referred to as the *constraint density*), which equals 0.05, 0.1, 0.15, 0.2, 0.25, 0.3, 0.5, 0.7, and 0.9, respectively, in the 9 groups. In each tuple of each constraint, we assign the weight with an integer randomly chosen between 0 and 100. We set a running time limit of 30 seconds for each benchmark instance.

Table 4.3: Comparison of suboptimalities of the two MSMP algorithms on small random benchmark instances. Each row represents the experimental results of one group. "Original" indicates the median suboptimality of the solutions produced by the original MSMP algorithm. "Lifted" indicates the median suboptimality of the solutions produced by the lifted MSMP algorithm.

| $p$ | Original | Lifted |
|---|---|---|
| 0.05 | 0.63 | 0.00 |
| 0.1 | 0.39 | 0.08 |
| 0.15 | 0.30 | 0.21 |
| 0.2 | 0.26 | 0.23 |
| 0.25 | 0.22 | 0.21 |
| 0.3 | 0.19 | 0.20 |
| 0.5 | 0.16 | 0.16 |
| 0.7 | 0.12 | 0.13 |
| 0.9 | 0.11 | 0.10 |

Table 4.3 shows the suboptimalities of the solutions produced by the original and lifted MSMP algorithms. On these benchmark instances, the lifted MSMP algorithm in general produced solutions that are closer to optimal than the original MSMP algorithm, especially for $p \leq 0.1$. In addition, the lifted MSMP algorithm produced more optimal solutions when $p$ is closer to either 0 or 1. On the other hand, the original MSMP algorithm produced more optimal solutions as $p$ increases.

Table 4.4 shows our experimental results of comparing the original and lifted MSMP algorithms directly. The lifted MSMP algorithm works better in terms of solution qualities when the constraint density is small, but this difference becomes smaller as the constraint density increases. In terms of running times, the original MSMP algorithm overall has an advantage (for $p \geq 0.1$) over the lifted MSMP algorithm due to the fact that the original MSMP algorithm converged on more benchmark instances. This advantage increases as $p$ increases. This is consistent with the experimental results shown in Tables 4.1 and 4.2.

Table 4.4: Comparison of solution qualities and running times of the two MSMP algorithms on small random benchmark instances. In both tables, each row represents the experimental results of one group. (Upper Panel): "Original" indicates the number of benchmark instances on which the original MSMP algorithm won. "Lifted" indicates the number of benchmark instances on which the lifted MSMP algorithm won. "Tie" indicates the number of benchmark instances on which the two MSMP algorithms reached a tie. (Lower Panel): "Original" indicates the number of benchmark instances on which the original MSMP algorithm converged within the running time limit. "Lifted" indicates the number of benchmark instances on which the lifted MSMP algorithm converged within the running time limit.

| $p$ | Solution Quality | | | Running Time | | |
|---|---|---|---|---|---|---|
| | Original | Lifted | Tie | Original | Lifted | Tie |
| 0.05 | 0 | 100 | 0 | 24 | 22 | 54 |
| 0.1 | 1 | 99 | 0 | 56 | 4 | 40 |
| 0.15 | 13 | 87 | 5 | 60 | 0 | 40 |
| 0.2 | 19 | 76 | 5 | 58 | 0 | 42 |
| 0.25 | 25 | 54 | 21 | 54 | 0 | 46 |
| 0.3 | 28 | 35 | 37 | 55 | 0 | 45 |
| 0.5 | 28 | 29 | 43 | 44 | 0 | 56 |
| 0.7 | 28 | 25 | 47 | 54 | 0 | 46 |
| 0.9 | 26 | 28 | 46 | 50 | 0 | 50 |

| $p$ | Original | Lifted |
|---|---|---|
| 0.05 | 92 | 89 |
| 0.1 | 67 | 32 |
| 0.15 | 60 | 4 |
| 0.2 | 58 | 0 |
| 0.25 | 54 | 0 |
| 0.3 | 55 | 0 |
| 0.5 | 44 | 0 |
| 0.7 | 54 | 0 |
| 0.9 | 50 | 0 |

## 4.5 Discussion

Similar to other message passing algorithms, a theoretical analysis on the original and lifted MSMP algorithms is hard to carry out. Therefore, we provide some intuitions that might explain why the lifted MSMP algorithm works better than the original MSMP algorithm.

- Direct effects: Equation (4.21) is much simpler than Equations (4.19) and (4.20). First, the lifted MSMP algorithm needs only one equation, instead of two equations, for each message update. Second, the size of each message is only one real number in the lifted MSMP algorithm instead of two real numbers. These significantly increase the efficiency for updating messages.

- Structural effects: The CCG provides a much simpler topological structure for message updating. The original MSMP algorithm creates an additional constraint vertex in the factor graph for each constraint, which needs to handle its internal constraint table during message updates. On the contrary, the lifted MSMP algorithm creates auxiliary vertices for each constraint as in the CCG, which do not by themselves consist of internal constraint tables. In other words, the lifted MSMP algorithm breaks the complexity of constraint vertices into multiple vertices in a vertex-weighted graph.

## 4.6 Conclusion

We revived the MSMP algorithm for solving the Boolean WCSP by applying it on its CCG instead of its original form. We observed not only that the lifted MSMP algorithm produced solutions that are close to optimal for a large fraction of benchmark instances, but also that, in general, it produced significantly better solutions than the original MSMP algorithm. Although the lifted MSMP algorithm requires slightly more work in each iteration since the CCG is constructed using auxiliary variables, the size of the CCG is only linear in the size of the tabular representation of the WCSP (Kumar 2008a; Kumar 2008b; Kumar 2016), and the lifted MSMP algorithm has the benefit of producing better solutions. Finally, we experimentally compared the two MSMP algorithms on small random benchmark instances with different constraint densities. We found that the lifted MSMP algorithm is more advantageous on benchmark instances with smaller constraint densities, and has almost the same effectiveness as the original MSMP algorithm when the constraint density becomes larger.

There are a number of implications of a better MSMP algorithm for the (Boolean) WCSP. (a) In distributed optimization such as DCOPs, the MSMP algorithm is a state-of-the-art algorithm[3]. While there exist a number of improved variants of the MSMP algorithm such as splitting (Ruozzi and Tatikonda 2013) and damping (Cohen and Zivan 2017), to the best of our knowledge, their changes to the standard MSMP algorithm are relatively minor. On the contrary, the idea of applying the MSMP algorithm on the CCG of a given WCSP instance affects a major change to the standard MSMP algorithm. If it can be further proved to be more useful than the standard MSMP algorithm, we will not only create a new direction for improving and understanding the MSMP algorithm, but also potentially advance the MSMP algorithm to a new paradigm. (b) Even in a centralized setting, unlike branch-and-bound search, the MSMP algorithm has the advantage of being able to easily parallelize WCSP solving. Due to the simplicity of message update rules, they can be further implemented on GPUs as well (Grauer-Gray, Kambhamettu, and Palaniappan 2008). In an era when GPUs are getting popular quickly, a revolution to the MSMP algorithm can potentially revolutionize solving the WCSP and other optimization problems where the MSMP algorithm is applicable.

[3]The MSMP algorithm is referred to as the “Max-sum” algorithm in the DCOP community.

# Chapter 5

# Integer Linear Programming Encoding of the Boolean WCSP via the CCG

## 5.1 Introduction

There are many ways to solve a given WCSP instance. The state-of-the-art methods include best-first AND/OR search (Marinescu and Dechter 2007) and branch-and-bound search algorithms that exploit soft arc consistencies (Hurley et al. 2016). Unfortunately, none of these WCSP solvers make use of the power of *integer linear programming* (integer LP, ILP) solvers, such as the Gurobi Optimizer (Gurobi Optimization, Inc. 2018) and lp_solve (Berkelaar, Eikland, and Notebaert 2004). ILP solvers are highly optimized and are extensively used for solving problems in operations research. An efficient ILP encoding of the WCSP would therefore create a connection between constraint programming and operations research.

An ILP encoding of the WCSP can be borrowed from the probabilistic reasoning community. Here, the WCSP arises as the *max-a-posteriori* (MAP) problem.[1] Although this ILP encoding is popularly used in probabilistic reasoning (Koller and Friedman 2009, Section 13.5), it does not scale to large instances since it creates an unwieldy number of variables and constraints.

[1]A MAP problem instance on a probabilistic graphic model, such as a Belief Network, can be formulated as a WCSP instance by taking the negative logarithm on the individual probabilities.

In this chapter, we introduce a new ILP encoding of the WCSP that is based on the idea of the CCG. We refer to this encoding as the *CCG-based ILP encoding.* We compare it with the ILP encoding in (Koller and Friedman 2009, Section 13.5) and an improved version thereof for the Boolean WCSP. We refer to these two ILP encodings as the *direct and improved direct ILP encodings*, respectively. We first derive and compare the theoretical bounds on the number of variables, the number of constraints, and the number of variables in each constraint in the ILPs generated by these three ILP encodings. We show that the CCG-based ILP encoding is more advantageous in terms of these theoretical bounds. In addition, experimentally, we found that the CCG-based ILP encoding is often more efficient than the direct and improved direct ILP encodings. Finally, we establish an important theoretical property of the CCG-based ILP encoding.

## 5.2 ILP Encodings of the WCSP

In this section, we describe three methods to encode a given WCSP instance as an ILP: The direct ILP encoding, the improved direct ILP encoding, and our proposed CCG-based ILP encoding. For notational convenience, throughout this section, we consider the WCSP instance $\mathcal{B} = \langle \mathcal{X} = \{X_1, X_2, \ldots, X_N\}, \mathcal{D} = \{\mathcal{D}(X_1), \mathcal{D}(X_2), \ldots, \mathcal{D}(X_N)\}, \mathcal{C} = \{C_1, C_2, \ldots, C_M\}\rangle$.

### 5.2.1 Direct ILP Encoding

For each $C \in \mathcal{C}$ and $a \in A(S(C))$, we introduce an ILP variable $q_a^C$. Here, $A(S(C))$ is the set of all assignments of values to variables in constraint $C$ (therefore $|A(S(C))| = \prod_{X \in S(C)} |\mathcal{D}(X)|$). $q_a^C$ is either 0 or 1: If $q_a^C = 1$, then the assignment $a$ to the variables in $C$ is part of the to-be-determined optimal solution $a^*$, i.e., $a^*|S(C) = a$; otherwise it is not. The direct ILP encoding of $\mathcal{B}$ is

$$\underset{q_a^C : q_a^C \in \boldsymbol{q}}{\text{minimize}} \quad \sum_{C \in \mathcal{C}} \sum_{a \in A(S(C))} w_a^C q_a^C \tag{5.1}$$

$$\text{s.t.} \quad q_a^C \in \{0,1\} \quad \forall q_a^C \in \boldsymbol{q} \tag{5.2}$$

$$\sum_{a \in A(S(C))} q_a^C = 1 \quad \forall C \in \mathcal{C} \tag{5.3}$$

$$\sum_{a \in A(S(C)) : a|S(C) \cap S(C') = s} q_a^C = \sum_{a' \in A(S(C')) : a'|S(C) \cap S(C') = s} q_{a'}^{C'} \quad \forall C, C' \in \mathcal{C} \text{ and } s \in A(S(C) \cap S(C')), \tag{5.4}$$

where $\boldsymbol{q} = \{q_a^C \mid C \in \mathcal{C} \wedge a \in A(S(C))\}$, $w_a^C$ denotes the weight of assignment $a$ specified by constraint $C$, and $a|S(C) \cap S(C')$ is the projection of the complete assignment $a$ onto the set of common variables in $C$ and $C'$. The cardinality of $\boldsymbol{q}$ is $\sum_{C \in \mathcal{C}} \prod_{X \in S(C)} |\mathcal{D}(X)|$. Here,

- Equation (5.2) represents the ILP constraints that enforce the Boolean property for all $q_a^C$'s. It consists of $\sum_{C \in \mathcal{C}} \prod_{X \in S(C)} |\mathcal{D}(X)| = \mathcal{O}\left(|\mathcal{C}|\hat{D}^{\hat{C}}\right)$ ILP constraints, where $\hat{C} = \max_{C \in \mathcal{C}} |S(C)|$ and $\hat{D} = \max_{X \in \mathcal{X}} |\mathcal{D}(X)|$.

- Equation (5.3) represents the ILP constraints that enforce a unique assignment of values to variables in each WCSP constraint. It consists of $|\mathcal{C}|$ ILP constraints, each of which has $|A(S(C))| = \prod_{X \in S(C)} |\mathcal{D}(X)| = \mathcal{O}\left(\hat{D}^{\hat{C}}\right)$ variables.

- Equation (5.4) represents the ILP constraints which enforce that every two assignments in two WCSP constraints must be consistent on their shared variables. It consists of $\mathcal{O}\left(|\mathcal{C}|^2\hat{D}^{\hat{C}}\right)$ ILP constraints. Each of these ILP constraints has $\mathcal{O}\left(\hat{D}^{\hat{C}}\right)$ variables.

Therefore, if $\mathcal{B}$ is a Boolean WCSP instance, the direct ILP encoding has $|\boldsymbol{q}| = \mathcal{O}\left(|\mathcal{C}|\hat{D}^{\hat{C}}\right) = \mathcal{O}\left(|\mathcal{C}|2^{\hat{C}}\right)$ variables and $\mathcal{O}\left(|\mathcal{C}|^2\hat{D}^{\hat{C}}\right) = \mathcal{O}\left(|\mathcal{C}|^2 2^{\hat{C}}\right)$ ILP constraints. Each of these ILP constraints has $\mathcal{O}\left(\hat{D}^{\hat{C}}\right) = \mathcal{O}\left(2^{\hat{C}}\right)$ variables.

### 5.2.2 Improved Direct ILP Encoding

The improved direct ILP encoding is similar to the direct ILP encoding, except that Equation (5.4) is replaced by

$$\sum_{a\in A(S(C)):a|S(C')=a'} q_a^C = q_{a'}^{C'} \quad (5.5)$$
$$\forall C \in \mathcal{C}, \forall C' : |S(C')| = 1 \wedge S(C') \subset S(C), \forall a' \in A(S(C')),$$

with a dummy unary constraint—a constraint that has zero weights in all its tuples—imposed on each variable on which there is no unary constraint.

Similar to Equation (5.4), Equation (5.5) also represents the ILP constraints which enforce that every two assignments in two WCSP constraints must be consistent on their shared variables. It consists of $\mathcal{O}\left(|\mathcal{C}|\cdot|\mathcal{X}|\cdot\hat{D}^{\hat{C}}\right)$ ILP constraints. Each of these ILP constraints has $\mathcal{O}\left(\hat{D}^{\hat{C}}\right)$ variables. However, unlike Equation (5.4), which enforces this ILP constraint by considering every pair of WCSP constraints, here, it only considers each WCSP constraint with all its relevant unary WCSP constraints. This improvement effectively reduces the number of ILP constraints from $\mathcal{O}\left(|\mathcal{C}|^2\hat{D}^{\hat{C}}\right)$ to $\mathcal{O}\left(|\mathcal{C}|\cdot|\mathcal{X}|\cdot\hat{D}^{\hat{C}}\right)$.

In summary, if $\mathcal{B}$ is a Boolean WCSP instance, the improved direct ILP encoding has $|\boldsymbol{q}| = \mathcal{O}\left(|\mathcal{C}|\hat{D}^{\hat{C}}\right) = \mathcal{O}\left(|\mathcal{C}|2^{\hat{C}}\right)$ variables and $\mathcal{O}\left(|\mathcal{C}| \cdot |\mathcal{X}| \cdot \hat{D}^{\hat{C}}\right) = \mathcal{O}\left(|\mathcal{C}| \cdot |\mathcal{X}| \cdot 2^{\hat{C}}\right)$ ILP constraints. Each of these ILP constraints has $\mathcal{O}\left(\hat{D}^{\hat{C}}\right) = \mathcal{O}\left(2^{\hat{C}}\right)$ variables.

### 5.2.3 CCG-Based ILP Encoding

We can encode a WCSP instance as an ILP after transforming it to an equivalent MWVC problem instance on its CCG $G = \langle V, E, w \rangle$. The resulting CCG-based ILP encoding is

$$\underset{x_i : v_i \in V}{\text{minimize}} \quad \sum_{i=1}^{|V|} w_i x_i \tag{5.6}$$

$$\text{s.t.} \quad x_i \in \{0, 1\} \qquad \forall\, v_i \in V \tag{5.7}$$

$$x_i + x_j \geq 1 \qquad \forall\, (v_i, v_j) \in E, \tag{5.8}$$

where variable $x_i$ represents the presence of $v_i$ in the MWVC, i.e., $x_i = 1$ and $x_i = 0$ indicate that $v_i$ is and is not in the MWVC, respectively (H. Xu, Kumar, and Koenig 2016). The numbers of ILP variables and constraints are determined by the CCG. We now assume that $\mathcal{B}$ is a Boolean WCSP instance. We can compute the number of vertices and edges in the CCG by following the CCG construction procedure in (Kumar 2008a). A constraint $C$ can be represented by the multivariate polynomial

$$\sum_{T \in \mathcal{P}(S(C))} \left[ c_T \prod_{X \in T} X \right], \tag{5.9}$$

where $\mathcal{P}(S(C))$ is the power set of $S(C)$ and the $c_T$'s are constants. The CCG gadget corresponding to term $c_T \prod_{X \in T} X$ has at most 2 auxiliary vertices and $\mathcal{O}(|T|)$ edges. The CCG gadget corresponding to constraint $C$ has at most $|S(C)|$ variable vertices. Therefore, it has an upper bound of

$$\mathcal{O}\left(|S(C)| + 2 \cdot |\mathcal{P}(S(C))|\right) = \mathcal{O}\left(2^{|S(C)|+1}\right) \tag{5.10}$$

vertices and

$$\mathcal{O}\left(\sum_{T \in \mathcal{P}(S(C))} |T|\right) = \mathcal{O}\left(\sum_{|T|=0}^{|S(C)|} \binom{|S(C)|}{|T|} |T|\right) = \mathcal{O}\left(2^{|S(C)|-1}|S(C)|\right) \tag{5.11}$$

edges. Therefore, if $\mathcal{B}$ is a Boolean WCSP instance, the CCG has $\mathcal{O}\left(|\mathcal{C}|2^{\hat{C}}\right)$ vertices and $\mathcal{O}\left(|\mathcal{C}|2^{\hat{C}}\hat{C}\right)$ edges constituting the ILP variables (Equation (5.7)) and constraints (Equation (5.8)), respectively, with each of these ILP constraints having at most 2 variables.

Table 5.1: Shows the numbers of variables, constraints, and variables per constraint in the three ILP encodings of Boolean WCSP instance $\mathcal{B} = \langle \mathcal{X}, \mathcal{D}, \mathcal{C} \rangle$.

| Encoding | Direct | Improved Direct | CCG-Based |
|---|---|---|---|
| Number of Variables | $\mathcal{O}\left(\lvert\mathcal{C}\rvert 2^{\hat{C}}\right)$ | $\mathcal{O}\left(\lvert\mathcal{C}\rvert 2^{\hat{C}}\right)$ | $\mathcal{O}\left(\lvert\mathcal{C}\rvert 2^{\hat{C}}\right)$ |
| Number of Constraints | $\mathcal{O}\left(\lvert\mathcal{C}\rvert^2 2^{\hat{C}}\right)$ | $\mathcal{O}\left(\lvert\mathcal{C}\rvert \cdot \lvert\mathcal{X}\rvert \cdot 2^{\hat{C}}\right)$ | $\mathcal{O}\left(\lvert\mathcal{C}\rvert 2^{\hat{C}} \hat{C}\right)$ |
| Number of Variables per Constraint | $\mathcal{O}\left(2^{\hat{C}}\right)$ | $\mathcal{O}\left(2^{\hat{C}}\right)$ | $\leq 2$ |

### 5.2.4 Comparison

We compare various parameters of the three ILP encodings for the Boolean WCSP in Table 5.1. For any non-trivial Boolean WCSP instances, the CCG-based ILP encoding has a huge advantage over the other two ILP encodings with respect to the number of variables per constraint. This is true even if $\hat{C}$ is bounded because, in the other two ILP encodings, the number of variables in an ILP constraint corresponding to a WCSP constraint $C$ in Equation (5.3) is $2^{|S(C)|} \geq 2$. For the number of constraints, while different values of the parameters lead to different trade-offs, the most interesting real-world applications of the WCSP have a large number $|\mathcal{C}|$ of constraints and a bounded arity $\hat{C}$ of the individual constraints. Under such assumptions, the CCG-based ILP encoding is more advantageous than the other two ILP encodings with respect to the number of constraints as well. The CCG-based ILP encoding has the same asymptotic number of variables as the other two ILP encodings. In general, when $\hat{C}$ is bounded, the CCG-based ILP encoding retains the same order of the number of variables as the other two ILP encodings and significantly wins on the number of constraints and the number of variables per constraint.

## 5.3 Experimental Evaluation

In this section, we experimentally evaluate the efficiencies of solving the Boolean WCSP using the three ILP encodings. We refer to the three algorithms that use these ILP encodings as the direct algorithm, the improved direct algorithm, and the CCG-based algorithm.

We used two sets of Boolean WCSP benchmark instances for our experiments. The first set of benchmark instances is from the UAI 2014 Inference Competition[2]. Here, maximum a posteriori (MAP) inference queries with no evidence on the PR and MMAP benchmark instances can be reformulated as Boolean WCSP instances by first taking the negative logarithms of the probabilities in each factor and then normalizing them. The second set of benchmark instances is from (Hurley et al. 2016)[3]. This set includes the Probabilistic Inference Challenge 2011, the Computer Vision and Pattern Recognition OpenGM2 benchmark, the Weighted Partial MaxSAT Evaluation 2013, the MaxCSP 2008 Competition, the MiniZinc Challenge 2012 & 2013 and the CFLib (a library of cost function networks). The experiments were performed on those benchmark instances that have only Boolean variables. We set a running time limit of 120 seconds for each algorithm on the first set of benchmark instances and 15 seconds on the second set of benchmark instances.

[2] http://www.hlt.utdallas.edu/~vgogate/uai14-competition/index.html

We used the Gurobi Optimizer (Gurobi Optimization, Inc. 2018) as the ILP solver. All default settings of the Gurobi Optimizer were kept except that it was configured to use only one CPU thread. The ILP encoding procedures and the CCG construction algorithm were implemented in C++ and were compiled by the GNU Compiler Collection (GCC) 6.3.0 with the "-O3" option. We used the Boost graph library (Siek, Lee, and Lumsdain 2002) to implement the graph representations and operations. We performed our experiments on a GNU/Linux workstation with an Intel Xeon processor E3-1240 v3 (8MB Cache, 3.4GHz) and 16GB RAM.

Table 5.2 shows the number of benchmark instances on which all three algorithms terminated within the running time limits. We compare the CCG-based encoding with the other two ILP encodings individually on each set of benchmark instances. The number of benchmark instances on which only the CCG-based algorithm terminated is much larger than the number of benchmark instances on which only the direct or improved direct algorithm terminated. We also examine

[3] `http://genoweb.toulouse.inra.fr/~degivry/evalgm/`

Table 5.2: Shows the number of benchmark instances on which the direct algorithm/improved direct algorithm and the CCG-based algorithm terminated within the running time limits.

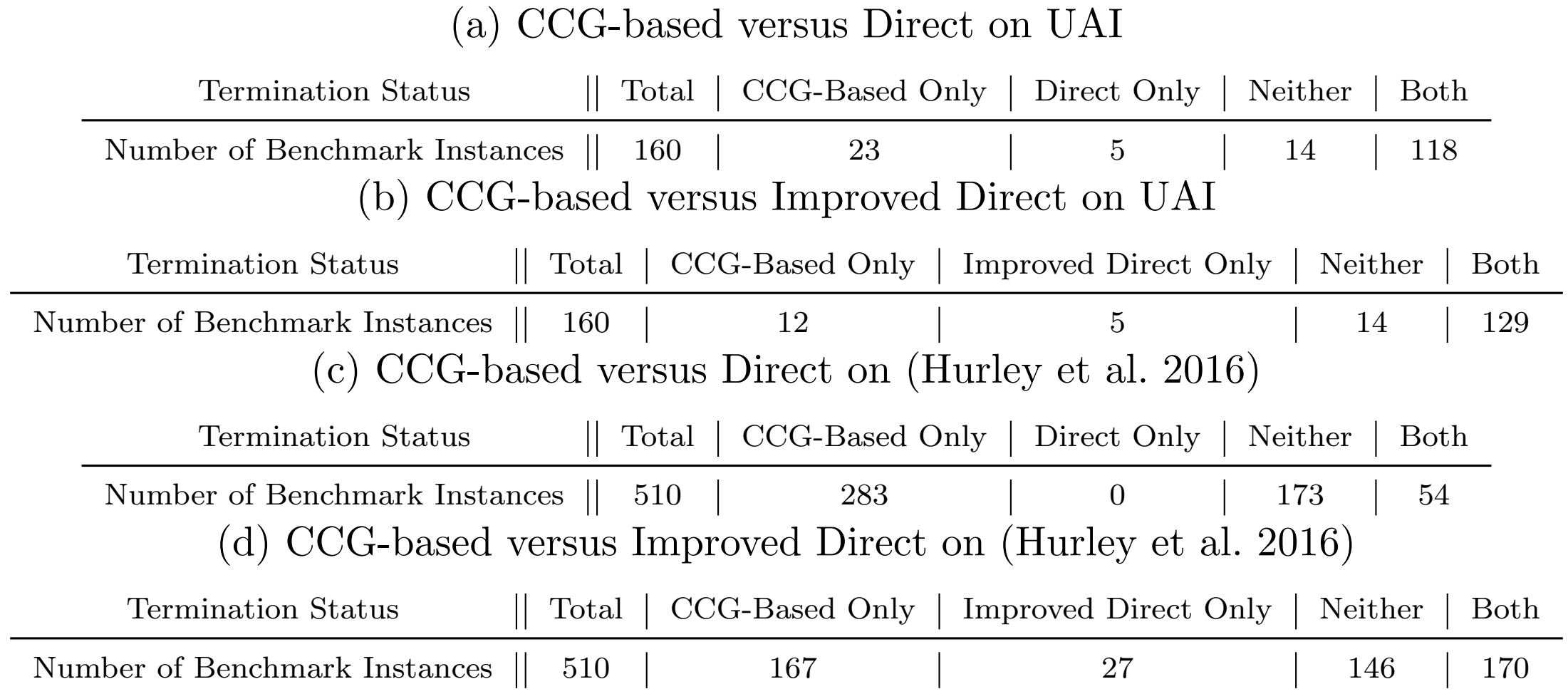

(a) CCG-based versus Direct on UAI

| Termination Status | Total | CCG-Based Only | Direct Only | Neither | Both |
|---|---|---|---|---|---|
| Number of Benchmark Instances | 160 | 23 | 5 | 14 | 118 |

(b) CCG-based versus Improved Direct on UAI

| Termination Status | Total | CCG-Based Only | Improved Direct Only | Neither | Both |
|---|---|---|---|---|---|
| Number of Benchmark Instances | 160 | 12 | 5 | 14 | 129 |

(c) CCG-based versus Direct on (Hurley et al. 2016)

| Termination Status | Total | CCG-Based Only | Direct Only | Neither | Both |
|---|---|---|---|---|---|
| Number of Benchmark Instances | 510 | 283 | 0 | 173 | 54 |

(d) CCG-based versus Improved Direct on (Hurley et al. 2016)

| Termination Status | Total | CCG-Based Only | Improved Direct Only | Neither | Both |
|---|---|---|---|---|---|
| Number of Benchmark Instances | 510 | 167 | 27 | 146 | 170 |

the benchmark instances on which both the CCG-based algorithm and the direct algorithm (or the improved algorithm) terminated.

Figure 5.1 reports the comparison of efficiencies of the directed, improved directed, and CCG-based algorithms on the benchmark instances on which both algorithms, the CCG-based and the directed or improved directed algorithms, terminated within the running time limits. The two left panels of Figure 5.1 compare the efficiencies of the CCG-based and direct algorithms on the benchmark instances on which both of them terminated within the running time limits. On the UAI benchmark instances, the CCG-based algorithm was more efficient on 110 benchmark instances (red points), and the direct algorithm was more efficient on 8 benchmark instances (blue points). On the (Hurley et al. 2016) benchmark instances, the CCG-based algorithm was more efficient on 54 benchmark instances (red points), and the direct algorithm was more efficient on no benchmark instance. For both sets of benchmark instances, most red points are far from the dashed diagonal line, meaning that the gap between the running times of the two algorithms

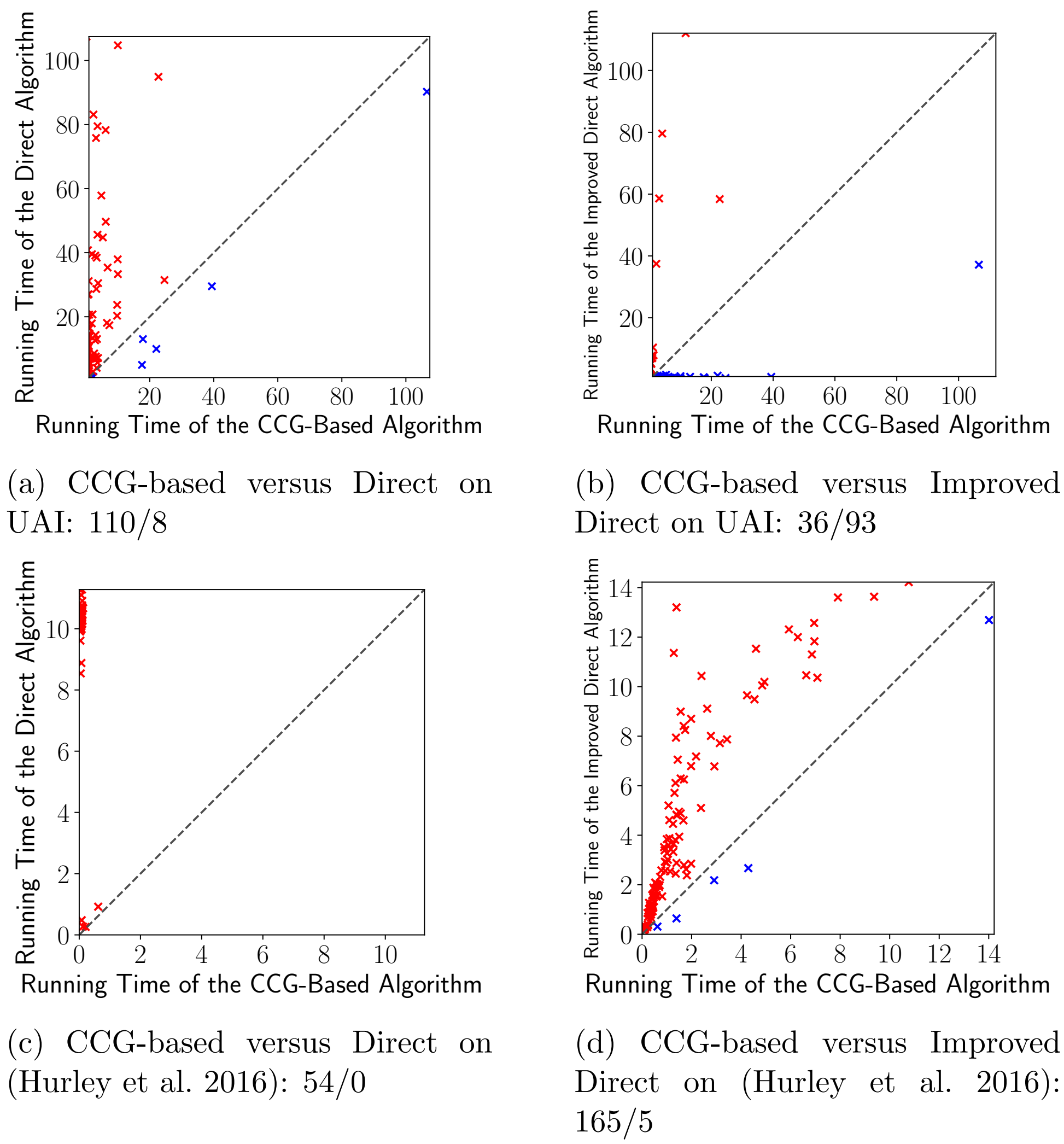

(a) CCG-based versus Direct on UAI: 110/8

(b) CCG-based versus Improved Direct on UAI: 36/93

(c) CCG-based versus Direct on (Hurley et al. 2016): 54/0

(d) CCG-based versus Improved Direct on (Hurley et al. 2016): 165/5

Figure 5.1: Compares the efficiencies of the direct, improved direct, and CCG-based algorithms on the benchmark instances on which both algorithms, the CCG-based and the direct or improved direct algorithms, terminated within the running time limits. Each point represents a benchmark instance. The x and y coordinates of each point show the running times of the CCG-based and (improved) direct algorithms on its corresponding benchmark instance, respectively. The dashed diagonal line represents equal running times. Points above and below this line are colored red and blue, respectively. Red and blue points represent benchmark instances on which the CCG-based and (improved) direct algorithms terminated more quickly, respectively. The caption of each plot shows the number of red/blue points.

was very large for those benchmark instances on which the CCG-based algorithm was more efficient. On the other hand, all blue points are close to the dashed diagonal line, meaning that the direct algorithm only marginally outperformed the CCG-based algorithm on these benchmark instances in terms of running time.

The two right panels of Figure 5.1 compare the efficiencies of the CCG-based and improved direct algorithms on the benchmark instances on which both of them terminated within the running time limit. On the UAI benchmark instances, the CCG-based algorithm was more efficient on 36 benchmark instances (red points), and the improved direct algorithm was more efficient on 93 benchmark instances (blue points). Here, contrary to the theoretical results, experimentally, the CCG-based algorithm was less efficient than the improved direct algorithm. Nevertheless, on the (Hurley et al. 2016) benchmark instances, the CCG-based algorithm was more efficient on 165 benchmark instances (red points), and the improved direct algorithm was more efficient on 5 benchmark instances (blue points). Here, the CCG-based algorithm was significantly more efficient than the improved direct algorithm.

## 5.4 A Theoretical Property of the CCG-Based ILP Encoding

Since an ILP itself can be interpreted as a WCSP instance with an infinite weight marking the violation of an ILP constraint and unary constraints representing the ILP objective function, the concept of the CCG is well defined for ILPs. It can be constructed in polynomial time for an ILP and can be used to generate the CCG-based ILP encoding of the given ILP. A desirable property of the CCG-based ILP encoding is therefore its ability to preserve the integrality of the vertices of the feasible region of its LP relaxation.

ILPs can be relaxed to LPs by removing all integrality constraints on their variables. LPs have convex feasible regions and can therefore be solved efficiently (in polynomial time). If the feasible region of the LP relaxation of an ILP has only integer vertices (equivalent to an ILP having a *totally unimodular* (TUM) constraint matrix (Sierksma 2001)), an optimal solution of the LP also yields an optimal solution of the ILP.

An ILP can be viewed as a WCSP instance as follows. Each ILP constraint translates to a WCSP constraint with weights of values zero or infinity. The ILP objective function translates to a set of unary WCSP constraints. The CCG-based ILP encoding of an ILP produces a new ILP. If the original ILP has only integer vertices in the feasible region of its LP relaxation, it is desirable for the new ILP to also have the same property. This would mean that, if the original ILP is solvable through LP relaxation, the new ILP is also solvable through LP relaxation. In this section, we show that this property of the CCG-based ILP encoding in fact holds for an important subclass of such ILPs, namely, ILPs that model MWVC problem instances on bipartite graphs.

The MWVC problem on a given vertex-weighted graph $G = \langle V, E, w \rangle$ is formulated as an ILP of the same form of Equations (5.6) to (5.8), where we simply associate a 0/1 variable $x_i$ with each vertex $v_i \in V$ of non-negative weight $w_i$ indicating the presence of $v_i$ in the MWVC. If $G$ is bipartite, its constraint matrix is TUM. Therefore, the LP relaxation of this ILP has only integer vertices in its feasible region (Sierksma 2001). We can formulate this ILP as a WCSP instance with the two types of constraints shown in Table 5.3.

| $x_i$ \ $x_j$ | 0 | 1 |
|---|---|---|
| 0 | $+\infty$ | 0 |
| 1 | 0 | 0 |

(a) The binary constraint that represents the requirement of covering each edge $(v_i, v_j) \in E$

| $x_i$ | 0 | 1 |
|---|---|---|
| Value | 0 | $w_i$ |

(b) The unary constraint for each vertex $v_i$ that represents a term in the objective function of minimizing the total weight of the vertex cover

Table 5.3: Shows the two types of WCSP constraints for the MWVC problem.

Now we show that the CCG created for the MWVC problem on any given bipartite graph is also bipartite, which establishes that the LP relaxation of the CCG-based ILP encoding has only integer vertices in its feasible region. Consider an edge $(v_i, v_j) \in E$. The CCG gadget that represents the constraint of covering this edge involves auxiliary vertices $A$ and $A'$ (Kumar 2008a). The CCG gadget itself has the edges $(v_i, A)$, $(A, A')$ and $(A', v_j)$. If the original graph is bipartite, then its vertices can be colored using either of two colors, red and blue, such that every edge connects a red vertex and a blue vertex. Without loss of generality, we assume that $v_i$ is colored red and $v_j$ is colored blue. We then color $A$ blue and $A'$ red. Such a coloring of the vertices ensures that the edges of the CCG gadgets also always connect a red vertex and a blue vertex. This means that the CCG is also bipartite. Hence, we establish the desired property of the CCG-based ILP encoding for the MWVC problem on any given bipartite graph.

## 5.5 Conclusion

In this chapter, we introduced the CCG-based ILP encoding of the WCSP. We compared it to the direct and improved direct ILP encodings adapted from the probabilistic reasoning community. We showed that the CCG-based ILP encoding has several theoretical advantages over the direct and improved direct ILP encodings. We experimentally showed that the CCG-based ILP encoding was more efficient than the direct ILP encoding. While it is less efficient than the improved ILP encoding on the UAI benchmark instances, it is more efficient on the (Hurley et al. 2016) benchmark instances. Finally, we showed that MWVC problem instances on bipartite graphs, whose corresponding ILPs have only integer vertices in the feasible regions of their LP relaxations, preserve this property in their CCG-based ILP encodings as well.

In theory, the CCG-based ILP encoding is asymptotically more efficient than the (improved) direct ILP encoding. This may be beneficial when growing problem scales in the future require us to solve extremely large WCSP instances: The difference in the asymptotic expressions may be better represented in practice as the problem sizes increase. Furthermore, branch-and-bound search algorithms for large-scale WCSP instances do not exist yet, whilst ILP solvers have been and will continue to be actively researched in large-scale settings. Therefore, reformulating a large-scale WCSP instance as an ILP and solving it can potentially become a more viable solution than branch-and-bound search for the WCSP.

Furthermore, improving ILP encoding of the WCSP can have implications for developing branch-and-bound search algorithms for the WCSP. Similar to the WCSP, the current mainstream algorithms for solving the ILP are also based on branch-and-bound search, and they have been studied for decades and have a much richer literature than branch-and-bound search for the WCSP. By building the bridge between the ILP and the WCSP, we introduced a new ILP-based perspective of WCSP solving, which can potentially inspire researchers to borrow branch-and-bound search techniques from ILP solving to solve the WCSP.

# Chapter 6

# Quantum Annealing for the WCSP via the Constraint Composite Graph

## 6.1 Introduction

Theoretical studies in physics suggest that quantum computers are inherently more efficient than classical computers, due to the unique features in *quantum processes*, such as superposition, interference and entanglement. For example, the integer factorization problem can be solved comfortably in polynomial time by Shor's algorithm (Shor 1994) on quantum computers, but is not known to admit an efficient classical algorithm.

Among all types of quantum computer hardware, the *quantum annealer* is perhaps the most widely used type nowadays due to the commercial availability of its physical realization. The quantum annealer solves combinatorial optimization problems using a quantum process called *quantum annealing*. It has been shown that quantum annealing is more advantageous than certain classical algorithms on certain classes of problems (Rieffel and Polak 2014).

In reality, quantum annealing processors have only been built by D-Wave Systems Inc. These so-called "D-Wave processors" (D-Wave Systems Inc. 2017) can only take the *Ising problem*, equivalent to the *quadratic unconstrained binary*[1] *optimization*(QUBO) problem, as its input. Therefore, to solve a combinatorial optimization problem other than the Ising problem, such as the WCSP, a reformulation process on classical computers is required. Such an algorithm is called a *hybrid quantum-classical algorithm* (HQCA). Currently, there exist many HQCAs for constraint optimization problems, such as the *maximum weighted independent set* (MWIS) problem (Choi 2008), the graph partition problem (Hen and Spedalieri 2016), the graph isomorphism problem (Hen and Sarandy 2016), and the set cover problem (Lucas 2014) as well as its generalization (Cao et al. 2016). However, due to the short history of the quantum annealer, HQCAs for constraint optimization problems still remain understudied in general. Therefore, developing HQCAs for the WCSP, a very general type of constraint optimization problem, not only facilitates WCSP solving, but also introduces HQCAs to other constraint optimization problems. In addition, HQCAs for the WCSP can enhance branch-and-bound search algorithms for solving combinatorial optimization problems (Tran et al. 2016).

In this chapter, we propose the first three HQCAs for approximately solving the WCSP. One HQCA is specifically for the binary Boolean WCSP based on the polynomial forms of constraints. The other two are for the general WCSP (where there may exist non-Boolean variables and non-binary constraints), one based on *integer linear programming* (ILP) and the other based on the CCG (Kumar 2008a; Kumar 2008b; Kumar 2016). We experimentally compare these approaches and show that while the simple polynomial reformulation works well on the binary Boolean WCSP, the CCG-based HQCA works better on the non-binary Boolean WCSP compared to the ILP-based HQCA. We note that these HQCAs are still far behind solvers on classical computers in terms of both runtime and solution optimality. Therefore, this chapter serves as a feasibility study on HQCAs for the Boolean WCSP and we hope that they can be more useful as the quantum annealer evolves and that they can intrigue future studies in this direction.

[1] Here, binary means Boolean in our terminology.

## 6.2 Quantum Annealing

Quantum annealing is a physical process that can be used to approximately solve combinatorial optimization problems. Naively, it can be understood as a meta-heuristic algorithm that makes use of features in quantum processes, such as superposition, interference, and entanglement. The expected solution optimality of quantum annealing for a given problem can be theoretically analyzed, albeit requiring methods that are too sophisticated and derivation that is too complicated to be within the scope of this chapter. In particular, the minimum gap between the energies of the ground state and the first-excited state during the quantum annealing is indicative of its solution optimality.

In practice, the D-Wave processor, a physical realization (and perhaps the most widely used realization) of the quantum annealer, solves the Ising problem, i.e., computes

$$\operatorname*{arg\,min}_{\boldsymbol{x}=\langle x_1,\ldots,x_n\rangle} \mathcal{E}(\boldsymbol{x}) = \sum_{x_i \in \boldsymbol{x}} h_i x_i + \sum_{\{x_i,x_j\}\in\mathcal{J}} J_{ij} x_i x_j, \tag{6.1}$$

where $h_i$ and $J_{ij}$ are input parameters, $\boldsymbol{x} \in \{-1,+1\}^n$, and $\mathcal{J}$ is a subset of the set of all pairs of variables in $\boldsymbol{x}$ determined by the D-Wave processor. Variables $\boldsymbol{x}$ are mapped to qubits in the processor, and parameters $h_i$ and $J_{ij}$ are mapped to interactions of each qubit with the external field and every other qubit, respectively.

| $X_i$ \ $X_j$ | 0 | 1 |
|---|---|---|
| 0 | 1 | 3 |
| 1 | 2 | 6 |

$$E_{C_{ij}}(\{X_i = (x'_i + 1)/2, X_j = (x'_j + 1)/2\}) = c_{-1,-1} + c_{+1,-1}x'_i + c_{-1,+1}x'_j + c_{+1,+1}x'_i x'_j$$

$$E_{C_{ij}}(\{X_i = 0, X_j = 0\}) = 1 \qquad E_{C_{ij}}(\{X_i = 0, X_j = 1\}) = 3$$
$$E_{C_{ij}}(\{X_i = 1, X_j = 0\}) = 2 \qquad E_{C_{ij}}(\{X_i = 1, X_j = 1\}) = 6$$

$$c_{-1,-1} = 3.0 \quad c_{+1,-1} = 1.0 \quad c_{-1,+1} = 1.5 \quad c_{+1,+1} = 0.5$$

Figure 6.1: Shows the polynomial form of the binary constraint $C_{ij}$ on the left. The top-right panel shows the polynomial form, where $c_{-1,-1}$, $c_{+1,-1}$, $c_{-1,+1}$ and $c_{+1,+1}$ are to-be-determined coefficients. The middle-right panel shows the system of linear equations that determines all coefficients. The bottom-right panel shows the coefficients after solving the system of linear equations.

## 6.3 Polynomial-based HQCA for the Binary Boolean WCSP

We can reduce the binary Boolean WCSP to the Ising problem through the construction of polynomial forms of unary and binary constraints. A unary constraint $C_i$ involving one variable $X_i$ can be rewritten in a polynomial form

$$E_{C_i}(\{X_i = (x'_i + 1)/2\}) = \frac{k_1 + k_0}{2} + \frac{k_1 - k_0}{2}x'_i, \tag{6.2}$$

where $k_0 = E_{C_i}(\{X_i = 0\})$, $k_1 = E_{C_i}(\{X_i = 1\})$ and $x'_i \in \{-1, +1\}$. A binary constraint $C_{ij}$ involving two variables $X_i$ and $X_j$ can be rewritten in a polynomial form

$$E_{C_{ij}}(\{X_i = (x'_i+1)/2, X_j = (x'_j+1)/2\}) = c_{-1,-1} + c_{+1,-1}x'_i + c_{-1,+1}x'_j + c_{+1,+1}x'_i x'_j \tag{6.3}$$

by simply solving a system of linear equations, where $x'_i, x'_j \in \{-1, +1\}$. Figure 6.1 illustrates this procedure. Finding an assignment of values to all $x'_i$'s so as to minimize the sum of the polynomial forms of all constraints is an Ising problem that is equivalent to solving the binary Boolean WCSP.

## 6.4 ILP-based HQCA

As shown in Equations (5.1) to (5.3) and (5.5), we have an ILP encoding of the WCSP improved upon (Koller and Friedman 2009, Section 13.5). Since a hybrid quantum-classical approach for a special class of ILP is known (Lucas 2014), this yields a possible HQCA for the WCSP. For notational convenience, in this subsection, we assume that, for each variable $X \in \mathcal{X}$, there exists a unary constraint $C$ such that $S(C) = \{X\}$.

We adapt the Ising formulation of a special class of ILPs (Lucas 2014) to our case as follows. The Ising formulation is divided into two parts

$$\min_{p_a^C : p_a^C \in \boldsymbol{p}} H = \alpha H_\alpha + \beta H_\beta, \tag{6.4}$$

where $p_a^C = 2q_a^C - 1 \in \{-1, +1\}$ and $\boldsymbol{p} = \{p_a^C \mid q_a^C \in \boldsymbol{q}\}$. Here, $H_\alpha$ represents ILP constraints and $H_\beta$ represents the ILP optimization goal, and $\alpha$ and $\beta$ are positive numbers.

For each ILP constraint, we add a squared term to $H_\alpha$ to represent it. The value of $H_\alpha$ is zero if all constraints are satisfied and otherwise positive:

$$\begin{aligned} H_\alpha = \sum_{C \in \mathcal{C}} \left[ \sum_{a \in A(S(C))} q_a^C - 1 \right]^2 + \sum_{\substack{C, C' \in \mathcal{C} : |S(C')| = 1 \wedge S(C') \subset S(C) \\ a' \in A(S(C'))}} \\ \left[ \sum_{a \in A(S(C)) : a|S(C') = a'} q_a^C - q_{a'}^{C'} \right]^2 . \end{aligned} \tag{6.5}$$

Here, after polynomial expansion, the quadratic terms $\left(q_a^C\right)^2$ can be merged into linear terms due to their Boolean nature, i.e., $c\left(q_a^C\right)^2 = c q_a^C$.

To characterize the ILP optimization goal (objective function), we have

$$H_\beta = \sum_{C\in\mathcal{C}} \sum_{a\in A(S(C))} w_a^C q_a^C. \tag{6.6}$$

$\alpha$ and $\beta$ only need to satisfy

$$\frac{\alpha}{\beta} > \sum_{C\in\mathcal{C}} \left[ \sum_{a\in A(S(C))} w_a^C \right]. \tag{6.7}$$

This guarantees that the minimum positive value of $\alpha H_\alpha$ is greater than the maximum value of $\beta H_\beta$, and therefore any assignment leading to a non-zero $H_\alpha$, i.e., violating at least one ILP constraint, cannot be optimal.

Combining Equations (6.4) to (6.7) and making the substitution of $q_a^C = (p_a^C + 1)/2$, we have an Ising formulation of the WCSP.

## 6.5 CCG-Based HQCA

The outline of the CCG-based HQCA is as follows: We first (a) convert the WCSP to the MWVC problem on its CCG, and then (b) approximately solve this MWVC problem using an HQCA as follows.

### 6.5.1 An HQCA for the MWVC Problem

An Ising formulation of the MWVC problem on a vertex-weighted graph $G = \langle V, E, w\rangle$ is as follows (Choi 2008). This formulation is adapted from that of the MWIS problem. For each vertex $v_i$, we associate a variable $x_i$ with it. $x_i = 0$ and $x_i = 1$ represent the presence and absence of $v_i$ in the MWVC, respectively. Then the QUBO formulation is to minimize

$$H(x_1, \ldots, x_{|V|}) = -\sum_{v_i \in V} w_i x_i + \sum_{(v_i, v_j) \in E} J_{ij} x_i x_j, \tag{6.8}$$

where $w_i$ is the weight associated with vertex $v_i$, and $J_{ij}$'s satisfy $\forall (v_i, v_j) \in E$ : $J_{ij} > \min\{w_i, w_j\}$. The minimum $H(x_1, \ldots, x_{|V|})$ (denoted by $H^*$) is the negative total weight of the MWIS on $G$—that is, $\sum_{v_i} w_i + H^*$ is the total weight of the MWVC on $G$. By making the substitution $x_i = (x'_i + 1)/2$ and $x_j = (x'_j + 1)/2$, where $x'_i, x'_j \in \{-1, +1\}$, we have an Ising formulation:

$$H'(x'_1, \ldots, x'_{|V|}) = -\sum_{v_i \in V} \frac{w_i}{2}(x'_i + 1) + \sum_{(v_i, v_j) \in E} \frac{J_{ij}}{4}(x'_i x'_j + x'_i + x'_j + 1). \tag{6.9}$$

## 6.6 Experimental Evaluation

We experimentally evaluated the efficiency and effectiveness of these three HQCAs using a D-Wave 2X processor. It is based on a physical lattice of qubits (variables in the Ising problem) and the couplers (coefficients in the Ising problem) that connect them. These qubits and couplers together are called the *Chimera graph*, as illustrated in Figure 6.2. In a D-Wave 2X processor (or any other currently available D-Wave processor), the Chimera graph is sparse. Therefore, it may not be possible to feed many Ising problem instances with dense connectivity directly into the D-Wave 2X processor. In this case, the process of *embedding* is necessary, which is to find an equivalent Ising problem instance that can be directly fed into the D-Wave 2X processor. In our experiments, for a proof of concept, we simply used the D-Wave library (D-Wave Systems Inc. 2017) to find embeddings.

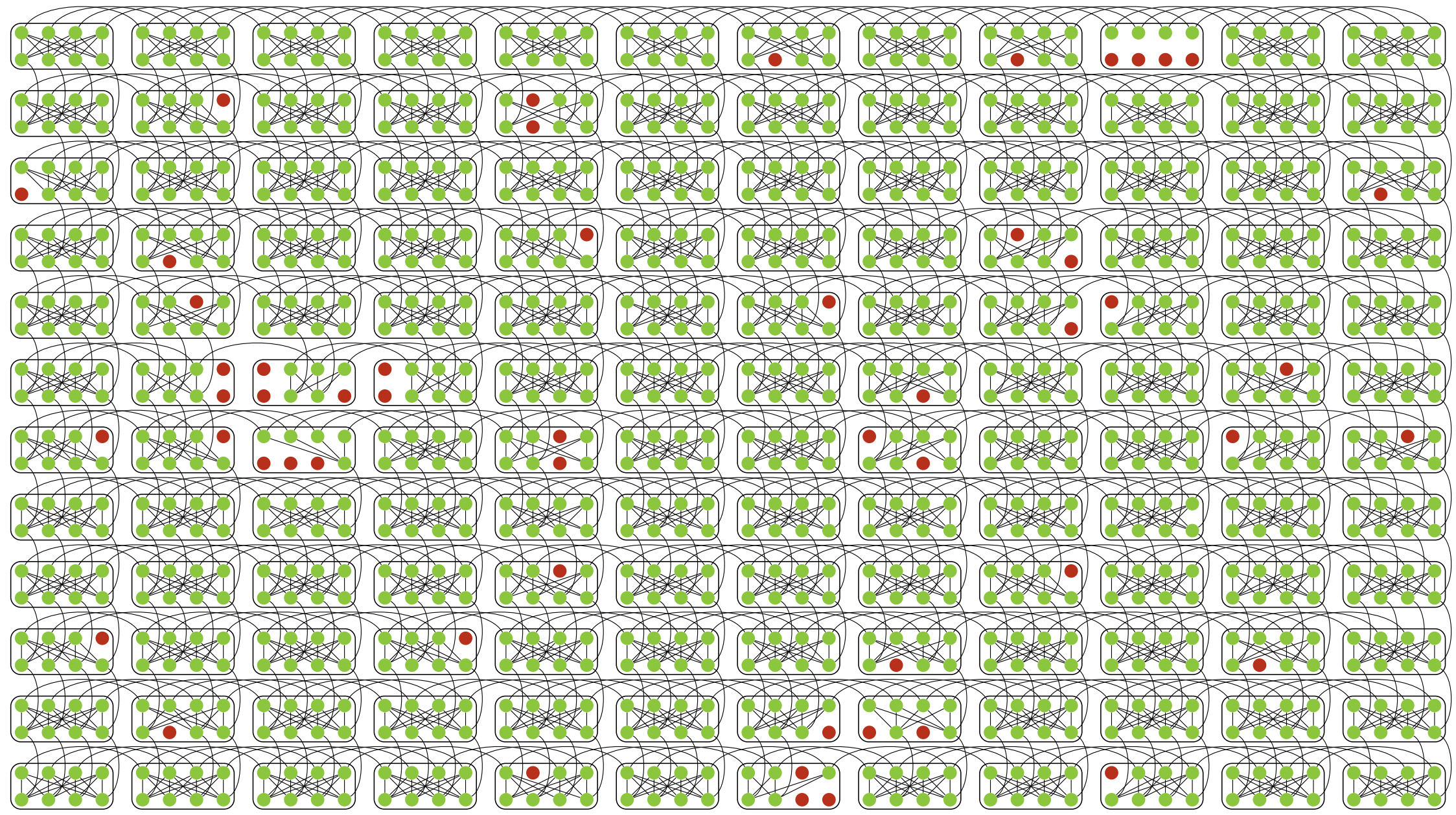

Figure 6.2: Shows the Chimera graph in a D-Wave 2X processor. The Chimera graph consists of a lattice of “imperfect” $K_{4,4}$ bipartite graph units. The green dots represent qubits and edges represent couplers. The red dots represent missing qubits in the $K_{4,4}$ units.

In our experiments, we selected real-world benchmark instances from the industrial weighted partial Max-SAT category of the Eleventh Max-SAT Evaluation[2] and reformulated them as the Boolean WCSP. We selected benchmark instances that have numbers of variables less than 30 to accommodate the limited number of qubits of the D-Wave 2X processor. Of these benchmark instances, only two (`wcsp/spot5/dir/8.wcsp.dir.wcnf` and `wcsp/spot5/log/8.wcsp.log.wcnf`) satisfy this criteria. The polynomial-based HQCA is not applicable to them because they have non-binary constraints. In addition, our experiments showed that the ILP-based HQCA could not embed any of them within the 5-minute time limit. The solutions produced by the CCG-based HQCA are 96 and 5, respectively, while the optimal solutions for both benchmark instances are 2.

[2]`http://www.maxsat.udl.cat/16/benchmarks/index.html`

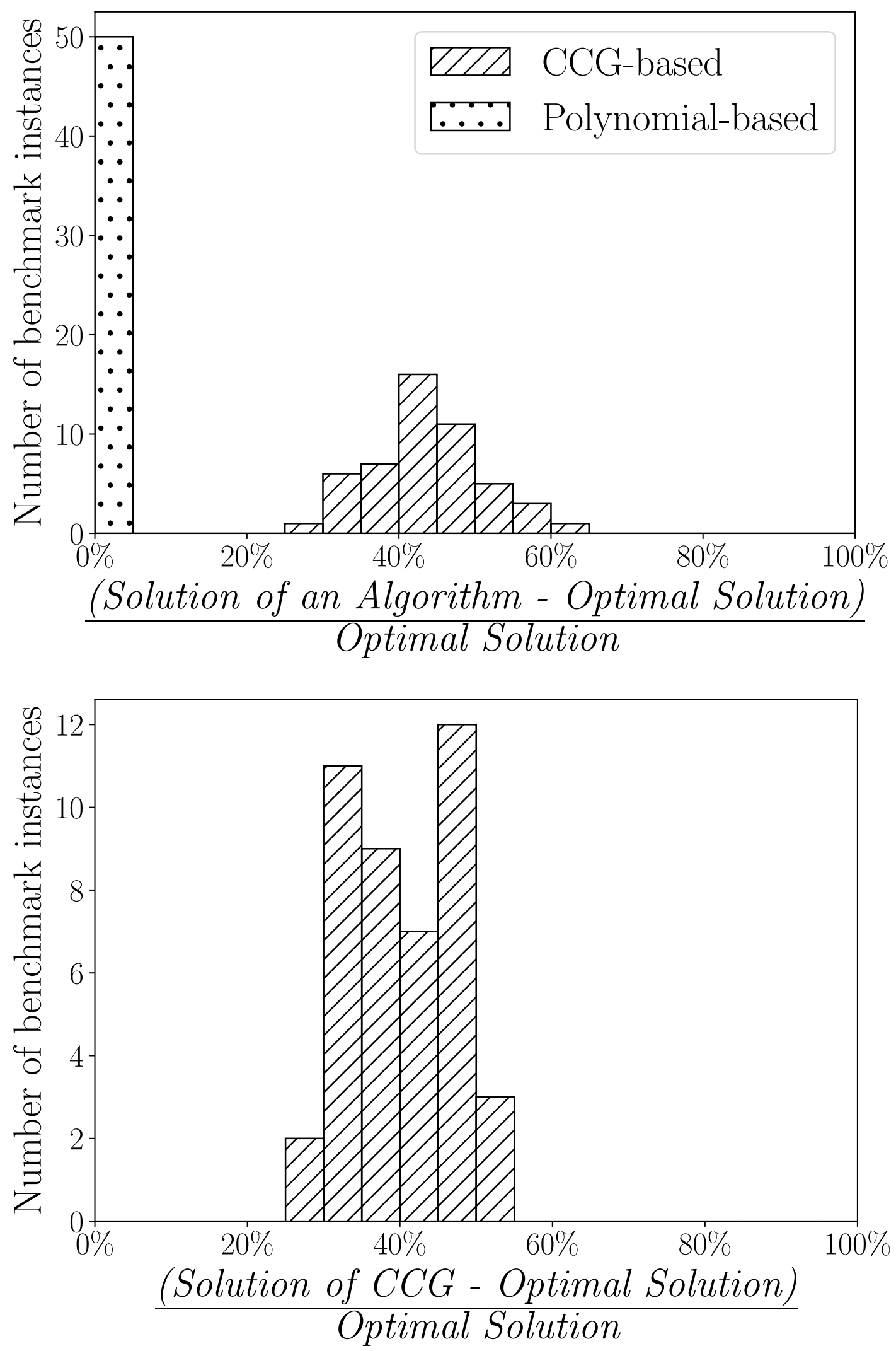

Figure 6.3: Compares suboptimalities of solutions produced by HQCAs on the two benchmark instance sets. The x-axes show the suboptimalities of the solutions produced by the CCG-based and the polynomial-based HQCAs. The y-axes show the number of benchmark instances in a range of suboptimality. The upper figure compares qualities of solutions produced by the CCG-based and the polynomial-based HQCAs with optimal solutions on the first benchmark instance set with only binary constraints. The polynomial-based HQCA produced optimal solutions on 23 out of 50 benchmark instances and the suboptimalities on all other benchmark instances are less than 10%. The lower figure compares suboptimalities of solutions produced by the CCG-based HQCA on WCSP benchmark instances from the second benchmark instance set.

Popular real-world benchmark instances, such as those in the Eleventh Max-SAT evaluation as well as those used in (Hurley et al. 2016), are too large to be embedded into a D-Wave 2X processor. For this reason, we also generated two random Boolean WCSP benchmark instance sets. (HQCAs that work on the D-Wave 2X processor will also work on larger benchmark instances on more advanced quantum annealing processors in the future.) In each benchmark instance in the first benchmark instance set, for every pair of variables, we generated a binary constraint between them with probability $p = 0.1$. We assigned a random integer weight between 0 and 100 to each tuple in these constraints. In each benchmark instance in the second benchmark instance set, we generated both binary and ternary constraints. Binary constraints were generated in the same way as in the first benchmark instance set, except with $p = 0.12$. For every triplet of variables, we also generated a ternary constraint between them with probability 0.0001. The number of variables in all benchmark instances is 50. Given the way the benchmark instances were generated, the average number of constraints that each variable participates in is about 3. We used functions `find_embedding` and `unembed_answer` from the D-Wave Python library to find embeddings and restore solutions to the original benchmark instances, respectively. For `find_embedding`, we set the time-out limit to 1000 seconds, and turned on the `fast_embedding` option for trading off fast embedding against embedding quality. For each benchmark instance, we requested the D-Wave 2X processor to run for 1000 times[3]. For all benchmark instances, we also obtained optimal solutions using `toulbar2` (Hurley et al. 2016), a state-of-the-art WCSP solver. The process of solving Ising instances was performed on a D-Wave 2X processor while other processes including finding embeddings and unembedding solutions were performed on a GNU/Linux workstation with an Intel Xeon processor E3-1240 v3 (8MB Cache, 3.4GHz) and 16GB RAM.

[3]While this may seem odd for algorithms on classical computers, it is common practice to run the quantum annealing procedure for thousands of times (and they normally terminate very quickly).

Figure 6.3 compares the qualities of solutions produced by the polynomial-based and CCG-based HQCAs with optimal solutions on the benchmark instances from both benchmark instance sets. The ILP-based HQCA could not embed any benchmark instances into the Chimera graph within the time limit and is thus not shown. The CCG-based HQCA terminated between 4 to 85 seconds on all benchmark instances. Despite 1000 runs, the running time of the D-Wave 2X processor on each benchmark instance is within 450 milliseconds. The Ising formulation processes in both HQCAs also cost insignificant amounts of time (within 60 milliseconds). The majority of the time was consumed by functions `find_embedding` and `unembed_answer`. In fact, if `find_embedding` and `unembed_answer` are not required, the efficiency and effectiveness of HQCAs can be outstanding for approximately solving the Boolean WCSP. To verify this, we generated 50 Ising problem instances, which can be seen as special cases of Boolean WCSP instances, by randomly selecting 50% of edges of the Chimera graph as constraints of Boolean WCSP instances with random integer weights. We used the D-Wave 2X processor, and `toulbar2` to solve them. We also reformulated them as weighted Max-SAT and solved them using `open-wbo` (Martins, Manquinho, and Lynce 2014). The experimental results showed that the quantum annealer produced solutions within 0.4 seconds and the qualities of the solutions were better than those produced by `toulbar2` and `open-wbo` within a 5-minute time limit.

## 6.7 Conclusion

In this chapter, we proposed the first three HQCAs for solving the WCSP: Polynomial-based, ILP-based, and CCG-based HQCAs. We evaluated them on the Boolean WCSP using experiments on a D-Wave 2X processor, a physical realization of the quantum annealer. We showed that the polynomial-based HQCA works well on the binary Boolean WCSP, but the CCG-based HQCA is not only more widely applicable, but also works better than the ILP-based HQCA on the general Boolean WCSP (where the polynomial-based HQCA is not applicable). While these HQCAs are still far behind solvers such as `toulbar2` on classical computers in terms of both runtime and solution optimality, we hope that these HQCAs become more useful as the quantum annealer evolves, and that they can serve as a starting point for future developments in using the quantum annealer for solving the WCSP.

# Chapter 7

# Promising Direction: Extending the Concept of the Constraint Composite Graph to the WCSP with Non-Boolean Variables

While we have demonstrated the advantages of the CCG in the previous few chapters, many real-world problems are hard to be efficiently modeled as the Boolean WCSP. Unfortunately, the existing literature only studied the CCG for the Boolean WCSP. This primarily stems from the fact that it is easy to reduce the Boolean WCSP to the MWVC problem on the CCG. The presence/absence of a vertex in the MWVC is used to represent a Boolean variable.

In this chapter, we extend the concept of the CCG to the WCSP with non-Boolean variables. We first give a formal definition of the CCG for the WCSP with non-Boolean variables. We then review the non-Boolean variable encoding from (Kumar 2008b), which we refer to as the *high-degree polynomial-based encoding*. We then propose three new—and more efficient—non-Boolean variable encodings, i.e., the *binary number-based encoding*, the *direct symmetric encoding*, and the *clique-based encoding*. Finally, experimentally, we preliminarily demonstrate the promisingness of the CCG for the WCSP with non-Boolean variables via quantum annealers.

| $Y_i$ \ $Y_j$ | 0 | 1 |
|---|---|---|
| 0 | 1 | 3 |
| 1 | 3 | 6 |
| 2 | 7 | 1 |

$$E_C(\{Y_i = y_i, Y_j = y_j\}) = c_{0,0} + c_{1,0}y_i + c_{2,0}y_i^2 + c_{0,1}y_j + c_{1,1}y_iy_j + c_{2,1}y_i^2y_j$$

$$E_C(\{Y_i = 0, Y_j = 0\}) = 1 \quad E_C(\{Y_i = 1, Y_j = 0\}) = 3 \quad E_C(\{Y_i = 2, Y_j = 0\}) = 7$$
$$E_C(\{Y_i = 0, Y_j = 1\}) = 3 \quad E_C(\{Y_i = 1, Y_j = 1\}) = 6 \quad E_C(\{Y_i = 2, Y_j = 1\}) = 1$$

$$c_{0,0} = 1 \quad c_{1,0} = 1 \quad c_{2,0} = 1 \quad c_{0,1} = 2 \quad c_{1,1} = 6 \quad c_{2,1} = -5$$

Figure 7.1: Shows the polynomial form of the constraint $C$ on the left. The top-right panel shows the polynomial form, where $c_{0,0}$, $c_{1,0}$, $c_{2,0}$, $c_{0,1}$, $c_{1,1}$, and $c_{2,1}$ are to-be-determined coefficients. The middle-right panel shows the system of linear equations that determines all coefficients. The bottom-right panel shows the coefficients after solving the system of linear equations.

## 7.1 Formal Definitions

We now define the CCG for the WCSP with non-Boolean variables.

**Definition 7.1.** A vertex-weighted undirected graph $G = \langle V, E, w \rangle$ is a CCG of a WCSP instance $\langle \mathcal{X}, \mathcal{D}, \mathcal{C} \rangle$ (with non-Boolean variables) if and only if there exists a subset $S \subseteq V$, to whose elements we refer as *variable vertices*, such that their presences and absences in any VC of $G$ correspond to an assignment of values to variables in the WCSP, and there exists a function $f : \mathcal{R} \to \mathcal{R}$ that maps the minimum possible weight of all VCs respecting these presences and absences to the weight of the corresponding assignment of values to variables in the original WCSP.

## 7.2 Construction of the CCG for the WCSP with Non-Boolean Variables

In this section, we review and formally describe one non-Boolean variable encoding and propose three new and more efficient non-Boolean variable encodings.

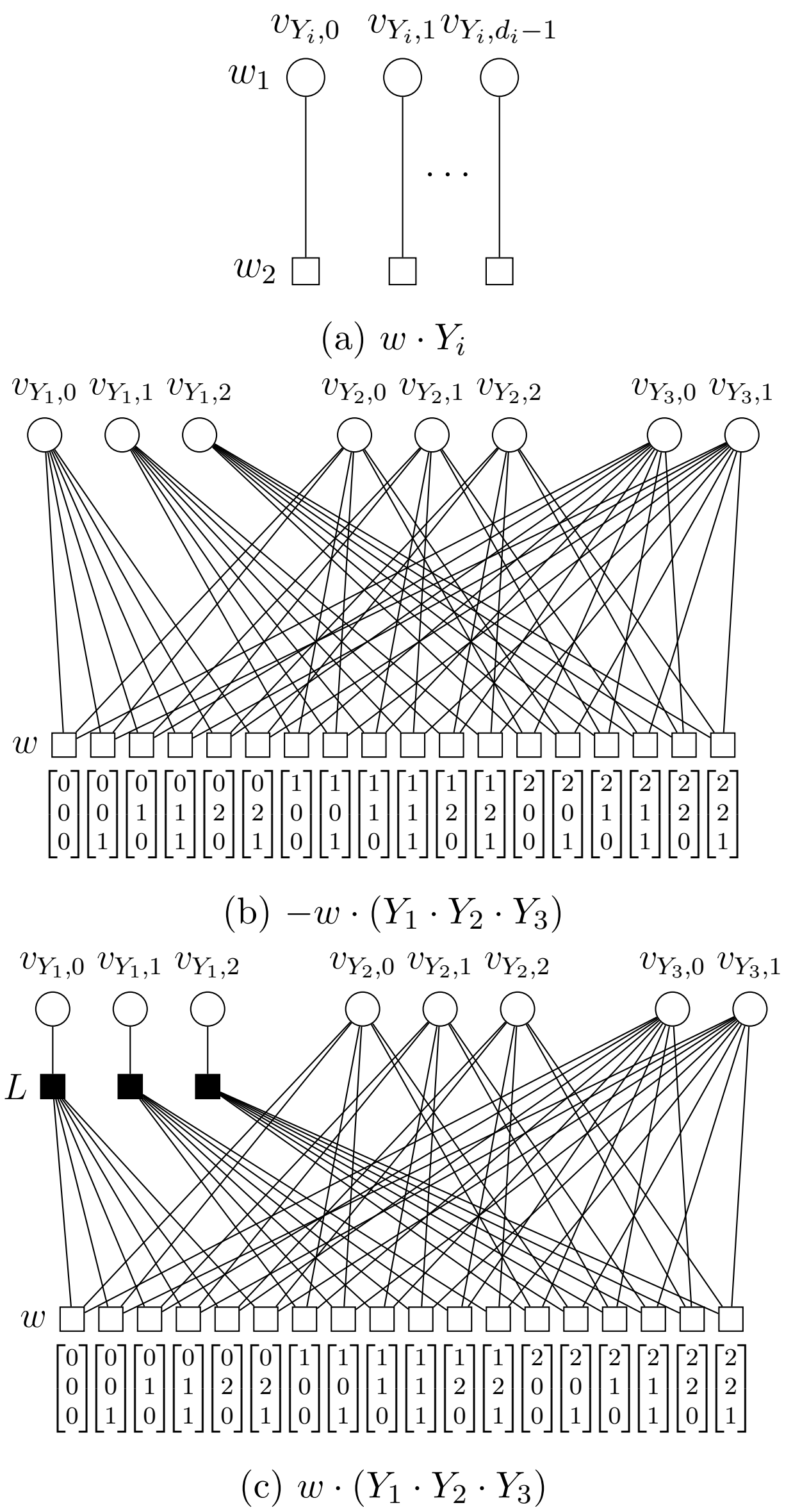

(a) $w \cdot Y_i$

(b) $-w \cdot (Y_1 \cdot Y_2 \cdot Y_3)$

(c) $w \cdot (Y_1 \cdot Y_2 \cdot Y_3)$

Figure 7.2: Illustrates the high-degree polynomial-based encoding. In (b) and (c), the variables $Y_1$, $Y_2$, and $Y_3$ are assumed to have domain sizes 4, 4, and 3, respectively. Circles represent variable vertices. Their weights are 0 in (b) and (c) (not explicitly shown). Empty and filled squares represent the auxiliary vertices that encode the coefficients and negation of variables, respectively. The triplet of numbers below each empty square indicates the variable vertices that it connects to.

### 7.2.1 High-Degree Polynomial-Based Encoding

The high-degree polynomial-based encoding was first proposed in (Kumar 2008b). It uses a high-degree polynomial to represent a constraint with non-Boolean variables (as illustrated in Figure 7.1). In this subsection, we let $d_i$ denote the domain size of a non-Boolean variable $Y_i$. Each non-Boolean variable $Y_i$ is represented by $d_i$ vertices $V_{Y_i} = \{v_{Y_i,0}, v_{Y_i,1}, \dots, v_{Y_i,d_i-1}\}$, referred to as *variable vertices*, in the CCG gadgets. The number of these vertices in the computed MWVC indicates the value of $Y_i$.

A linear term $w \cdot Y_i$, where $w$ may be either positive or negative, can be represented by $d_i - 1$ connected components, where each connected component consists of 2 connected vertices with weights of $w_1$ and $w_2$ respectively such that $w_1 - w_2 = w$. The vertices with weight $w_1$ represent $Y_i$. Figure 7.2a illustrates this.

For a negative non-linear term $-w \cdot (Y_1 \cdot Y_2 \cdot \ldots \cdot Y_m)$, where $w > 0$, we construct the CCG gadget as follows. We create a bipartite graph. The first partition consists of all and only variable vertices. In the second partition, we add $\prod_{i=1}^{m}(d_i - 1)$ auxiliary vertices with weight $w$, with each of these vertices representing an assignment of values to the variables in the term. Each auxiliary vertex connects to exactly one variable vertex of each variable. It connects to variable vertices that constitute its corresponding assignment. For example, for the term $-w \cdot (Y_1 \cdot Y_2 \cdot Y_3)$, an auxiliary vertex connected to $v_{Y_{1,0}}$, $v_{Y_{2,2}}$, and $v_{Y_{3,1}}$ corresponds to the assignment $\{Y_1 = 0, Y_2 = 2, Y_3 = 1\}$. This CCG gadget represents the term $w \cdot (\prod_{i=1}^{m}(d_i - 1) - \prod_{i=1}^{m} Y_i)$. Figure 7.2b illustrates this. Intuitively, this can be seen as follows. The $i^{\text{th}}$ variable leaves $\frac{Y_i}{d_i-1}$ of all auxiliary vertices to be potentially excluded from the MWVC, i.e., at least one adjacent edge is already covered. This leaves $\prod_{i=1}^{m} \frac{Y_i}{d_i-1}$ of all auxiliary vertices, i.e., $\prod_{i=1}^{m} Y_i$ auxiliary vertices, to be excluded from the MWVC. That is, the total weight of vertices selected in the MWVC is $w \cdot (\prod_{i=1}^{m}(d_i - 1) - \prod_{i=1}^{m} Y_i)$.

For a positive non-linear term $w \cdot (Y_1 \cdot Y_2 \cdot \ldots \cdot Y_m)$, where $w > 0$, we construct the CCG gadget as follows. We first create the CCG gadget as if $w < 0$. Then, to accommodate the positive coefficient, we split each edge that is adjacent to any variable vertex of $Y_1$ into two parts by inserting a vertex of a large weight $L$. These newly introduced vertices form a third partition, and are meant to represent the negation of the variable $Y_1$. For this reason, $L$ should be chosen such that it is greater than the sum of the weights of the vertices that it connects to, i.e., $L > w \cdot \prod_{i=2}^{m}(d_i - 1)$. This CCG gadget represents $w \cdot [\prod_{i=1}^{m}(d_i - 1) - (d_1 - 1 - Y_1) \prod_{i=2}^{m} Y_i] + L \cdot (d_1 - 1 - Y_1)$, in which the highest-degree term is the non-linear term of interest and the CCG gadgets for lower-degree terms are recursively constructed (Kumar 2008b). An illustration is shown in Figure 7.2c, which represents $w \cdot (18 - (3 - Y_1)Y_2Y_3) + L \cdot (3 - Y_1)$.

### 7.2.2 Binary Number-Based Encoding

For each non-Boolean variable $Y$ with domain size $d$, the binary number-based encoding uses $\lceil \log_2 d \rceil$ Boolean variables $\mathcal{X}_Y = \{X_{Y,1}, X_{Y,2}, \ldots, X_{Y,\lceil \log_2 d \rceil}\}$ to represent it. This converts any constraint involving this variable into a *Boolean constraint*, i.e., a constraint with only Boolean variables. The binary representation of the value of $Y$ corresponds to the values of these Boolean variables. For example, if $d = 6$, then $Y = 3$ corresponds to $X_{Y,1} = 1$, $X_{Y,2} = 1$, and $X_{Y,3} = 0$. If $\log_2 d$ is not an integer, then some assignments of values to variables in $\mathcal{X}_Y$ are forbidden, since they may represent values larger than what $Y$ can take. Continuing the above example, $\{X_{Y,1} = 1, X_{Y,2} = 1, X_{Y,3} = 1\}$ is forbidden since $Y = 7$ is not allowed. To forbid such assignments, we impose a high weight corresponding to them in the Boolean constraint. The binary number-based encoding is similar to the "log encoding" used in converting the CSP to SAT (Walsh 2000).

### 7.2.3 Direct Symmetric Encoding

For each non-Boolean variable $Y$ with domain size $d$, the direct symmetric encoding uses $d$ Boolean variables $\mathcal{X}_Y = \{X_{Y,0}, X_{Y,1}, \ldots, X_{Y,d-1}\}$ to represent it. $X_{Y,i} = 1$ and $\forall j \in \{0, 1, \ldots, d-1\} \setminus \{i\} : X_{Y,j} = 0$ together indicate $Y = i$. All other assignments of values to $X_{Y,0}, X_{Y,1}, \ldots, X_{Y,d-1}$ are forbidden via a global constraint on these $d$ variables. Constraints over non-Boolean variables are converted to constraints over these Boolean variables. This encoding is similar to the "direct encoding" used in converting the CSP to SAT (Walsh 2000).

### 7.2.4 Clique-Based Encoding

The clique-based encoding exploits the unique structure of the MWVC problem. To the best of our knowledge, this encoding does not have a counterpart in SAT encoding of the CSP. For each non-Boolean variable $Y$ with domain size $d$, the clique-based encoding uses $(d-1)$ Boolean variables $\mathcal{X}_Y = \{X_{Y,1}, X_{Y,2}, \ldots, X_{Y,d-1}\}$ to represent it. Similar to the binary number-based and direct symmetric encodings, the clique-based encoding converts any constraint $C$ involving non-Boolean variables into a Boolean constraint $C'$. $Y = 0$ corresponds to all these Boolean variables equal to 1, and $Y = y$, where $y \in \{1, 2, \ldots, d-1\}$, corresponds to $X_{Y,y} = 0$ and all other Boolean variables equal to 1. All other possible assignments of values to variables in $S(C')$ are forbidden. Since they are forbidden for representational reasons, they are referred to as being *variable-representationally forbidden*. We impose zero weight to variable-representationally forbidden assignments in $C'$, but forbid them with additional edges in the CCG gadget. In particular, in the CCG gadget, for each non-Boolean variable $Y$, we connect every pair of vertices representing Boolean variables in $\mathcal{X}_Y$ to form a clique. It is easy to see that all variable-representationally forbidden assignments of values to variables in $\mathcal{X}_Y$ correspond to only invalid VCs of the CCG gadget, but all other assignments have valid corresponding VCs.

Consider the polynomial form of $C'$ as illustrated in Figure 7.1. Since all variables in $C'$ are Boolean, this polynomial is only multi-linear, i.e., the highest-degree of any variable in $C'$ is 1. Furthermore, the polynomial form of $C'$ has only terms with degrees no less than $|S(C')| - |S(C)|$. This significantly simplifies the construction of the CCG gadget, since the procedure, as shown in (Kumar 2008a), to construct the CCG gadget considers each term of the polynomial one at a time. This property of the polynomial form of $C'$ is further leveraged in the clique-based encoding as follows. While the construction procedure in (Kumar 2008a) is straightforward for linear and negative non-linear terms, for a positive non-linear term $T$, we need to introduce a lower order term $T'$ by removing a variable from $T$. To minimize the size of the resulting CCG gadget, we always choose the variable to remove from a preset order on all variables. We create this preset order by (a) fixing an order on all variables in the WCSP instance and all Boolean variables representing each of them, and (b) concatenating these groups of ordered Boolean variables according to the order of variables in $S(C)$. Compared to arbitrary choices of the variable to remove, this scheme usually decreases the number of introduced lower order terms.

### Comparing Non-Boolean Variable Encodings

We consider a constraint consisting of $n$ variables $\{Y_1, Y_2, \ldots, Y_n\}$ in which in general no two weights are equal and all variables have domain size $d$. For the sake of theoretical analysis, we examine the asymptotic size of the CCG gadget for it with respect to either large $n$ or large $d$ (with $n \geq 2$). The number of vertices, dominated by the number of auxiliary vertices, and the number of edges can be counted as follows (for generality, we assume that lower order terms introduced during the construction of the CCG gadget do not cancel existing lower order terms):

**High-Degree Polynomial-Based Encoding** The high-degree polynomial has $n$ types of terms, among which each type involves $i = 1, 2, \ldots, n$ variables. The type of term that involves $i$ variables has $\binom{n}{i}$ combinations of participating variables. For $i$ given variables, there are $(d-1)^i$ terms, among which each term corresponds to $\Theta\left((d-1)^i\right)$ auxiliary vertices and $\Theta\left(i(d-1)^i\right)$ edges. Therefore, the numbers of vertices and edges of the CCG gadget produced by the high-degree polynomial-based encoding are $\Theta\left(\sum_{i=1}^{n}(d-1)^{2i}\binom{n}{i}\right) = \Theta\left(\left((d-1)^2+1\right)^n\right)$ and $\Theta\left(\sum_{i=1}^{n} i(d-1)^{2i}\binom{n}{i}\right) = \Theta\left(n(d-1)^2\left((d-1)^2+1\right)^{n-1}\right)$, respectively.

**Binary Number-Based Encoding** There are $\lceil\log_2 d\rceil$ Boolean variables representing each variable, and therefore there are $n\lceil\log_2 d\rceil$ variables in total. By enumerating the presence and absence of these variables in each term, we have $\binom{n\lceil\log_2 d\rceil}{i}$ terms that involve $i$ variables, among which each term consists of $\Theta(1)$ auxiliary vertices and $\Theta(i)$ edges. Therefore, there are $\Theta\left(\sum_{i=1}^{n\lceil\log_2 d\rceil}\binom{n\lceil\log_2 d\rceil}{i}\right) = \Theta\left(\bar{d}^n\right)$ vertices and $\Theta\left(\sum_{i=1}^{n\lceil\log_2 d\rceil} i\binom{n\lceil\log_2 d\rceil}{i}\right) = \Theta\left(n\lceil\log_2 d\rceil\bar{d}^n\right)$ edges in total, where $\bar{d} \geq d$ is the smallest integer such that $\log_2 \bar{d}$ is an integer.

**Direct Symmetric Encoding** There are $d$ Boolean variables representing each non-Boolean variable and each Boolean constraint consists of exactly one Boolean variable representing each non-Boolean variable, and therefore there are $d^n$ Boolean constraints of arity $n$. Now we consider the worst case, i.e., all these Boolean constraints have positive terms in their polynomial forms. For each of these Boolean constraints, we have $\mathcal{O}(n)$ auxiliary vertices and $\mathcal{O}(n^2)$ edges. Therefore, we have $\mathcal{O}(nd^n)$ vertices and $\mathcal{O}(n^2d^n)$ edges. We note that the number of vertices and edges introduced by the global constraint on the Boolean variables representing each non-Boolean variable can be neglected since they are only polynomial with respect to $d$. This is because the global constraint can be seen as an *exact* 1-*out-of-d function*, which in turn can be converted to $\mathcal{O}(d^2)$ binary Boolean constraints (Anthony et al. 2016). This leads to only $\mathcal{O}(nd^2)$ auxiliary vertices and edges.

**Clique-Based Encoding** There are $(d-1)$ Boolean variables representing each non-Boolean variable, and therefore there are $n(d-1)$ Boolean variables in total. Now we consider the worst case, i.e., all terms have positive coefficients. We follow the recursive algorithm to introduce lower order terms described in Section 7.2.4 and, without loss of generality, assume that non-Boolean variables are in the order $Y_n, Y_{n-1}, \dots, Y_1$. There are $\mathcal{O}\left((j+1)d^{i-1}\right)$ terms which consist of exactly $1 \le j \le d-1$ Boolean variables representing $Y_i$ and no Boolean variable representing $Y_{i'}$, where $i' > i$. Each term corresponds to $\Theta(1)$ auxiliary vertices and $\Theta((i-1)(d-1)+j)$ edges. Therefore, the total numbers of vertices and edges equal $\mathcal{O}\left(\sum_{i=1}^{n}\sum_{j=1}^{d-1}(j+1)d^{i-1}\right) = \mathcal{O}(d^{n+1})$ and $\mathcal{O}\left(\sum_{i=1}^{n}\sum_{j=1}^{d-1}((i-1)(d-1)+j)(j+1)d^{i-1}\right) = \mathcal{O}(nd^{n+2})$, respectively. We note that we neglect the edges connecting Boolean variables representing each non-Boolean variable, since the number $n(d-1)(d-2)/2$ of these edges is far less than $nd^{n+2}$. We also note that, in practice, since it is common that some terms have negative coefficients, the number of vertices and edges can be much lower.

Table 7.1: Compares the sizes of the CCG gadget using different non-Boolean variable encodings. We construct the CCG gadget corresponding to a constraint with $n \geq 2$ variables $\{Y_1, Y_2, \ldots, Y_n\}$ in which in general no two weights are equal and all variables have domain size $d$.

| Encoding | Vertices | Edges |
|---|---|---|
| High-Degree Polynomial-Based Encoding | $\Theta\left(\left((d-1)^2+1\right)^n\right)$ | $\Theta\left(n(d-1)^2\left((d-1)^2+1\right)^{n-1}\right)$ |
| Binary Number-Based Encoding | $\Theta\left(\bar{d}^n\right) = \mathcal{O}\left(2^n d^n\right)$ | $\Theta\left(n\lceil\log_2 d\rceil\bar{d}^n\right) = \mathcal{O}\left(n(\log_2 d)2^n d^n\right)$ |
| Direct Symmetric Encoding | $\mathcal{O}(nd^n)$ | $\mathcal{O}(n^2 d^n)$ |
| Clique-Based Encoding | $\mathcal{O}\left(d^{n+1}\right)$ | $\mathcal{O}\left(nd^{n+2}\right)$ |

Table 7.2: Shows the most advantageous non-Boolean variable encoding for different bounding on $n$ or $d$ in terms of the asymptotic numbers of vertices and edges of the produced CCG gadgets.

| Bounded | Vertices | Edges |
|---|---|---|
| $n$ but not $d$ | Binary Number-Based; Direct Symmetric | Direct Symmetric |
| $d$ but not $n$ | Clique-Based | Clique-Based |

Table 7.1 summarizes the numbers of vertices and edges in the CCG gadget for each of the four non-Boolean variable encodings. If either $d$ or $n$ is bounded, the high-degree polynomial-based encoding is the least favorable. Table 7.2 shows the advantages (in terms of asymptotic numbers of vertices and edges of the produced CCG gadgets) of the other three non-Boolean variable encodings under different settings. We see that none of them is always the best: There are trade-offs among them. The clique-based encoding has another advantage over the binary number-based and direct symmetric encodings: It does not introduce very large weights. This can be potentially helpful for avoiding numerical accuracy issues in practice.

All these non-Boolean variable encodings except for the direct symmetric encoding reduce to the same encoding for the Boolean WCSP. Despite the numbers being exponential with respect to $n$, they are all polynomial with respect to the actual input size, since the input size of the constraint, i.e., the number of input bits required to represent it, is $\Theta(d^n)$.

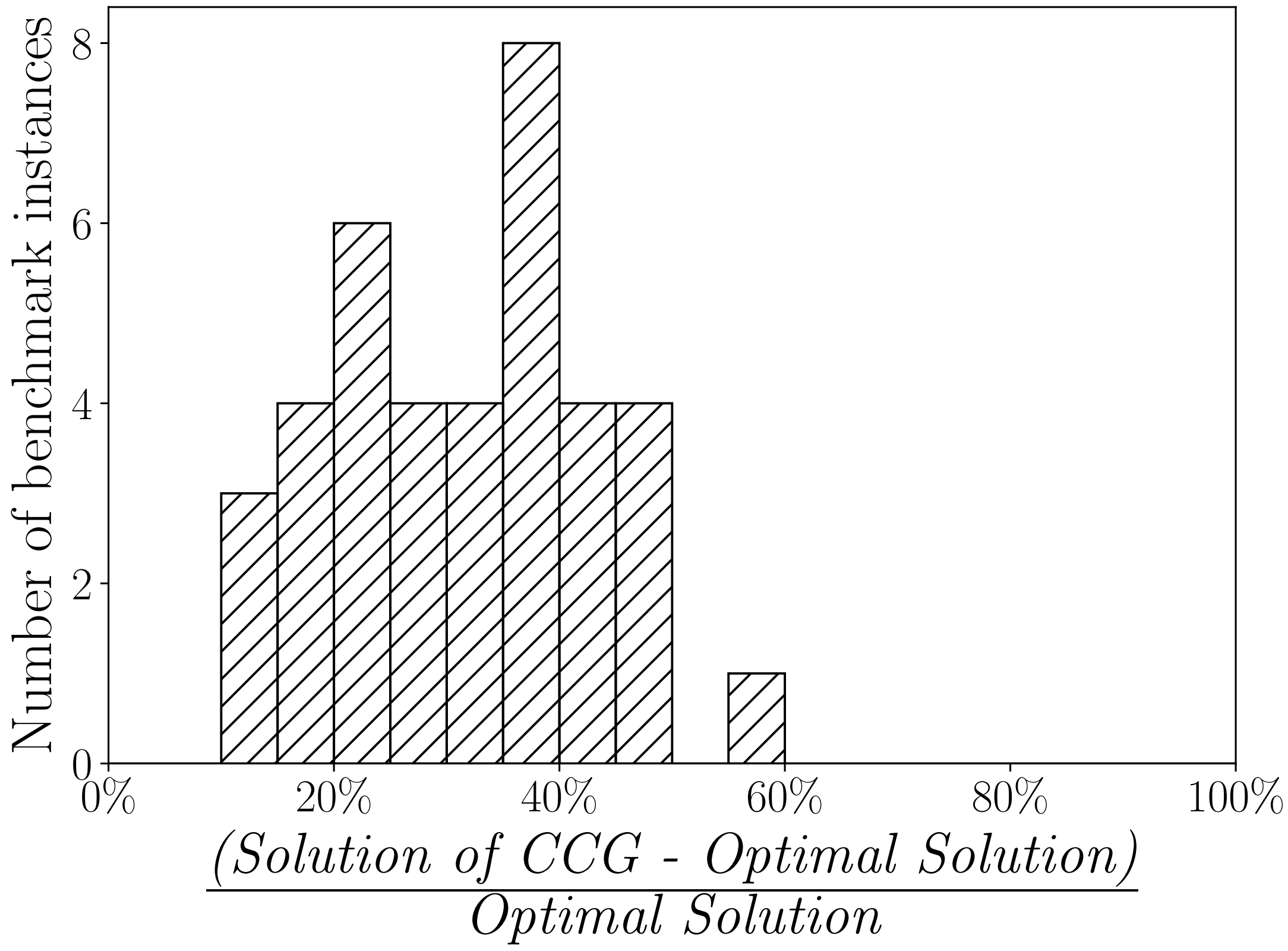

Figure 7.3: Compares suboptimalities of solutions produced by the CCG-based HQCA. The x-axis shows the suboptimalities of the solutions produced by the CCG-based HQCA. The y-axis shows the number of benchmark instances in a range of suboptimality.

## 7.3 Experimental Evaluation of the CCG-Based HQCA

We now repeat the experiments in Chapter 6 with clique-based encoding enabled in the CCG-based HQCA on a benchmark instance set with WCSP instances with non-Boolean variables. For similar reasons as in Chapter 6, we generated a random WCSP benchmark instance set. In each benchmark instance, the number of variables is 20 and the domain size of each variable is randomly set to be 2 or 3. Constraints are generated in the same way as in Chapter 6, except with $p = 0.2$.

Given the way the benchmark instances were generated, the average number of constraints that each variable participates in is about 3.

Figure 7.3 compares the qualities of solutions produced by the CCG-based HQCA with optimal solutions on the benchmark instances. Once again, within the time limit, the ILP-based HQCA could not embed any benchmark instances into the Chimera graph and is thus not shown. The polynomial-based HQCA is not shown since it is not applicable. The CCG-based HQCA successfully embedded 38 out of 50 benchmark instances between 3 and 292 seconds. Since its advantage over the ILP-based HQCA carries over to the non-Boolean WCSP, we believe that the CCG for the non-Boolean WCSP can be promising.

## 7.4 Conclusion

In this chapter, we extended the concept of the CCG to the WCSP with non-Boolean variables. We reviewed one non-Boolean variable encoding and proposed three new non-Boolean variable encodings. We compared all of them, and we concluded that, in theory, the binary number-based encoding, the direct symmetric encoding, and the clique-based encoding are more advantageous than the high-degree polynomial-based encoding, while there are trade-offs among the former three non-Boolean variable encodings under different settings. In practice, the clique-based encoding can potentially be the most preferable non-Boolean variable encoding due to its better numerical accuracy. Finally, experimentally, we preliminarily demonstrated the promisingness of the CCG for the WCSP with non-Boolean variables. We reran the experiments in Chapter 6 on WCSP benchmark instances with non-Boolean variables by introducing the clique-based encoding to the CCG construction procedure.

# Chapter 8

# Conclusion

## 8.1 Conclusion of Contributions

The WCSP is a general mathematical framework for COPs. It is known by different names in different research communities and therefore studying the WCSP brings them together. In this dissertation, we demonstrated that the CCG can be useful in both theory and practice for improving some algorithms for solving the Boolean WCSP. We also extended the concept of the CCG to the WCSP with non-Boolean variables. This dissertation makes the following contributions:

- In Chapter 3, we experimentally studied the effects of enabling the *Nemhauser-Trotter reduction* (NT reduction) on the Boolean WCSP via the CCG. This leads to a polynomial-time preprocessing algorithm that fixes the optimal values of a subset of variables in a WCSP instance. This subset can often be the set of all variables: We observed that the NT reduction could determine the optimal values of all variables for about $1/8^{\text{th}}$ of the benchmark instances without search. The enabling of the NT reduction can also be potentially meaningful for improving branch-and-bound search for the WCSP if we view the NT reducibility as a kind of implicit local consistency.

- In Chapter 4, we experimentally studied the advantages of applying the *min-sum message passing* (MSMP) algorithm to the CCG of the Boolean WCSP. We observed not only that the lifted MSMP algorithm produced solutions that are close to optimal for a large fraction of benchmark instances, but also that, in general, it produced significantly better solutions than the original MSMP algorithm. Although the lifted MSMP algorithm requires slightly more work in each iteration since the CCG is constructed using auxiliary variables, the size of the CCG is only linear in the size of the tabular representation of the Boolean WCSP (Kumar 2008a; Kumar 2008b; Kumar 2016), and the lifted MSMP algorithm has the benefit of producing better solutions. We experimentally compared the two MSMP algorithms on small random benchmark instances with different constraint densities. We found that the lifted MSMP algorithm is more advantageous on benchmark instances with smaller constraint densities, and has almost the same effectiveness as the original MSMP algorithm when the constraint density becomes larger. Furthermore, this lifted MSMP algorithm non-trivially altered the standard MSMP algorithm and may inspire, or even directly advance, the message passing algorithms to a new generation in the future. In addition, due to the parallel nature of the MSMP algorithm, it has the advantage of being able to make use of GPUs, which is harder for branch-and-bound search.

- In Chapter 5, We compared the CCG-based ILP encoding with the direct and improved direct ILP encodings adapted from the probabilistic reasoning community. We showed that the CCG-based ILP encoding has several theoretical advantages over the direct and improved direct ILP encodings. We experimentally showed that the CCG-based ILP encoding was more efficient than the direct ILP encoding. While it is less efficient than the improved ILP encoding on the UAI benchmark instances, it is more efficient on the (Hurley et al. 2016) benchmark instances. Finally, we showed that MWVC problem instances on bipartite graphs, whose corresponding ILPs have only integer vertices in the feasible regions of their LP relaxations, preserve this property in their CCG-based ILP encodings as well. Having an efficient ILP encoding for the WCSP may potentially lead to more viable large-scale WCSP solving due to the fact that ILP solvers have been actively developed and advanced for decades for large problem instances. Having an efficient ILP encoding for the WCSP may also facilitate the development of branch-and-bound search algorithms for the WCSP by introducing an ILP perspective, since the major class of algorithms for solving the ILP is also based on branch-and-bound search and has been actively advanced for decades.

- In Chapter 6, we demonstrated the advantages of solving the Boolean WCSP using the quantum annealer via the CCG. We evaluated the CCG-based HQCA on the Boolean WCSP using experiments on a D-Wave 2X processor, a physical realization of the quantum annealer. We showed that the polynomial-based HQCA works well on the binary Boolean WCSP, but the CCG-based HQCA is not only more widely applicable, but also works better than the ILP-based HQCA on the general Boolean WCSP (where the polynomial-based HQCA is not applicable). While these HQCAs are still far behind solvers such as `toulbar2` on classical computers in terms of both runtime and solution optimality, we hope that these HQCAs become more useful as the quantum annealer evolves, and that they can serve as a starting point for future developments in using the quantum annealer for solving the WCSP.

- In Chapter 7, we extended the concept and the construction of the CCG to the WCSP with non-Boolean variables and showed that it can be a potentially promising future direction. We reviewed one non-Boolean variable encoding and proposed three new non-Boolean variable encodings. We compared all of them, and we concluded that, in theory, the binary number-based encoding, the direct symmetric encoding, and the clique-based encoding are more advantageous than the high-degree polynomial-based encoding, while there are trade-offs among the former three non-Boolean variable encodings under different settings. In practice, the clique-based encoding can potentially be the most preferable non-Boolean variable encoding due to its better numerical accuracy. Experimentally, we preliminarily demonstrated the promisingness of the CCG for the WCSP with non-Boolean variables. We reran the experiments in Chapter 6 on WCSP benchmark instances with non-Boolean variables by introducing the clique-based encoding to the CCG construction procedure. If we are able to present more evidence supporting the usefulness of the CCG for the WCSP with non-Boolean variables, we would have a handy tool for exploiting the structure of the WCSP with non-Boolean variables.

## 8.2 Further Discussion

We have demonstrated that the CCG can be useful in both theory and practice for improving some algorithms for solving the Boolean WCSP. We discuss answers to the following natural questions:

**Is the CCG the holy grail for solving the Boolean WCSP?** In terms of the bipartitivity of the CCG, the short answer is no. For example, the maximum matching problem is in P, but the corresponding CCG is not necessarily bipartite and the MWVC problem on it cannot be easily identified as belonging to P.

**There exists a polynomial-time factor-2 approximation algorithm for solving the MWVC problem. Does the CCG carry over this approximability property to the WCSP?** In general, the CCG does not. The reason lies in the procedure that constructs CCG gadgets. CCG gadgets represent WCSP constraints but usually with an additional constant (such as $w$ in the case of negative nonlinear terms). This additional constant inhibits the approximability property from being carried over to the WCSP.

**For the CSP, does the CCG have any advantages or disadvantages?** Advantages: The CCG can also be used for discovering tractable subclasses of the CSP, and all methods discussed in this dissertation are also applicable. Disadvantages: The CSP is a satisfaction problem. However, using the CCG would effectively cast the problem as an optimization problem, which is in general much harder than satisfaction problems.

**How do the various aforementioned CCG-based algorithms experimentally scale with the problem size?** This dissertation does not discuss how various aforementioned CCG-based algorithms experimentally scale with the problem size. While this looks simple on the surface, here, we argue that such experiments face multiple difficulties.

- The availability of real-world WCSP instances with different sizes from the same application domain is quite limited. Therefore, experimenting this scalability on real-world problem instances is quite challenging, and requires systematic compilation of more real-world WCSP instances for meaningful results.

- When real-world problem instances are unavailable, many researchers would turn to random problem instances. However, this practice has been criticized in experimental algorithm studies in general, because random problem instances usually bear properties that are very different from real-world problem instances and the applications of a specific algorithm on these problem instances also exhibit different scalabilities. For example, different ways to generate random Boolean satifiability problem (SAT) instances lead to very different runtimes for each type of algorithm (Balyo and Chrpa 2018). Unless the random problem instances can be generated in a way that simulates real-world problem instances, which by itself commonly requires dedicated research, this disparity between real-world and random problem instances often diminishes the meaningfulness of this practice. This is also one of the major reasons why researchers carefully compile and build benchmarks for various problems.

## 8.3 Future Work

In addition to Chapter 7, concerning the essence of the CCG, in the future, we can also do the following:

- Further experiment and improve on non-Boolean variable encodings. Currently, as shown in Chapter 7, we only preliminarily experimented on our newly proposed non-Boolean variable encodings. To further study non-Boolean variable encodings, we should perform more thorough experiments on them. In addition, there may be potentially new non-Boolean variable encodings that are more advantageous on one or many aspects.

- Explore the usefulness of constructing the CCG recursively, i.e., constructing the CCG of the MWVC problem instance on the CCG associated with a WCSP instance, and so on. From the dissertation, we have already seen that the CCG offers benefits regarding the structure of COPs. Since the MWVC problem is also a kind of COP, constructing the CCG for the MWVC problem may lead to more interesting results and may potentially introduce new concepts such as a new kind of higher orders of structure.

- Explore the crown reduction (Chlebík and Chlebíková 2008), another kernelization algorithm for the MWVC problem. In this dissertation, while we have only discussed how the CCG helps solve and understand the WCSP via the NT reduction, other kernelization algorithms for the MWVC problem can also be interesting and are worth further attention.

- Develop efficient CCG representations dedicated for more specialized types of COPs such as weighted Max-SAT and weighted Max-Cut problems. The current CCG construction procedure is made for the general mathematical framework of COPs, i.e., the WCSP, and we have already demonstrated its usefulness for it. Therefore, it would be interesting to see whether specific new structure can be understood using the CCG for specialized types of COPs.

For specific algorithms applied on the WCSP, we can further:

- Develop a distributed version of the lifted MSMP algorithm using grid/cloud computing facilities. As discussed in Chapter 4, one advantage of the MSMP algorithm is that it can be easily adapted to be used in distributed settings. Therefore, it would be interesting to study and experiment the lifted MSMP algorithm in real-world distributed settings.

- Prove properties of the CCG-based ILP encoding for ILPs with TUM constraint matrices. ILPs with TUM constraint matrices have a number of nice properties, which make them tractable. Therefore, it would be useful to discover tractable subclasses of the WCSP by finding those whose CCG-based ILP encoding leads to ILPs with TUM constraint matrices.

- Use our techniques to make ILP-based approaches competitive with other approaches for solving the WCSP. As discussed in Chapter 5, the ILP-based approaches have several potential advantages over the currently mainstream branch-and-bound search algorithms. Therefore, it would be interesting to study and experiment the ILP-based approaches to attempt to demonstrate their true advantages.

- Combine the powers of state-of-the-art WCSP solvers on classical computers and the quantum annealer. For solving the WCSP, algorithms on classical computers are often inefficient but have theoretically guaranteed solution qualities, and HQCAs often produce low-quality solutions but are fast. It would be interesting if we can bring the best of both worlds together by combining these two kinds of algorithms.

- Theoretically understand HQCAs for the WCSP. While we have already experimentally showed that the CCG-based HQCA has an advantage over other HQCAs, a theoretical study of them is certainly useful in further understanding them.